\documentclass[fleqn,usenatbib]{mnras}

\usepackage{txfonts}

\usepackage[T1]{fontenc}
\usepackage{ae,aecompl}

\usepackage{graphicx}	
\usepackage{amssymb}	
\usepackage{dcolumn}
\usepackage{epsfig}
\usepackage{color}
\usepackage{longtable,lscape}
\usepackage{pdflscape}
\usepackage{upgreek}
\usepackage{mathptmx}
\usepackage{grffile}
\usepackage{array,float}
\usepackage{multirow}
\usepackage{lscape}
\usepackage[normalem]{ulem}
\usepackage{wrapfig}

\newcolumntype{d}[1]{D{.}{\cdot}{#1}}
\definecolor{mygray}{gray}{0.6}
\newcolumntype{.}{D{.}{.}{-1}}

\bibpunct[; ]{(}{)}{,}{a}{}{;}

\newcommand{\lsun}{L$_\odot$}
\newcommand{\msun}{M$_\odot$}

\newcommand{\lbol}{\emph{L}$_{\rm{bol}}$}
\newcommand{\mclump}{\emph{M}$_{\rm{clump}}$}

\newcommand{\reff}{\emph{R}$_{\rm{eff}}$}

\newcommand{\vlsr}{$v_{\rm{lsr}}$}

\newcommand{\mum}{$\umu$m}

\newcommand{\kms}{km\,s$^{-1}$}

\newcommand{\hi}{H{\sc i}}
\newcommand{\hii}{H{\sc ii}}
\newcommand{\uchii}{UC\,H{\sc ii}}

\newcommand{\poi}{Poisson}

\newcommand{\lm}{$L_{\rm{bol}}/M_{\rm{clump}}$}

\newcommand{\submm}{submillimetre}

\newcommand{\KS}{Kolmogorov-Smirnov}

\newcommand{\sag}{Sagittarius}

\newcommand{\per}{Perseus}
\newcommand{\scu}{Scutum-Centaurus}

\title[ATLASGAL --- properties of dense clumps]{ATLASGAL --- properties of a complete sample of Galactic clumps\thanks{The full version of Tables\,2, 5, 8 and A2, and Figs.\,6, 8 and A1 are only available in electronic form at the CDS via anonymous ftp to cdsarc.u-strasbg.fr (130.79.125.5) or via http://cdsweb.u-strasbg.fr/cgi-bin/qcat?J/MNRAS/.}}
\author[J.\,S.\,Urquhart et al.]{J.\,S.\,Urquhart$^{1,2}$\thanks{E-mail: j.s.urquhart@gmail.com}, C.\,K\"onig$^{2}$, A.\,Giannetti$^{3,2}$, S.\,Leurini$^{4,2}$, 
T.\,J.\,T.\,Moore$^{5}$, D.\,J.\,Eden$^{5}$, \newauthor T.\,Pillai$^{2}$, M.\,A.\,Thompson$^{6}$, C.\,Braiding$^{7}$, M.\,G.\,Burton$^{8,7}$, T.\,Csengeri$^{2}$, J.\,T.\,Dempsey$^{9}$, \newauthor C.\,Figura$^{10}$,  D.\,Froebrich$^{1}$, K.\,M.\,Menten$^{2}$, F.\,Schuller$^{11,2}$, M.\,D.\,Smith$^{1}$, F.\,Wyrowski$^{2}$\\
\\
$^{1}$ Centre for Astrophysics and Planetary Science, University of Kent, Canterbury, CT2\,7NH, UK \\
$^{2}$ Max-Planck-Institut f\"ur Radioastronomie, Auf dem H\"ugel 69, D-53121 Bonn, Germany \\
{$^{3}$ INAF - Istituto di Radioastronomia \& Italian ALMA Regional Centre, Via P. Gobetti 101, I-40129 Bologna, Italy  }\\
{\color{black}$^{4}$ INAF-Osservatorio Astronomico di Cagliari, Via della Scienza 5, I-09047, Selargius (CA)} \\
{\color{black}$^{5}$ Astrophysics Research Institute, Liverpool John Moores University, Liverpool Science Park, 146 Brownlow Hill, Liverpool, L3\,5RF, UK}\\ 
{\color{black}$^{6}$ Science and Technology Research Institute, University of Hertfordshire, College Lane, Hatfield, AL10 9AB, UK}\\
{\color{black}$^{7}$  School of Physics, University of New South Wales, Sydney, NSW 2052, Australia}\\
{\color{black}$^{8}$ Armagh Observatory, College Hill, Armagh BT61 9DG. Northern Ireland}\\
{\color{black}$^{9}$ East Asian Observatory, 660 North A'Ohoku Pl, Hilo, Hawaii, USA}\\
{\color{black}$^{10}$ Wartburg College, Waverly, IA, 50677 USA}\\
$^{11}$ IRFU, CEA, Universit\'e Paris-Saclay, F-91191 Gif-sur-Yvette, France
}

\date{Accepted XXX. Received YYY; in original form ZZZ}

\pubyear{2017}

\begin{document}
\label{firstpage}
\pagerange{\pageref{firstpage}--\pageref{lastpage}}
\maketitle

\begin{abstract}
The APEX Telescope Large Area Survey of the Galaxy (ATLASGAL) is an unbiased 870\,\mum\ submillimetre survey of the inner Galactic plane ($|\ell| < 60\degr$ with $|b|< 1.5\degr$). It is the largest and most sensitive ground-based submillimetre wavelength Galactic survey to date and has provided a large and systematic inventory of \emph{all} massive, dense clumps in the Galaxy ($\ge$1000\,\msun\ at a heliocentric distance of 20\,kpc) and includes representative samples of all of the earliest embedded stages of high-mass star formation. Here we present the first detailed census of the properties (velocities, distances, luminosities and masses) and spatial distribution of a \emph{complete} sample of $\sim$8000 dense clumps located in the Galactic disk ($5\degr < |\ell|< 60\degr$). We derive highly reliable velocities and distances to $\sim$97\,per\,cent of the sample and use mid- and far-infrared survey data to develop an evolutionary classification scheme that we apply to the whole sample. Comparing the evolutionary subsamples reveals trends for increasing dust temperatures, luminosities and line-widths as a function of evolution indicating that the feedback from the embedded proto-clusters is having a significant impact on the structure and dynamics of their natal clumps. We find that the vast majority of the detected clumps are capable of forming a massive star  and 88\,per\,cent are already associated with star formation at some level. We find the clump mass to be independent of evolution suggesting that the clumps form with the majority of their mass in-situ. We estimate the statistical lifetime of the quiescent stage to be $\sim$5$\times10^4$\,yr for clump masses $\sim$1000\,\msun\ decreasing to $\sim$1$\times10^4$\,yr for clump masses $>$10000\,\msun. We find a strong correlation between the fraction of clumps associated with massive stars and  peak column density. The fraction is initially small at low column densities but reaching 100\,per\,cent for column densities above 10$^{23}$\,cm$^{-2}$; there are no clumps with column density clumps above this value that are not already associated with massive star formation. All of the evidence is consistent with a dynamic view of star formation wherein the clumps form rapidly and are initially very unstable so that star formation quickly ensues. 

\end{abstract}




\begin{figure*}
\centering
\includegraphics[width=0.98\textwidth, trim= 0 20 0 0, clip]{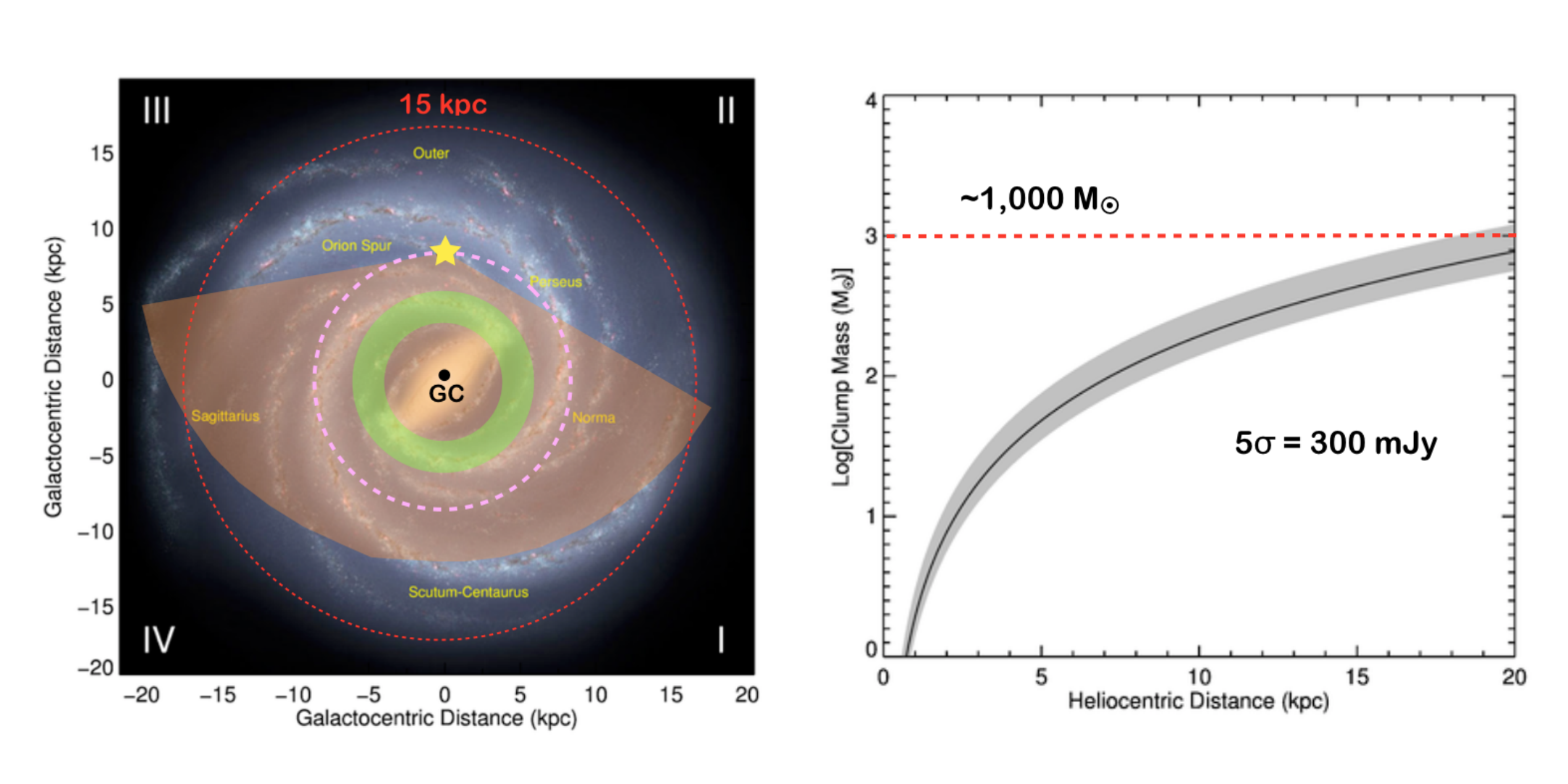}

\caption{The image shown in the left panel gives an overview of the coverage of the Galactic mid-plane provided by the ATLASGAL survey. The background is an artist's impression of how the Milky Way would appear to an external observer looking down from the Northern Galactic Pole (courtesy of NASA/JPL-Caltech/R. Hurt (SSC/Caltech)). The yellow star indicates the position of the Sun while the  orange shaded region shows the coverage of the survey out to a heliocentric distance of 20\,kpc. The Solar circle and the 5\,kpc molecular ring are shown by the pink dashed line and the light green band centred on the Galactic centre (GC).  The right panel shows the mass sensitivity limits of the survey as a function of heliocentric distance for a 5$\sigma$ flux sensitivity of 300\,mJy\,beam$^{-1}$. The black curve shows the mass sensitivity assuming a dust temperature of 20\,K while the grey shaded region shows the uncertainty, allowing $\pm$5\,K in temperature.   \label{fig:overview} } 

\end{figure*}

\section{Introduction}
\label{sect:intro}

Star formation is a Galaxy-wide phenomenon that takes place within the densest parts of giant molecular clouds (GMCs), which are themselves subject to a wide range of physical and environmental conditions (\citealt{heyer2015}). Observations of nearby spiral galaxies have revealed that their molecular gas and star formation are both tightly correlated and trace a well-defined spiral pattern (e.g., M51; \citealt{columbo2014}). Analysis of the distribution of molecular gas and star formation in the Milky Way has revealed significant peaks at specific Galactocentric radii (e.g., \citealt{moore2012,urquhart2014_rms}) and found to be concentrated along loci in longitude-velocity space (\citealt{urquhart2014_atlas}), both of which confirm the spiral structure of the Milky Way and its connection to the ongoing star formation in the Galaxy. Establishing the statistical properties of star-forming regions and their connection with Galactic structures are the aim of several programmes based on large-scale surveys of the Galactic plane (e.g.\,Hi-GAL; \citealt{Molinari2010,elia2017}).

Although there is a strong correlation between the spiral arms, the molecular material and star formation, it is not currently clear how these are connected and what role the spiral arms play in the star formation process. Theoretical models of the spiral arms predict that they play a role in the formation of GMCs and enhance the overall star formation efficiency (\citealt{dobbs2006}). The first point is supported by \citet{moore2012} who found an increase in the density of molecular clouds and star formation in the spiral arm regions.  However, these authors found that the star formation efficiency (SFE), as measured from the luminosity to mass ratio ($L/M$), was not significantly enhanced in the spiral arms compared to inter-arm regions once two of the most extreme star forming regions in the Galaxy were excluded (i.e., W49 and W51). This led  \citet{moore2012} to conclude that the increase in star formation found in the spiral arms was the result of source crowding rather than any direct influence from the arms themselves. This conclusion has been supported by \citet{eden2013,eden2015} who found little variation between the $L/M$ ratio of star forming clumps in the spiral arm and inter-arm locations along two different lines of sight.

\citet{eden2012} also estimated the efficiency with which clouds in the spiral arms and inter-arm regions convert their mass into dense clumps, which is the first stage in the star formation process. This is known as the clump formation efficiency (CFE), and was calculated from the ratio of dense gas as determined from the dust emission mapped by the Bolocam Galactic Plane Survey (BGPS; \citealt{aquirre2011_bgps}) and the cloud mass as calculated from the $^{13}$CO (1-0) data provided by the Galactic Ring Survey (GRS; \citealt{jackson2006}). Comparisons of the spiral arms and inter-arm regions revealed no significant variations between the two regions. Another more recent study investigated the star formation fraction (SFF; \citealt{ragan2016}) as a function of Galactic position. This parameter is simply the ratio of the number of Hi-GAL clumps associated with star formation (i.e., a 70\,\mum\ point source) and the total number of Hi-GAL clumps. Analysis of the distribution of the SFF as a function of Galactocentric distance revealed a modest decrease in the SFF with increasing distance from the Galactic centre but no significant increases in the SFF were observed towards the locations of the spiral arms. \citet{elia2017} find no significant differences between the average evolutionary status of sources in locations within and between arms.

The absence of any significant variations in the CFE, SFF, SFE or trend as a function of Galactocentric position suggests that although the spiral arms may play an important role in the formation and/or concentration of molecular clouds, they do not play a role in the conversion of the cloud mass into dense clumps or in the subsequent star formation associated with the clumps. However, many of the studies discussed are averaging these parameters on kiloparsec scales, and so it may transpire that there are significant underlying localised variations in the star formation efficiency measures that are largely washed out when averaged over the large segments of the spiral arms; indeed, \citet{moore2012} found significantly higher SFEs towards two of the most intense star forming regions in the Galaxy (i.e., W49 and W51; \citealt{urquhart2014_rms}).

Recent studies have revealed that a  significant fraction of the star formation taking place in our Galaxy is concentrated in 18-30 star-forming complexes  (e.g., $\sim$50\,per\,cent --- \citealt{murray2010} and $\sim$30\,per\,cent --- \citealt{urquhart2014_rms}, respectively). These have very different physical properties and environmental conditions, including some of the most extreme in the Galaxy. The remaining star formation is more evenly distributed across the Galactic plane, and while the physical properties and local environments tend to be less extreme, these regions are still responsible for a  large fraction of the overall Galactic star formation rate (50-70\,per\,cent)). In order to obtain a clear picture of the role of the spiral arms and the influence of different environments on star formation, we need to conduct a comprehensive study of the star-forming properties of a significant fraction of the Galactic plane. The recent multi-wavelength surveys of the Galactic plane (e.g., Hi-GAL \citealt{Molinari2010}, WISE \citealt{Wright2010}, MSX \citealt{price2001}, GLIMPSE {\citealt{benjamin2003_ori} and CORNISH {\citealt{hoare2012}) provide the data to conduct just such a study.

The APEX Telescope Large Area Survey of the Galaxy (ATLASGAL) is the largest and most sensitive ground-based submillimetre survey of the inner Galactic Plane (\citealt{schuller2009}; see Fig.\,\ref{fig:overview} for survey coverage and sensitivity). The primary goal of this survey is to provide a large and systematic inventory of dense  molecular clumps that includes representative samples of sources in all of the early embedded evolutionary stages associated with high-mass star formation. The Galactic distribution of dense gas has been investigated using ATLASGAL and reported by \citet{beuther2012} and \citealt{csengeri2014}. The ATLASGAL Compact Source Catalogue (CSC; \citealt{contreras2013, urquhart2014_csc}) includes $\sim$10,000 dense clumps that have been extracted from the processed emission maps. At a distance of 20\,kpc the ATLASGAL coverage includes all of the compact\footnote{ATLASGAL is sensitive to angular scales up to 2.7\arcmin\ due to spatial filtering in the reduction process; this means that it is sensitive to different physical scales at different distances (i.e., cores within a kpc, clumps out to 4-5\,kpc and small clouds at larger distances given the typical clump FWHM size of 1\arcmin; see figure\,11 of \citealt{contreras2013}). high-column density molecular gas ($N$(H$_2$) $\gtrsim$ 10$^{22}$\,cm$^{-2}$) }located within the Solar Circle and $\sim$90\,per\,cent of all dense molecular gas in the Galaxy (\citealt{urquhart2014_csc}). Assuming a dust temperature of 15\,K\footnote{This temperature is the average kinetic temperature determined from ammonia observations of quiescent clumps (i.e., \citealt{wienen2012}).}, a distance of 20\,kpc and a flux density of 0.3\,Jy\,beam$^{-1}$ ($\sim$5$\sigma$) corresponds to a minimum mass sensitivity of $\sim$1000\,\msun\,beam$^{-1}$ or $\sim$700\,\msun\,pc$^{-2}$. The ATLASGAL CSC should therefore include the vast majority of all current and future high-mass star forming clumps ($M_{\rm{clump}}>$1000\,\msun) in the Galaxy (assuming a standard initial mass function and a typical star formation efficiency of $\sim$30\,per\,cent; \citealt{lada2003}). 

Although the ATLASGAL CSC provides a large and statistically representative sample of massive dense clumps, analysis is limited as a priori distances are unknown, and these are required to derive important physical properties such as the clump mass, luminosity and Galactic distribution. In this paper we derive distances and physical properties to nearly all ATLASGAL sources with a peak flux above 5$\sigma$ and that are located away from the Galactic centre region (300\degr\ $< \ell <$ 355\degr\ and 5\degr\ $< \ell <$ 60\degr). We determine distances to all of the sources using a combination \hi\ analysis, maser parallax and spectroscopic measurements taken from the literature and a clustering analysis to group clumps into associations. We combine this catalogue of clumps with mid- and far-infrared data (Hi-GAL, WISE, MSX and GLIMPSE) to determine their dust temperature from a fit to their spectral energy distribution (SED), masses and SFE ($L$/$M$). We used these parameters to investigate variations in the SFE as a function of Galactic position and environment on physical scales of individual clumps and whole complexes.

This is the fourth paper in a series that focuses on the properties of the Galactic population of dense clumps identified by ATLASGAL. The three previous papers have focused on clumps associated with massive star formation tracers: methanol masers (\citealt{urquhart2013_methanol}; hereafter Paper\,I); \uchii\ regions (\citealt{urquhart2013_cornish}; hereafter Paper\,II); and massive young stellar objects (\citealt{urquhart2014_atlas}; hereafter Paper\,III). Combined these three papers identified $\sim$1300 massive star forming (MSF) clumps, however, this sample only make up a relatively small fraction of the total population of clumps identified by ATLASGAL ($\sim$15\,per\,cent). 

In this paper we derive the physical properties for the whole population of clumps and investigate their star forming properties and Galactic distribution.  The structure of the paper is as follows: in Section\,\ref{sect:vlsr} we give an overview of the molecular line surveys used to assign radial velocities to the clumps and describe the APEX observations that were made to increase the completeness of the velocity measurements for the ATLASGAL sample of clumps; in Section\,\ref{sect:distances} we describe the methods used to determine distances to the clumps and describe the clustering analysis used to group clumps into star forming complexes; in Section\,\ref{sect:seds} we briefly discuss the method used to fit the spectral energy distribution and estimate their current level of star formation activity; in Section\,\ref{sect:physical_properties} we derive the physical properties of the clumps and in Section\,\ref{sect:gal_distribution} we investigate the Galactic distribution of clumps and complexes and their correlation with the spiral arms; in Section\,\ref{sect:empirical_relationships} we derive empirical star formation relationships and discuss what these can tell us about the star formation process; in Section\,\ref{sect:clump_evolution} we investigate the evolution of the clumps over their lifetimes;  in Section\,\ref{sect:active_regions} we compare the properties of the most active regions in the Galaxy and assess their contribution to the Galactic dense gas reservoir and star formation; and finally in Section\,\ref{sect:conclusions} we summarize our results and present our main conclusions.

\section{Radial velocity measurements}
\label{sect:vlsr}

A first step to determining a source's distance and physical properties is its radial velocity with respect to the local standard of rest (\vlsr). This velocity can be used in conjunction with a model of the Galactic rotation to obtain a kinematic distance (see Section\,\ref{sect:distances}). The radial velocities of molecular clumps can be measured from molecular line observations (e.g., CO, NH$_3$, CS etc) and these are readily available for many of the ATLASGAL clumps from a number of Galactic plane surveys (e.g., GRS, Mopra CO Galactic plane Survey (MGPS; \citealt{burton2013,braiding2015}), ThrUMMS, (\citealt{barnes2015}), SEDIGISM (\citealt{schuller2017}), COHRS (\citealt{dempsey2013}), CHIMPS (\citealt{rigby2016})) and large targeted observational programmes towards selected samples (e.g., MALT90 (\citealt{jackson2013}), RMS (\citealt{urquhart_13co_south,urquhart_13co_north, urquhart2011_nh3,urquhart2014_rms}), BGPS (\citealt{dunham2011,schlingman2011_bgps_v, shirley2013})) as well as dedicated ATLASGAL follow-up observations (e.g., \citealt{wienen2012, csengeri2016_sio,kim2017}).

\subsection{Archival molecular line surveys}
\label{sect:archival_surveys}

We began this process of assigning velocities by matching ATLASGAL clumps with all large molecular line catalogues reported in the literature and assigning a velocity to a source where the pointing centre of the observation was found to overlap with the structure of the source. Where multiple observations were available the transition with the highest critical density (e.g., NH$_3$  and HNC over CO) was preferred as these are less affected by multiple components arising from diffuse molecular clouds lying along the same line of sight as the target source. In Table\,\ref{tbl:molecular_surveys} we present a summary of the various surveys used to assign velocities to the vast majority of ATLASGAL sources of interest to this study.

For sources where a velocity was not already available we extracted spectra directly from the various survey data cubes (i.e., MALT90, ThrUMMS and the GRS) or from spectra which were provided by the survey teams (i.e., COHRS, MGPS and CHIMPS); these data are fully reduced and calibrated and so no further processing was required. The extracted spectra were fit with Gaussian profiles using an automatic routine. The spectra and resulting fits were inspected to ensure the data quality was good and the reliability of the resulting fits. Poor quality data (e.g., those contaminated by emission in the off-source positions or strong baseline ripples) were discarded and poor reliability fits to the emission profile were refit by hand (this tended to occur in complex regions where multiple velocity components overlap with each other, however, this only affected approximately 5\,per\,cent of sources). 

When the quality of the data and fits were considered reliable they were used to assign velocities to the clumps using the following criteria: 1)\,if only a single component was detected then the peak velocity of the molecular line was assigned to the source; 2)\,if multiple components are detected but all of the strongest components (integrated line intensity) are within 10\,\kms\ of each other then the velocity of the strongest component was selected; 3)\,if multiple components were detected then the velocity of the component with the largest integrated line intensity was used, provided it was at least twice as large as the next strongest component. If none of these criteria were satisfied then no velocity was allocated and additional observations were obtained (as described in the following subsection).  In Table\,\ref{tbl:summary_vlsr} we give a summary of the surveys used and number of velocities each one provides.

\setlength{\tabcolsep}{6pt}
\begin{table*}


\begin{center}
\caption{\label{tbl:summary_vlsr} Surveys used to assign radial velocities to 7809 ATLASGAL CSC sources. \label{tbl:molecular_surveys}}
\begin{minipage}{\linewidth}
\begin{tabular}{lccl}
\hline
Survey 	& \# of  	& Molecular &Reference \\
 	& sources   	& transitions & \\

\hline\hline
ATLASGAL	&	1101	&	$^{13}$CO (2-1)/C$^{18}$O (2-1)	&	This paper	 \\
ATLASGAL	&	693	&	NH$_3$ (1,1)	&	\citet{wienen2012}	\\
ATLASGAL	&	299	&	NH$_3$ (1,1)	&	\citet{wienen2017} \\
ATLASGAL	&	154	&	N$_2$H$^{+}$ (1-0)	&	{\color{red}Urquhart et al. 2017 (in prep.)} \\
ATLASGAL (70\,\mum\,dark)	&	50	&	C$^{18}$O (2-1)/N$_2$H$^+$  (1-0)	& {\color{red}Pillai et al. 2017  (in prep.)}\\
ATLASGAL	&	62	&	N$_2$H$^{+}$ (1-0)	&	\citet{csengeri2016_sio}	\\
\hline
MALT90$^{a}$	&	1205	&	HNC/N$_2$H$^{+}$ (1-0)	&	\citet{jackson2013}	\\
BGPS	&	959	&	N$_2$H$^{+}$/HCO$^{+}$ (1-0)	&	\citet{shirley2013}	\\
ThrUMMS$^{b}$	&	960	&	CO (1-0)	&	\citet{barnes2015}	\\
COHRS	&	580	&	$^{12}$CO (3-2)	&	\citet{dempsey2013}	\\
MSGPS	&	499	&	CO (1-0)	&	\citet{burton2013}	\\
RMS	&	407	&	$^{13}$CO/NH$_3$	&	\citet{urquhart_13co_south,urquhart_13co_north,urquhart2011_nh3,urquhart2014_rms}	\\
GRS$^{c}$	&	293	&	$^{13}$CO (1-0)	&	\citet{jackson2006}	\\
BGPS	&	153	&	NH$_3$ (1,1)	&	\citet{dunham2011_bgps_vii}	\\
UC\,\hii\ cand.	&	148	&	CS (2-1)	&	\citet{bronfman1996}	\\
CHIMPS	&	83	&	$^{13}$CO (3-2)	&	\citet{rigby2016}	\\
HOPS	&	60	&	NH$_3$ (1,1)	&	\citet{purcell2012}	\\
IRDCs	&	55	&	CS (2-1)	&	\citet{jackson2008}	\\
SEDIGISM	&	43	&	$^{13}$CO (2-1)	& \citet{schuller2017}	\\
M8	&	3	&	CO (3-2)	&	\citet{tothill2002}	\\
IR-quiet clumps	&	2	&	NH$_3$ (1,1)	&	\citet{hill2010}	\\
\hline
\end{tabular}

$^{a}$ Spectra were extracted toward the submillimetre peak of each ATLASGAL source from the 3\arcmin$\times$3\arcmin\ maps of the brightest four transitions covered by this survey. These maps have been downloaded from the reduced and calibrated data products available from the Australia Telescope Online Archive (ATOA; http://atoa.atnf.csiro.au/). \\ 
$^{b}$ Source:
http://www.astro.ufl.edu/$\sim$peterb/research/thrumms/\\
$^{c}$  Source:
http://www.bu.edu/galacticring/\\
\end{minipage}
\end{center}

\end{table*}
\setlength{\tabcolsep}{6pt}

\subsection{ATLASGAL follow-up line surveys}

There have been a number of follow-up molecular line surveys that have provided velocity information for $\sim$1000 sources (e.g., ammonia observations: \citealt{wienen2012} and {\citealt{wienen2017}; 3-mm line surveys: \citealt{csengeri2016_sio, kim2017,giannetti2017}). Velocity information for a further sixty-seven 70\,$\mu$m dark sources presented here was obtained from a dedicated follow up of 70\,$\mu$m dark sources with the APEX 12\,m (C$^{18}$O 2--1) and IRAM 30\,m (N$_2$H$^+$ 1--0) telescopes (the APEX and IRAM 30\,m telescopes have similar beam sizes at these frequencies; FWHM $\sim$ 28\arcsec\ and 26\arcsec, respectively). The observations are part of a separate follow-up programme and a more detailed description will be presented in a forthcoming paper ({\color{red}Pillai et al. in prep.}).

The results of these dedicated follow-up programmes and results taken from the literature provide velocity information to many thousands of ATLASGAL sources, however, there were still approximately 1200 sources for which  no velocity measurement were available. It is necessary to obtain velocities for as many sources as possible in order to improve the completeness of our velocity coverage and the reliability of any statistical analysis based on this sample. We therefore instigated a bad-weather backup programme on the APEX telescope that observed all sources with a missing velocity assignment. These observations are reported in Appendix\,\ref{sect:obs} and provide radial velocities for  1115 clumps of which 1101 were previously unknown.

\setlength{\tabcolsep}{3pt}

\begin{table*}


\begin{center}
\caption{Summary of the kinematic distance analysis.}
\label{tbl:source_vlsr}
\begin{minipage}{\linewidth}
\small
\begin{tabular}{l.l.....cll.}
\hline \hline
\multicolumn{1}{c}{}&  \multicolumn{1}{c}{}&\multicolumn{1}{c}{}&	\multicolumn{3}{c}{Reid Distances}  && \multicolumn{3}{c}{Adopted Kinematic Solution}  & \multicolumn{1}{c}{} & \multicolumn{1}{c}{} \\
\cline{4-6}\cline{8-10}
\multicolumn{1}{c}{ATLASGAL}&  \multicolumn{1}{c}{\vlsr}&\multicolumn{1}{c}{\vlsr\ Ref.}&	\multicolumn{1}{c}{Bayesian}  & \multicolumn{1}{c}{Near}  & \multicolumn{1}{c}{Far} && \multicolumn{1}{c}{Distance} & \multicolumn{1}{c}{Solution} & \multicolumn{1}{c}{Dist.} & \multicolumn{1}{c}{Cluster} & \multicolumn{1}{c}{Cluster Dist.} \\

\multicolumn{1}{c}{CSC name}&  \multicolumn{1}{c}{(\kms)}&\multicolumn{1}{c}{}&	\multicolumn{1}{c}{(kpc)}  & \multicolumn{1}{c}{(kpc)}  & \multicolumn{1}{c}{(kpc)} && \multicolumn{1}{c}{(kpc)} & \multicolumn{1}{c}{Flag$^a$} & \multicolumn{1}{c}{ Ref.$^b$} & \multicolumn{1}{c}{Name} & \multicolumn{1}{c}{(kpc)} \\

\hline
AGAL005.001$+$00.086	&	2.1	&	This paper	&	2.9	&	0.7	&	17.0	&&	17.0	&	(vi)	&	\multicolumn{1}{c}{$\cdots$}	&	\multicolumn{1}{c}{$\cdots$}	&	\multicolumn{1}{c}{$\cdots$}	\\
AGAL005.041$-$00.097	&	46.3	&	This paper	&	10.8	&	6.6	&	10.0	&&	10.8	&	(vi)	&	\multicolumn{1}{c}{$\cdots$}	&	G005.087$-$00.097	&	10.8	\\
AGAL005.049$-$00.192	&	6.1	&	This paper	&	2.9	&	1.7	&	15.4	&&	2.9	&	(vi)	&	\multicolumn{1}{c}{$\cdots$}	&	\multicolumn{1}{c}{$\cdots$}	&	\multicolumn{1}{c}{$\cdots$}	\\
AGAL005.076$-$00.091	&	45.1	&	\citet{jackson2013}	&	10.8	&	6.5	&	10.1	&&	10.8	&	(vi)	&	\multicolumn{1}{c}{$\cdots$}	&	G005.087$-$00.097	&	10.8	\\
AGAL005.094$-$00.104	&	45.7	&	\citet{jackson2013}	&	10.8	&	6.6	&	10.1	&&	10.8	&	(vi)	&	\multicolumn{1}{c}{$\cdots$}	&	G005.087$-$00.097	&	10.8	\\
AGAL005.139$-$00.097	&	44.1	&	This paper	&	10.8	&	6.4	&	10.2	&&	10.8	&	(vi)	&	\multicolumn{1}{c}{$\cdots$}	&	G005.087$-$00.097	&	10.8	\\
AGAL005.184$+$00.159	&	\multicolumn{1}{c}{$\cdots$}	&	\multicolumn{1}{c}{$\cdots$}	&	\multicolumn{1}{c}{$\cdots$}	&	\multicolumn{1}{c}{$\cdots$}	&	\multicolumn{1}{c}{$\cdots$}	&&	\multicolumn{1}{c}{$\cdots$}	&	(x)	&	\multicolumn{1}{c}{$\cdots$}	&	\multicolumn{1}{c}{$\cdots$}	&	\multicolumn{1}{c}{$\cdots$}	\\
AGAL005.192$-$00.284	&	8.0	&	This paper	&	2.9	&	2.0	&	14.9	&&	2.9	&	(vi)	&	\multicolumn{1}{c}{$\cdots$}	&	\multicolumn{1}{c}{$\cdots$}	&	\multicolumn{1}{c}{$\cdots$}	\\
AGAL005.202$-$00.036	&	0.7	&	This paper	&	16.0	&	0.1	&	17.7	&&	16.0	&	(vi)	&	\multicolumn{1}{c}{$\cdots$}	&	\multicolumn{1}{c}{$\cdots$}	&	\multicolumn{1}{c}{$\cdots$}	\\
AGAL005.321$+$00.184	&	\multicolumn{1}{c}{$\cdots$}	&	\multicolumn{1}{c}{$\cdots$}	&	\multicolumn{1}{c}{$\cdots$}	&	\multicolumn{1}{c}{$\cdots$}	&	\multicolumn{1}{c}{$\cdots$}	&&	\multicolumn{1}{c}{$\cdots$}	&	(x)	&	\multicolumn{1}{c}{$\cdots$}	&	\multicolumn{1}{c}{$\cdots$}	&	\multicolumn{1}{c}{$\cdots$}	\\

\hline\\
\end{tabular}\\

Notes: Only a small portion of the data is provided here, the full table is available in electronic form at the CDS via anonymous ftp to cdsarc.u-strasbg.fr (130.79.125.5) or via http://cdsweb.u-strasbg.fr/cgi-bin/qcat?J/MNRAS/.\\ 
$^a$ The distance solution flags refer to the different steps described in Sect.\,B1 and Table\,B2.\\
$^b$ References for distance solutions: (1) \citet{anderson2009a}: (2) \citet{araya2002}; (3) \citet{battisti2014}; (4) \citet{urquhart2013_cornish}; (5) \citet{dunham2011_bgps_vii}; (6) \citet{fish2003}, (7) \citealt{reid2014}, (8) \citet{bessel_sanna2014}, (9) \citet{bessel_wu2014}, (10) \citet{bessel_xu2009}, (11) \citet{fish2003}, (12) \citealt{downes1980}, (13) \citet{giannetti2015}, (14) \citet{green2011b}, (15) \citet{immer2012}, (16) \citet{Kolpak2003}; (17) \citet{pandian2009}; (18) \citet{sewilo2003}; (19) \citet{moises2011}; (20) \citet{pandian2008}; (21) \citet{roman2009}; (22) \citet{stead2010}; (23) \citet{bessel_sanna2009}; (24) \citet{bessel_sato2010_w51}; (25) \citet{urquhart2012_hiea}; (26) \citet{watson2003}; (27) \citet{xu2011}; (28) \citet{bessel_zhang2013}; (29) (\citet{nagayama2011}

\end{minipage}

\end{center}
\end{table*}
\setlength{\tabcolsep}{6pt}

\begin{figure}

\includegraphics[width=0.49\textwidth, trim= 0 0 0 0]{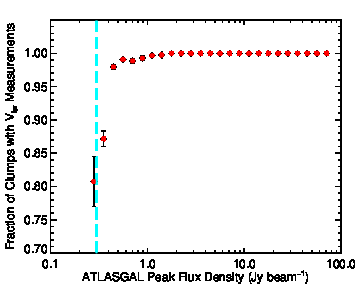}

\caption{\label{fig:flux_distribution} Plot of the ratio of sources with velocities as a function of their peak flux densities. The bin size for both plots is 0.1\.dex. The errors shown in the lower panel have been estimated using binomial statistics. The vertical dashed line indicates the 5$\sigma$ sensitivity limit of the ATLASGAL survey.} 

\end{figure}

\subsection{Completeness}
\label{sect:completenss}

\begin{figure}
\centering
\includegraphics[width=0.49\textwidth, trim= 0 0 0 0]{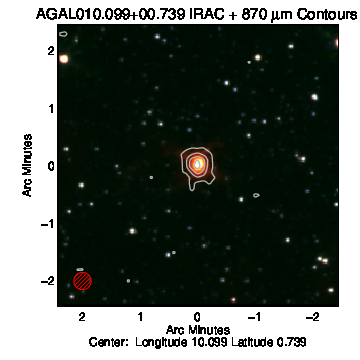}

\caption{Three colour image of an evolved star candidate (this is constructed using the GLIMPSE 3.6, 5.8 and 8\,\mum\ images, which are coloured blue, green and red, respectively). The contours show the submillimetre dust emission.  The infrared and dust emission both show a pointlike distribution and there is no $^{13}$CO (2--1) emission detected. \label{fig:evolved_star} } 

\end{figure}

\begin{figure*}

\includegraphics[width=0.98\textwidth, trim= 0 0 0 0]{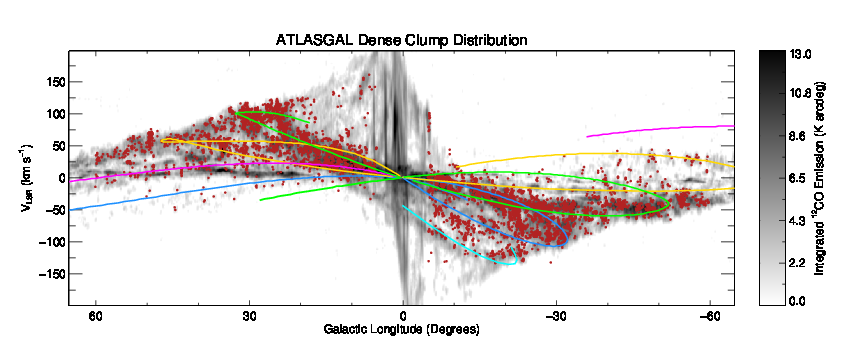}

\caption{Longitude-velocity distribution of all ATLASGAL source for which we have been able to assign a velocity. The greyscale image shows the distribution of molecular gas as traced by the integrated $^{12}$CO (1-0) emission for comparison (\citealt{dame2001}). The location of the spiral arms are shown as curved solid lines, coloured to identify the individual arms. The positions of the four main spiral and local arms have been taken from model by \citet{taylor1993} and updated by \citet{cordes2004}, while the position of the near 3-kpc arm has been taken from \citet{bronfman2000}. Colours: Norma/Outer $\rightarrow$ blue; Perseus $\rightarrow$ magenta; Sagittarius $\rightarrow$ yellow; Scutum-Centaurius $\rightarrow$ green; Near 3-kpc arm $\rightarrow$ cyan. \label{fig:vlsr_distribution} } 

\end{figure*}

There are 8002 ATLASGAL sources located in the region of interest considered here and through the steps described in the previous subsection we have determined a velocity to 7809 sources ($\sim$98\,per\,cent of the sample). In Table\,\ref{tbl:source_vlsr} we give the source names, velocities, molecular transition used to determine the velocity and the reference from where these data originated. In Fig.\,\ref{fig:flux_distribution} we show the proportion of sources with determined radial velocities as a function of the peak flux distribution. This plot shows that we are complete for nearly all sources with a peak flux density $\simeq$0.5\,Jy\,beam$^{-1}$. Visual inspection of infrared and submillimetre maps for the $\sim$200 weak sources for which we have not been able to assign a velocity reveals that many of them have a wispy structure and  often appear to be part of the diffuse outer envelopes of larger regions. Molecular line emission is actually detected towards many of these diffuse sources; however, this emission is normally in the $^{12}$CO and $^{13}$CO (1--0) transitions where multiple peaks of  approximately equal intensity are detected. It therefore seems likely that many of the clumps with undetermined velocities are of little interest in terms of their current star formation potential.

While the vast majority of sources identified in the ATLASGAL CSC are dense molecular clumps, the CSC is also sensitive to the submillimetre emission from the dusty envelopes that surround evolved stars. Although these dusty envelopes are warm enough to produce significant dust emission, they do not generally have sufficiently high column densities to be detected in molecular line observations ($^{12}$CO (1-0 and 2-1) typically less than 1\,K; \citealt{loup1993}, see also Fig.\,\ref{fig:evolved_star} for an example of an evolved star candidate). The possible contamination of the ATLASGAL catalogue by diffuse gas and warm envelopes of evolved stars suggests that our sample is likely as complete as possible in terms of velocity determination.

Fig.\,\ref{fig:vlsr_distribution} presents a longitude-velocity ($\ell v$)-diagram that shows the positions of all of the ATLASGAL CSC for which we have determined a velocity. The background image shown in this figure is the integrated $^{12}$CO (1-0) map from \citet{dame2001} that traces the large scale Galactic distribution of molecular gas. Comparison between the CO emission and dense gas traced by the dust emission reveals a good correlation between the two. This plot also shows the loci of the spiral arms from \citet{taylor1993} and \citet{cordes2004} and  the loci of the 3-kpc expanding arm from \citet{bronfman2000}. The majority of ATLASGAL sources are seen to be tightly clustered around the four-arm Milky Way model spiral arm loci. We will revisit the Galactic distribution in Section\,\ref{sect:gal_distribution}.

\section{Distances and associations}
\label{sect:distances}

A key element required for determining the physical properties of dense clumps identified by ATLASGAL is their heliocentric distance. This has been achieved using a multi-step approach that includes adopting maser parallax and spectroscopic distances where available, determining kinematic distances to all remaining sources for which we have derived a radial velocity, and resolving any resulting distance ambiguities using archival \hi\ data. These steps are described in detail in Appendix\,\ref{sect:append_distances} and outlined in the flow-chart presented in Fig.\,\ref{fig:distance_flowchart}. 

\begin{figure*}
\centering
\includegraphics[width=0.98\textwidth, trim= 0 1.5cm 0 2cm, clip]{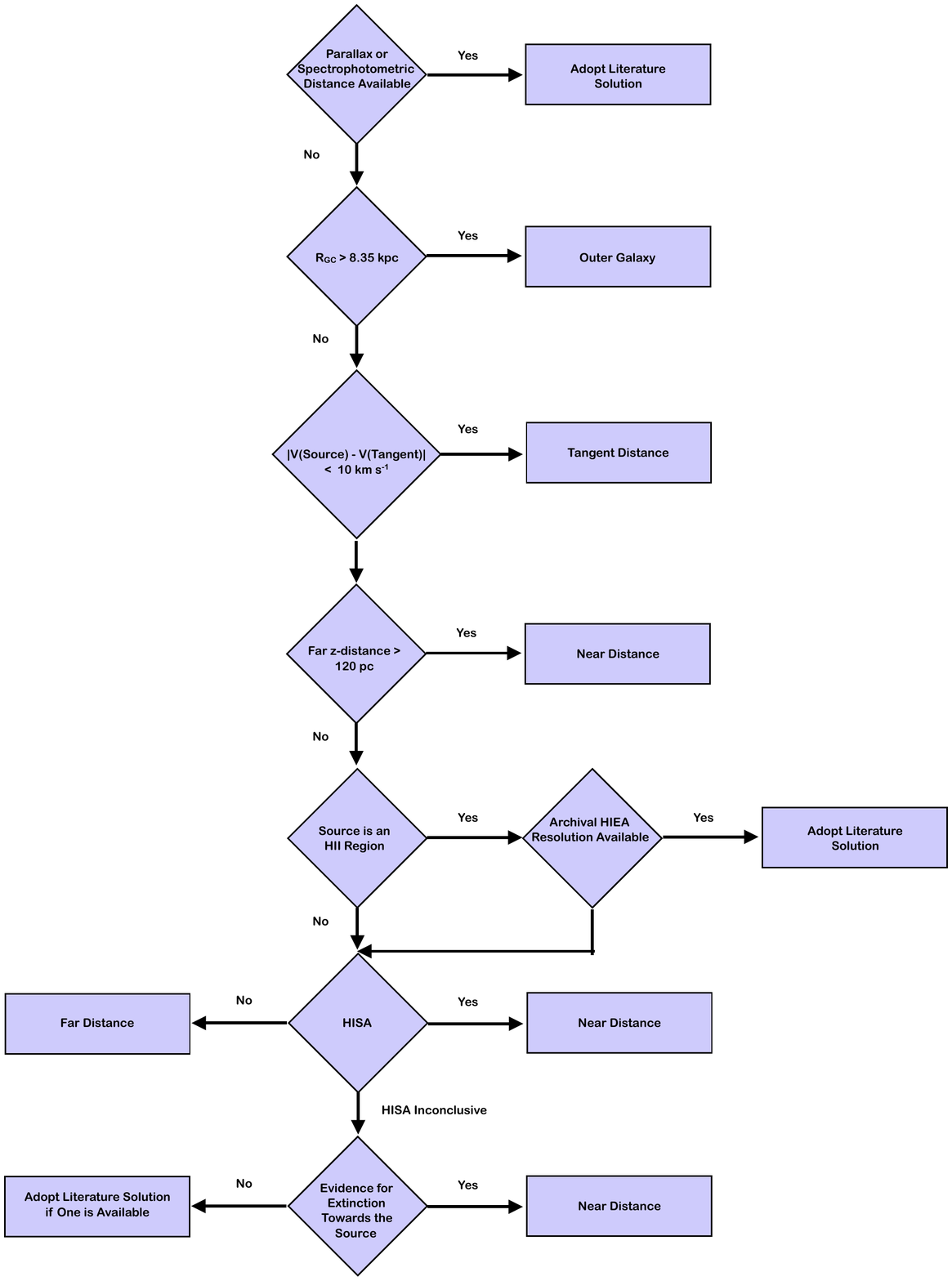}

\caption{Flow chart showing the criteria used to determine distance to ATLASGAL CSC clumps.  \label{fig:distance_flowchart} } 

\end{figure*}

Using these steps we have determined kinematic distance solutions to $\sim$7000 ATLASGAL clumps, with $\sim$80\,per\,cent being considered to be reliable. Comparing these with distances solutions recently published in the literature we have found agreement in 73\,per\,cent of cases (see Table\,\ref{tbl:dist_literature_comparion} for a detailed breakdown). This analysis has resulted in 78\,per\,cent of sources being placed at the near distance and the remaining 22\,per\,cent being placed at the far distance, which is consistent with other studies (cf. \citealt{eden2012}). We have not performed a direct comparison with the recent work presented by \citet{elia2017} as only $\sim$40\,per\,cent of their distances were the result of distance ambiguity resolution: the remainder are arbitrarily  placed at the far distance and so the samples are not comparable as their results are likely to be biased to systematically larger distances.

Although the steps outlined in Fig.\,\ref{fig:distance_flowchart} have been able to apply distances to almost 90\,per\,cent of the sample there are still a significant number for which we have been unable to determine a velocity, or for which we have been unable to resolve the distance ambiguity. In an effort to determine distances to these sources and to provide a consistency check on the kinematic solutions derived from the \hi\ data we have performed a friends-of-friends clustering analysis; this is described in detail in Appendix\,\ref{sect:fof_analysis}. This analysis has identified 776 clusters of ATLASGAL sources, many of the largest of which, are associated with some of the most well-known star forming regions in the Galaxy.

In Fig.\,\ref{fig:clusters} we present a few examples of the clusters identified by the friends-of-friends analysis; the distributions of clumps are overlaid on  8\,\mum\ Spitzer images taken as part of the GLIMPSE legacy survey. The 8\,\mum\ images are very sensitive to emission from polycyclic aromatic hydrocarbons (PAHs; e.g., \citealt{urquhart2003}) which are excited in photo-dominated regions (PDRs) often associated with \hii\ regions and are, therefore, excellent tracers of star forming regions. Areas of extinction are very useful for tracing the colder quiescent dense clumps where future star formation is expected to take place. These mid-infrared images are therefore able to trace both the large scale environments of active star formation regions and trace the distribution of dense clumps identified by ATLASGAL. The examples presented in Fig.\,\ref{fig:clusters} nicely illustrate these points and demonstrate the reliability of the friends-of-friends analysis.   

\begin{figure}
\centering
\includegraphics[width=0.49\textwidth, trim= 30 0 0 0,clip]{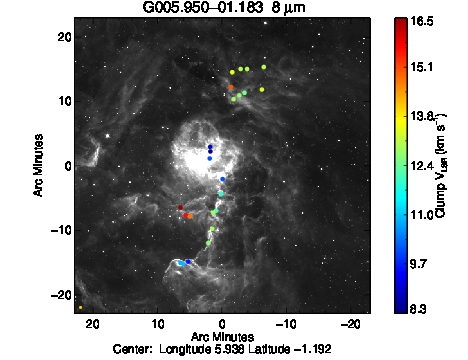}
\includegraphics[width=0.49\textwidth, trim= 30 0 0 0,clip]{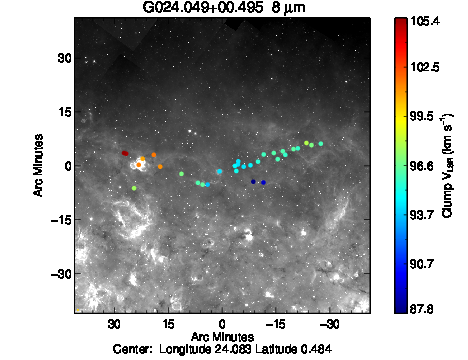}
\includegraphics[width=0.49\textwidth, trim= 30 0 0 0,clip]{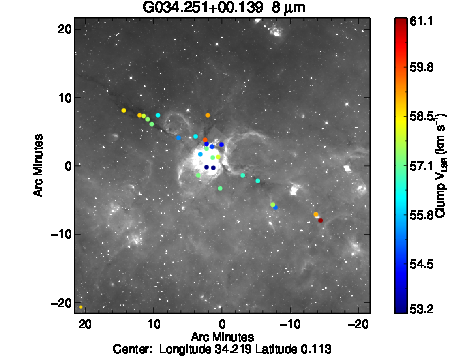}

\caption{Examples of clusters identified by the Friends-of-Friends analysis. The background image is the 8\,\mum\ emission taken from the GLIMPSE survey. The filled circles show the positions of ATLASGAL CSC objects, the colours of which provide an indication of their velocities (see colour bars for values). In the upper, middle and lower panels we show the Lagoon nebula (M8), the giant molecular filament (GMF) G023.985+0.479 (\citealt{li2016}); see also CFG024.00+0.48 \citealt{wang2015}) and G34.  \label{fig:clusters} } 

\end{figure}

\subsection{Distance summary}

In total, we have identified 776 clusters that are correlated in $\ell b v$ and distance; these are associated with 6620 clumps, including 75 clumps for which no velocity was available and a further 549 clumps for which we were previously unable to determine a distance. We have either used the \hi SA technique to solve the distance ambiguity or extracted more reliable distances from the literature for an additional 1150 sources. In total, this combination of methods has provided distances to 7770 clumps ($\sim$97\,per\,cent of the full CSC sample in the region covered here). In Fig.\,\ref{fig:galactic_cluster_distribution} we show the Galactic distribution of the clusters and isolated clumps that is derived from the distances discussed in this section.

    
\begin{figure*}
\centering
\includegraphics[width=0.49\textwidth, trim= 20 0 20 0]{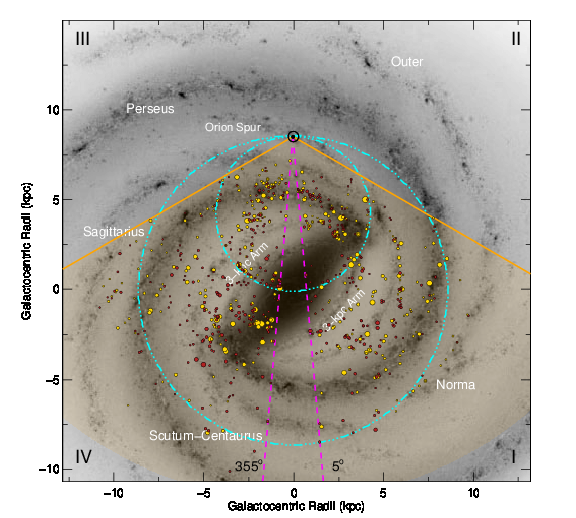}
\includegraphics[width=0.49\textwidth, trim= 20 0 20 0]{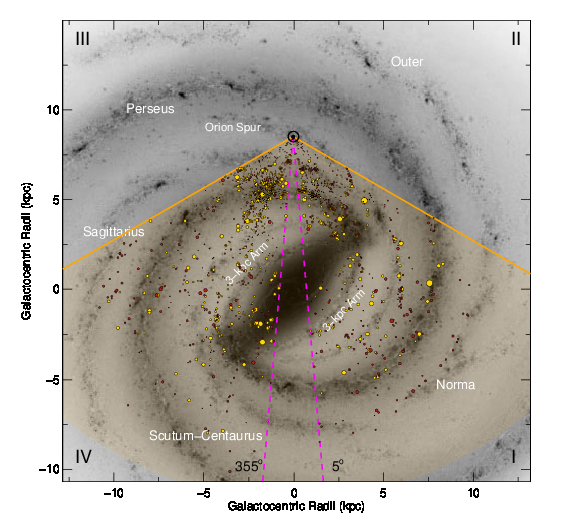}

\caption{2-D images showing expected large-scale features of the Milky Way as viewed from the Galactic pole. The positions of ATLASGAL sources have been overlaid to facilitate comparison of their distribution to the large-scale structure of the Galaxy. The yellow circles show positions of the clusters identified by the friends-of-friends analysis described in Sect.\,\ref{sect:fof_analysis} while the red circles show the positions of individual clumps. The sizes of the circles in the left and right panels give an indication of the masses and luminosities of the clumps and clusters, respectively. The background image is a schematic of the Galactic disc as viewed from the Northern Galactic Pole (courtesy of NASA/JPL-Caltech/R. Hurt (SSC/Caltech)). The Sun is located at the apex of the wedge and is indicated by the $\odot$ symbol. The smaller of the two cyan dot-dashed circles represents the locus of tangent points, while the larger circle traces the solar circle. The spiral arms are labeled in white and Galactic quadrants are given by the roman numerals in the corners of the image. The magenta line shows the innermost region toward the Galactic centre where distances are not reliable.  The distances of the ATLASGAL sources have been determined using a mixture of the \citet{reid2014} method to calculate the near and far distances while taking advantage of the additional constraints provided by the \citet{reid2016} Bayesian maximum likelihood to better constrain sources with velocities close to the Sun's, the tangent position and outside the Solar circle.}
\label{fig:galactic_cluster_distribution}

\end{figure*}

\section{Spectral energy distributions}
\label{sect:seds}

\begin{figure}
\centering
\includegraphics[width=0.49\textwidth, trim= 0 0 0 0]{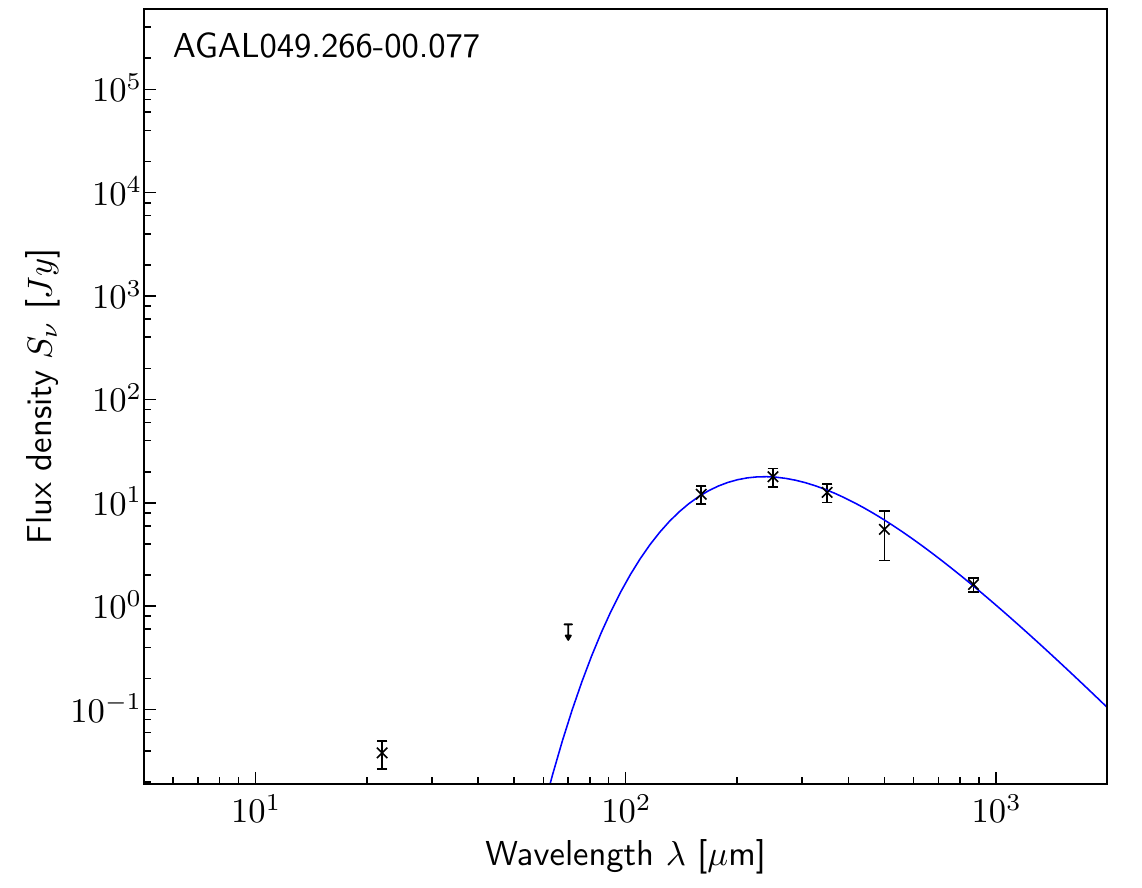}
\includegraphics[width=0.49\textwidth, trim= 0 0 0 0]{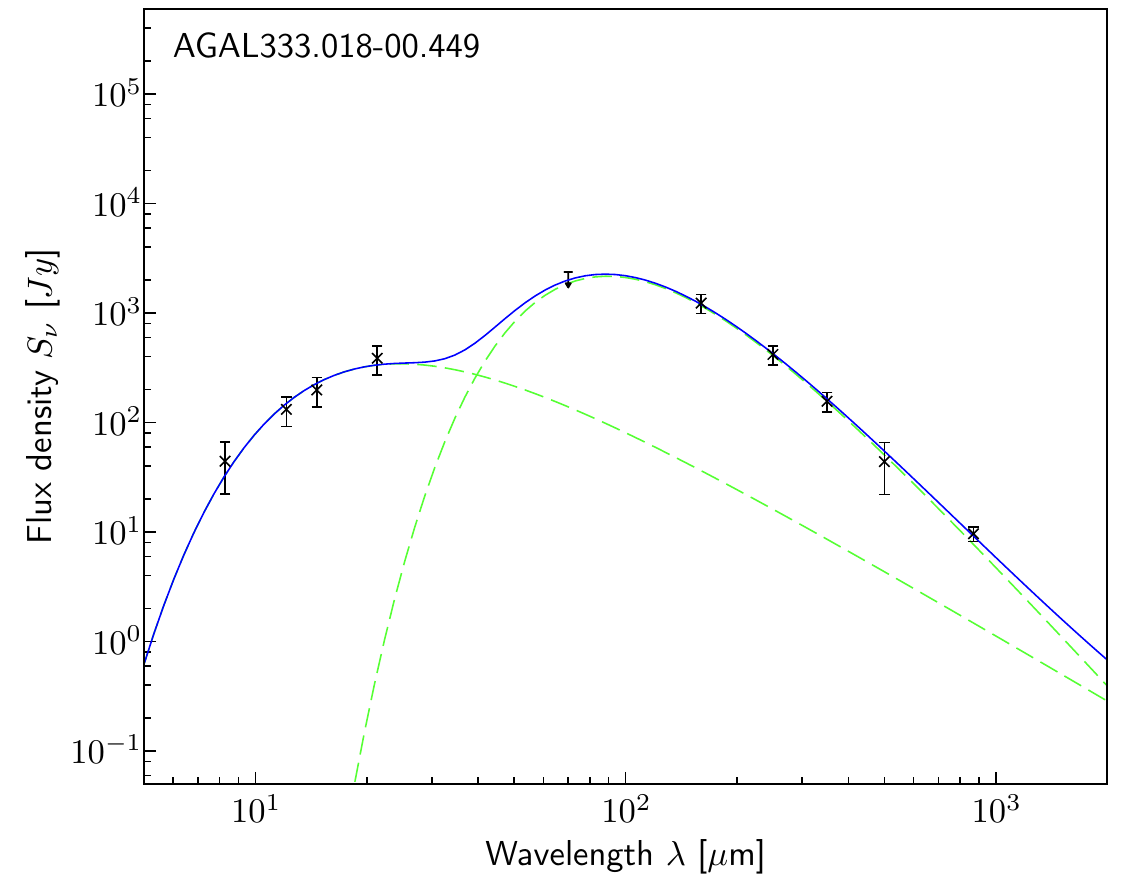}
\caption{Example SEDs showing a source fit with single-component greybody model (upper panel) and with a two-component model (lower panel).  The photometry measurements are shown as crosses with their associated uncertainties indicated by the vertical limits dissecting them. In the upper panel, the greybody fit to the data is shown in blue while, in the lower panel, the blue curve shows the combined fits to these data of the individual cold greybody and warm blackbody functions; these are shown by the two green dashed curves. \label{fig:sed_fits} } 

\end{figure}

We follow the general procedures of \citet{konig2017} to obtain the spectral energy distributions (SEDs) of the clumps, but apply some changes where necessary to be able to obtain the SEDs for the complete CSC in a fully automated way. We use the ATLASGAL \citep{schuller2009} 870\,$\mu$m emission maps as well as the Hi-GAL \citep{Molinari2010} images available for the PACS \citep{poglitsch2010} 70\,$\mu$m and 160\,$\mu$m and the SPIRE \citep{griffin2010} 250, 350 and 500\,$\mu$m bands to reconstruct the SEDs cold dust component. We estimate the emission from a hot and embedded component, likely to be associated with a more evolved part of the clump, using MSX \citep{egan2003} emission maps at 8, 12, 14 and 21\,$\mu$m, as well as images from the WISE \citep{Wright2010} 12 and 24\,$\mu$m bands. After reprojecting the images into the same projection as the ATLASGAL images, we converted all maps to Jy/pixel units and extracted $5 \times 5$ arcminute-sized images for each source. From these images we obtained the photometric data in all bands and subsequently fit the SEDs by a single- or two-component model. Examples of these fits are shown in Fig.\,\ref{fig:sed_fits} and a detailed description of the photometric measurements and fitting procedure are provided in Appendix\,\ref{sect:append_sed}.

\subsection{Evolutionary sequence}

Following the classification scheme introduced in \citet{konig2017} and confirmed in \citet{giannetti2017}, we find that the majority of sources are actively forming stars with 5275 clumps ($\sim$66\,per\,cent) being classified as mid-infrared bright (i.e. associated with a 21-24\,\mum\ point source with a flux $>$ 2.6\,mJy). We make a distinction in this sample between clumps previously associated with a massive star formation tracer (MSF clumps; i.e., radio bright \hii\ regions, massive young stellar objects and methanol masers; see Paper\,III for details). We will refer to the other mid-infrared bright clumps as YSO-forming clumps: these subsamples contribute $\sim$16 and 52\,per\,cent of the sample, respectively. Another 1640 sources ($\sim$21\,per\,cent) are classified as mid-infrared weak but far-infrared bright, making them likely to be in an earlier protostellar phase; we refer to these as protostellar clumps. The remaining 946 clumps ($\sim$12\,per\,cent) are classified as 70-$\mu$m weak, indicating that they are the youngest objects in the sample, likely to be in a starless or pre-stellar phase, however, a number of these 70\,\mum\ sources have been found to be associated with molecular outflows (e.g., \citealt{traficante2017}) and so this fraction should be considered as an upper limit. In Table\,\ref{tab:classes} we present a  summary of evolutionary types identified and the numbers of sources associated with each. On the face of it, this suggests that 88\,per\,cent of all ATLASGAL clumps are currently involved in the star-formation process and the pre-stellar phase is relatively short. 

\setlength{\tabcolsep}{4pt}
\begin{table}

\caption{Summary of evolutionary types identified from the SED analysis and our previous work (Paper\,III).}
\label{tab:classes}
\begin{tabular}{l c c  c c c}
\hline \hline
Evolutionary & \# of & \multicolumn{1}{c}{Fraction } & \multicolumn{3}{c}{\multirow{ 2}{*}{Notes}} \\
 type & sources & \multicolumn{1}{c}{of total}  \\
\hline
MSF & 1222  & 0.16 & \multirow{ 3}{*}{{\Bigg{\}}}} & \multirow{ 3}{*}{0.88} & \multirow{3}{*}{Star forming}\\
YSO & 4053  & 0.52 & \\
Protostellar & 1640  & 0.21\\
Quiescent & 946  & $<$0.12 &  \multirow{ 1}{*}{\}} &\multirow{ 1}{*}{$<$0.12} & Non-star forming\\
\hline

\end{tabular}
\end{table}

Taken together, these four subsamples represent an evolutionary sequence for the formation of massive stars and clusters (\citealt{konig2017}) and therefore comparing the physical properties of these different phases will provide some insight into the processes associated with star formation. 

\section{Physical properties}
\label{sect:physical_properties}

In this section we will use the distances assigned in Sect.\,\ref{sect:distances} and the parameters derived from the analysis of the SEDs presented in Sect.\,\ref{sect:seds} to determine physical properties for all of the clumps. 

\subsection{Analysis tools: Sample comparison}

We use the two-sample \KS\ (KS) test to compare the similarities between the different evolutionary types for the various derived properties. {\color{black} This is a non-parametric test that compares the empirical cumulative distribution functions for the two samples and measures the largest difference between them (this is referred to as the KS statistic ($D$)) and its associated confidence value referred to as the $p$-value. The null hypothesis is that both samples are drawn from the same parent population; however, this can be rejected if the $p$-value is smaller than 3$\sigma$ (i.e., $<$ 0.0013), allowing us to conclude that there is sufficient evidence to consider the samples to be drawn from different populations.} The KS test is useful as it is sensitive to differences in both the location and shape of the cumulative distributions of the different samples. 

\setlength{\tabcolsep}{6pt}
\begin{table*}

\begin{center}\caption{Summary of physical properties of the whole population of clumps and the four evolutionary subsamples identified. In Col.\,(2) we give the number of clumps in each subsample, in Cols.\,(3-5) we give the mean values, the error in the mean and the standard deviation, in Cols.\,(6-8) we give the median and minimum and maximum values of the samples. Note we do not breakdown the results of the scale height analysis into evolutionary groups as there is no significant difference between them.}
\label{tbl:derived_para_all}
\begin{minipage}{\linewidth}
\small
\begin{tabular}{lc......}
\hline \hline
  \multicolumn{1}{l}{Parameter}&  \multicolumn{1}{c}{\#}&	\multicolumn{1}{c}{$\bar{x}$}  &	\multicolumn{1}{c}{$\frac{\sigma}{\sqrt(N)}$} &\multicolumn{1}{c}{$\sigma$} &	\multicolumn{1}{c}{$x_{\rm{med}}$} & \multicolumn{1}{c}{$x_{\rm{min}}$}& \multicolumn{1}{c}{$x_{\rm{max}}$}\\
\hline
Temperature  (K) &         7861&19.52&0.07 & 5.80 & 18.60 & 7.90 & 56.10\\
\cline{1-1}
MSF Temperature  &         1222  &24.58&0.14 & 5.06 & 24.00 & 12.60 & 56.10\\
YSO Temperature  &         4053 &20.93&0.08 & 5.25 & 20.10 & 8.60 & 53.00\\
Protostellar Temperature   &         1640 &15.52&0.07 & 2.80 & 15.20 & 8.30 & 30.20\\
Quiescent Temperature  &          946 &13.88&0.12 & 3.64 & 13.30 & 7.90 & 48.70\\
\hline

Radius (pc) &         4836&0.71&0.01 & 0.62 & 0.52 & 0.01 & 7.73\\
\cline{1-1}
MSF Radius  &         1017&1.00&0.02 & 0.77 & 0.78 & 0.01 & 7.73\\
YSO Radius  &         2552&0.70&0.01 & 0.59 & 0.52 & 0.02 & 4.86\\
Protostellar Radius &          840&0.53&0.01 & 0.43 & 0.40 & 0.05 & 2.89\\
Quiescent Radius  &          427&0.47&0.02 & 0.37 & 0.38 & 0.03 & 2.25\\
\hline
Log[Luminosity (\lsun)] &         7614&2.95&0.01 & 1.03 & 2.89 & -0.30 & 6.91\\

\cline{1-1}
MSF &        1191&4.01&0.03 & 0.93 & 4.03 & -0.30 & 
6.91\\
YSO &            3922&3.12&0.01 & 0.84 & 3.10 & 0.43 & 
6.24\\
Protostellar &           1580&2.29&0.02 & 0.69 & 2.24 & 
-0.30 & 4.62\\

Quiescent &          921&1.99&0.02 & 0.72 & 1.93 & 
0.00 & 4.83\\
\hline
Log[Clump Mass (\msun)] &         7614&2.68&0.01 & 0.65 & 2.69 & -1.00 & 5.04\\
\cline{1-1}
MSF &         1191&2.95&0.02 & 0.71 & 3.02 & -1.00 & 5.04\\
YSO &         3922&2.62&0.01 & 0.66 & 2.64 & -0.40 & 4.72\\
Protostellar &         1580&2.64&0.01 & 0.58 & 2.62 & 0.18 & 4.36\\
Quiescent &          921&2.66&0.02 & 0.57 & 2.64 & 0.81 & 4.34\\
\hline
Log[$N$(H$_2$) (cm$^{-2}$)] &         7861&22.33&0.00 & 0.29 & 22.30 & 21.58
 & 24.02\\
\cline{1-1}
MSF &         1222&22.55&0.01 & 0.39 & 22.50 & 21.68 & 24.02\\
YSO &         4053&22.25&0.00 & 0.26 & 22.22 & 21.58 & 23.54\\
Protostellar &         1640&22.33&0.01 & 0.21 & 22.31 & 21.72 & 23.27\\
Quiescent &          946&22.40&0.01 & 0.19 & 22.39 & 21.91 & 23.20\\
\hline
\lm\ ratio (\lsun/\msun) &         7614&18.76&0.75 & 65.17 & 4.47 & 0.03 & 2519.50\\
\cline{1-1}
MSF &          1191&38.90&2.75 & 94.88 & 19.03 & 0.44 & 2519.50\\
YSO &         3922&22.61&1.11 & 69.75 & 7.30 & 0.08 & 1944.37\\
Protostellar &        1634&2.42&0.12 & 5.02 & 1.27 & 0.04 & 95.08\\
Quiescent &          933&6.65&1.58 & 48.37 & 0.58 & 0.03 & 833.62\\
\hline
Log[Mass Surface Density (\msun\,pc$^{-2}$)] &         4838&2.92&0.00 & 0.35 & 2.87 & 1.86 & 4.34\\
\cline{1-1}
MSF &         1019&2.80&0.01 & 0.26 & 2.78 & 1.92 & 4.05\\
YSO  &         2552&2.85&0.01 & 0.33 & 2.81 & 1.86 & 4.07\\
Protostellar &          840&3.10&0.01 & 0.34 & 3.05 & 2.30 & 4.34\\
Quiescent &          427&3.21&0.02 & 0.35 & 3.17 & 2.45 & 4.32\\
\hline
Scale height (z) (pc) &         7123&-4.34&0.36 & 30.63 & -5.50 & -345.90 & 235.40\\
\hline\\
\end{tabular}\\

\end{minipage}

\end{center}
\end{table*}

\setlength{\tabcolsep}{6pt}

\setlength{\tabcolsep}{1pt}

\begin{table*}

\begin{center}\caption{Derived clump parameters.}
\label{tbl:derived_clump_para}
\begin{minipage}{\linewidth}
\small
\begin{tabular}{ll..........}
\hline \hline
  \multicolumn{1}{c}{CSC}&  
  \multicolumn{1}{c}{Evolution}&
  \multicolumn{1}{c}{Peak flux}   &
  \multicolumn{1}{c}{Int. flux}   &
  \multicolumn{1}{c}{$V_{\rm{LSR}}$}   &	
  \multicolumn{1}{c}{Distance} &
  \multicolumn{1}{c}{R$_{\rm{GC}}$}&
  \multicolumn{1}{c}{Radius}&
  \multicolumn{1}{c}{T$_{\rm{dust}}$} & 
  \multicolumn{1}{c}{Log[$L_{\rm{bol}}$]}	&
  \multicolumn{1}{c}{Log[$M_{\rm{clump}}$]} & 
  \multicolumn{1}{c}{Log[$N$(H$_2$)]} \\
  
    \multicolumn{1}{c}{name }&  
    \multicolumn{1}{c}{type} & 
    \multicolumn{1}{c}{(Jy beam$^{-1}$)}  &
    \multicolumn{1}{c}{(Jy)}  &	
    \multicolumn{1}{c}{(km\,s$^{-1}$)}&	
    \multicolumn{1}{c}{(kpc)} &
    \multicolumn{1}{c}{(kpc)}&
    \multicolumn{1}{c}{(pc)}&
    \multicolumn{1}{c}{(K)} &	
    \multicolumn{1}{c}{(\lsun)} &
    \multicolumn{1}{c}{(\msun)} &
    \multicolumn{1}{c}{(cm$^{-2}$)} \\
\hline
AGAL006.264$-$00.506	&	Quiescent	&	0.44	&	1.49	&	22.8	&	3.0	&	5.4	&	0.35	&	12.7	&	1.607	&	2.186	&	22.361	\\
AGAL006.268$-$00.749	&	YSO	&	0.48	&	3.29	&	21.5	&	3.0	&	5.4	&	0.70	&	19.4	&	2.502	&	2.231	&	22.100	\\
AGAL006.339$-$00.746	&	YSO	&	0.50	&	2.77	&	21.7	&	3.0	&	5.4	&	0.42	&	18.5	&	2.312	&	2.187	&	22.148	\\
AGAL006.368$-$00.051	&	MSF	&	1.46	&	4.97	&	141.3	&	14.3	&	6.0	&	3.49	&	24.3	&	4.700	&	3.627	&	22.443	\\
AGAL006.393$-$00.166	&	Protostellar	&	0.44	&	2.84	&	14.6	&	3.0	&	5.4	&	0.35	&	16.9	&	2.189	&	2.258	&	22.153	\\
AGAL006.404$-$00.039	&	Protostellar	&	0.53	&	1.59	&	141.4	&	14.3	&	6.0	&	1.66	&	17.0	&	3.436	&	3.359	&	22.230	\\
AGAL006.441$-$00.524	&	YSO	&	0.45	&	2.45	&	160.1	&	9.2	&	1.3	&	1.07	&	16.6	&	3.285	&	3.185	&	22.175	\\
AGAL006.469$-$00.229	&	YSO	&	0.59	&	6.55	&	16.7	&	3.0	&	5.4	&	0.87	&	15.1	&	2.331	&	2.700	&	22.359	\\
AGAL006.479$-$00.251	&	YSO	&	0.60	&	5.03	&	16.8	&	3.0	&	5.4	&	0.80	&	17.6	&	2.723	&	2.479	&	22.260	\\
AGAL006.484$-$00.547	&	YSO	&	0.52	&	1.91	&	14.8	&	3.0	&	5.4	&	0.35	&	19.9	&	2.573	&	1.978	&	22.118	\\
\hline\\
\end{tabular}\\
Notes: Only a small portion of the data is provided here: the full table is available in electronic form at the CDS via anonymous ftp to cdsarc.u-strasbg.fr (130.79.125.5) or via http://cdsweb.u-strasbg.fr/cgi-bin/qcat?J/MNRAS/.
\end{minipage}

\end{center}
\end{table*}
\setlength{\tabcolsep}{6pt}

\subsection{Distance Distribution}
\label{sect:dist_dist}

\begin{figure}
\begin{center}

\includegraphics[width=0.49\textwidth, trim= 0 0 0 0]{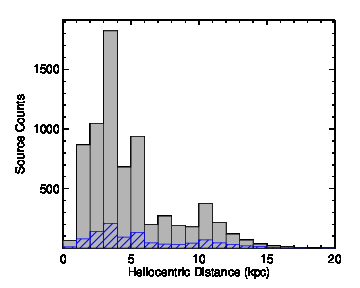}

\caption{\label{fig:distance_dist} Heliocentric distance distribution of all clumps (grey) and clumps associated with massive star formation (blue). The bin size used is 1\,kpc.  }

\end{center}
\end{figure}

We present the distance distribution for the full sample and the subsample of MSF clumps in Fig.\,\ref{fig:distance_dist}. The heliocentric distribution reveals that the vast majority of clumps are relatively nearby ($<$ 5\,kpc), with the strongest and second-strongest peaks likely to be associated with the segments of the Sagittarius and Scutum-Centaurus arms that lie between us and the Galactic centre. There is another significant peak between $\sim$10-11\,kpc, which coincides with the end of the far side of the Galactic bar and the far-side Scutum-Centaurus and Perseus arms. The distribution of the MSF clumps shows a similar distribution of peaks but the relative ratio of MSF clumps is much lower for the nearer distance bins, revealing that although the majority of clumps are located nearby, few are associated with MSF and are therefore likely to be predominantly lower-mass clumps. 

We note that the highest density of sources occurs between 2 and 4\,kpc for both the full sample of clumps and for the MSF-clumps, and so we define clumps in this range as the \emph{distance-limited sample}. In the following subsection we provide histograms of the full distribution and for the distance-limited samples, together with associated cumulative distribution plots to aid with comparison of the different evolutionary phases.  These show the distributions of parameters for the distance-limited subsample of each evolutionary phase (Figs.\,\ref{fig:size_parameter}, \ref{fig:temp_dist}, \ref{fig:luminosities} and  \ref{fig:mass_colden}). A summary of the derived physical parameters for the whole sample and the different evolutionary types is presented in Table\,\ref{tbl:derived_para_all} and measurements of individual clumps are given in Table\,\ref{tbl:derived_clump_para}.

\begin{figure}
\centering
\includegraphics[width=0.49\textwidth, trim= 0 0 0 0]{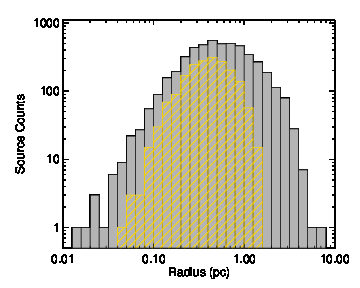}
\includegraphics[width=0.49\textwidth, trim= 0 0 0 0]{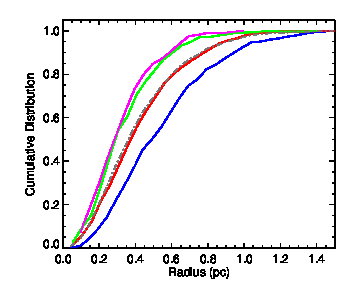}

\caption{The radius frequency distribution for all clumps (grey) and for the distance-limited sample (yellow) is shown in the upper panel. The bin size is 0.1\,dex. The lower panel shows the cumulative radius distribution for the whole sample as the grey dashed-dotted curve.  The distance-limited subsamples of the four different evolutionary types (pre-stellar, protostellar, mid-infrared bright and massive star forming -- MMB, Massive-YSO or \hii\ region) are represented by the magenta, green, red and blue curves, respectively.}
\label{fig:size_parameter}

\end{figure}

\subsection{Clump sizes}
\label{sect:sizes}

The physical sizes of the clumps have been calculated using their effective angular radii and the distances discussed in the previous section. A significant fraction of the ATLASGAL CSC is unresolved, and so we are only able to determine sizes for $\sim$60\,per\,cent of the sample. We show the radius distribution for the 4714 clumps that are resolved in the upper panel of Fig.\,\ref{fig:size_parameter}. The mean size of the clumps is 0.72$\pm0.01$\,pc.

In the lower panel of Fig.\,\ref{fig:size_parameter} we show the size distributions of the four evolutionary distance-limited subsamples. This plot reveals a trend of increasing clump radius as a function of advanced evolutionary stage. There is no significant difference between the sizes of the pre-stellar and protostellar clumps; however, there is a visible trend for more evolved clumps to be larger than the preceding phase: this is confirmed by the KS test. This difference in size is likely to be the result of an observational bias as the extended envelops of warmer and more evolved objects are more readily detectable (this is discussed in more detail in Sect.\,\ref{sect:clump_evolution}).

\begin{figure}
\centering
\includegraphics[width=0.49\textwidth, trim= 0 0 0 0]{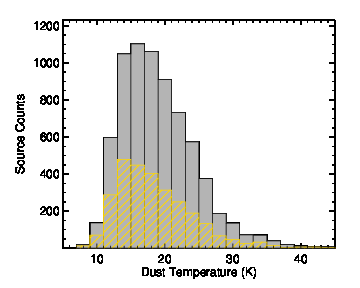}
\includegraphics[width=0.49\textwidth, trim= 0 0 0 0]{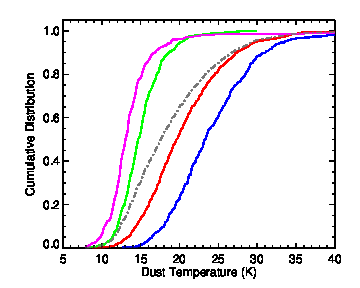}

\caption{Upper panel shows the temperature distribution for all clumps (grey) and the distance-limited sample (yellow). Lower panel shows the temperature distribution of the distance-limited subsamples of the four evolutionary types described in Sect.\,\ref{sect:seds} (colours as given in Fig.\,\ref{fig:size_parameter}). The bin size used in the upper panel is 2\,K. \label{fig:temp_dist} } 

\end{figure}

\begin{figure}
\centering
\includegraphics[width=0.49\textwidth, trim= 0 0 0 0]{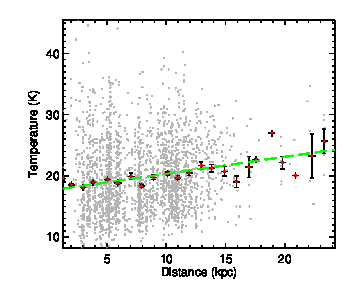}
\includegraphics[width=0.49\textwidth, trim= 0 0 0 0]{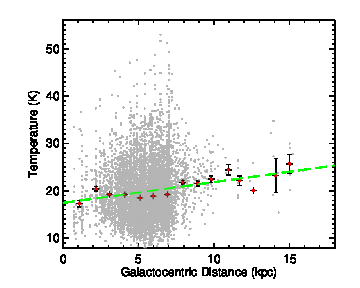}

\caption{Upper and lower panels show the temperature distribution as a function of heliocentric and Galactocentric distance, respectively. The red filled circles show the average temperatures and the error bars indicate the standard error of the mean; these are averaged in bins of 1\,kpc. The dashed green lines show the result of linear fits to these data. \label{fig:temp_distance_dist} } 

\end{figure}

\subsection{Dust temperature and bolometric luminosity}
\label{sect:lum}

\begin{figure}
\begin{center}

\includegraphics[width=0.49\textwidth, trim= 0 0 0 0]{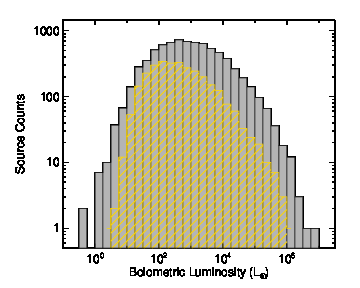}\\
\includegraphics[width=0.49\textwidth, trim= 0 0 0 0]{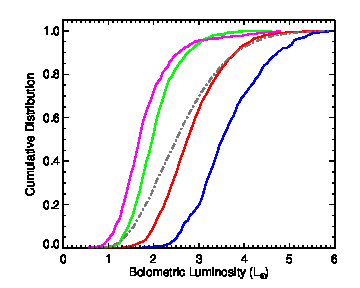}\\

\caption{\label{fig:luminosities} Upper panel shows the bolometric luminosity distribution for all clumps (grey) and the distance-limited sample (yellow). Lower panel shows the luminosity distribution of the distance-limited subsamples (colours as given in Fig.\,\ref{fig:size_parameter}). The bin size used in the upper panel is  0.1\,dex.} 

\end{center}
\end{figure}

The dust temperature is derived directly from the greybody fit to the \submm\ dust emission, as described in Sect.\,\ref{sect:seds}.  The distribution of these temperatures is shown in the upper and lower panels  of Fig.\,\ref{fig:temp_dist}. The dust temperatures are between 10 and 40\,K with a peak at $\sim$16\,K, but is skewed towards warmer temperatures. The mean error of the temperature measurements is 1.9\,K with a standard deviation of 1.9\,K. Comparison of the temperature distribution between the whole sample and distance-limited subsample reveals that more distant sources tend towards higher temperatures. The lower panel of Fig.\,\ref{fig:temp_dist} clearly shows that the evolutionary subsamples are well-separated in temperature, which increases in line with the expected evolutionary sequence (KS tests comparing all subsamples returns $p$-values $\ll 0.001$). 

In the upper panel of Fig.\,\ref{fig:temp_distance_dist} we show the temperature distribution as a function of heliocentric distance. This plot reveals a trend for increasing dust temperature with distance; however, this increase is quite modest, rising by only a few Kelvin between 1 and 15\,kpc, which is comparable with the standard deviation in the temperature. A linear fit to the data shows the trend continues beyond 15\,kpc although the statistics are quite low and the averaged values show a larger variation. The slope of the fit is 0.28 and the Spearman correlation coefficient is 0.2 with a $p$-value $\ll 0.001$.

We show the temperature distribution as a function of Galactocentric distance in the lower panel of Fig.\,\ref{fig:temp_distance_dist}. The majority of sources are located between 3-7\,kpc and, within this range, the temperature is relatively constant ($\sim$20\,K). A linear fit to the data reveals a general trend for an increase in clump temperatures with galactocentric distance that extends beyond the Solar circle. The slope of the fit is 0.44 and the Spearman correlation coefficient is 0.80 with a $p$-value = 0.0004 and so there is a stronger dependence on temperature with their distance from the Galactic centre than their heliocentric distance.

We show the luminosity distribution of the whole sample and the distance-limited subsample in the upper panel of Fig.\,\ref{fig:luminosities}. Both distributions have a similar shape, but the distance-limited sample is shifted to slightly lower luminosities. This shift is the result of the Marquis bias, reflecting that more luminous and massive sources tend to be located at greater distances. The lower panel of Fig.\,\ref{fig:luminosities} shows the luminosity distributions of the four evolutionary subsamples, and clearly exhibits a strong increase in luminosity as a function of evolutionary stage.

\begin{figure*}
\begin{center}

\includegraphics[width=0.49\textwidth, trim= 0 0 0 0]{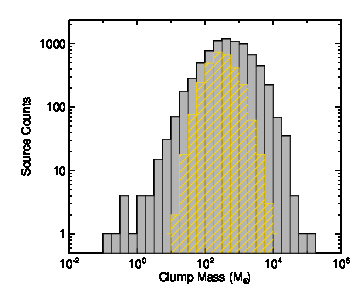}
\includegraphics[width=0.49\textwidth, trim= 0 0 0 0]{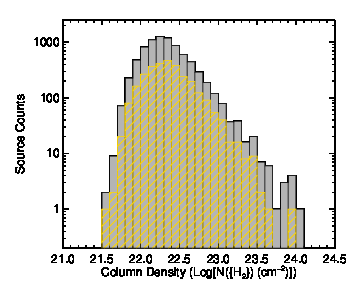}\\
\includegraphics[width=0.49\textwidth, trim= 0 0 0 0]{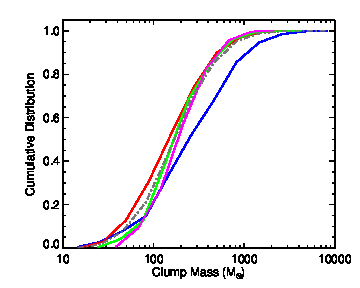}
\includegraphics[width=0.49\textwidth, trim= 0 0 0 0]{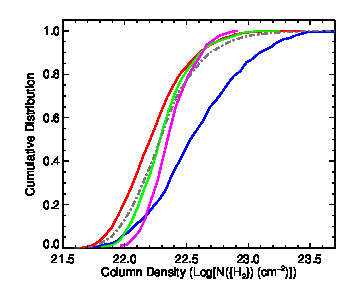}

\caption{\label{fig:mass_colden} Clump mass and peak column density distribution. In the upper left and right panels we present histograms of the whole sample (grey) and the distance-limited sample (yellow), respectively. Lower panels show the cumulative distributions of the distance-limited subsamples (colours as given in Fig.\,\ref{fig:size_parameter}). The bin sizes used in the upper left and right panels are 0.25\,dex and 0.1\,dex, respectively. } 

\end{center}
\end{figure*}

\subsection{Clump masses, peak column and mean surface densities}
\label{sect:mass}

Detailed descriptions of the determination of clump mass and column density are presented in Paper\,I, and so we will only provide a brief summary of the procedure here. The isothermal clump masses are estimated using the \citet{hildebrand1983} method assuming that the total clump mass is proportional to the integrated sub-millimetre flux density measured over the source and assuming that the emission is optically thin:

\begin{equation}
\label{eqn:mass}
M_{\rm{clump}} \, = \, \frac{D^2 \, S_\nu \, \gamma}{B_\nu(T_{\rm{dust}}) \, \kappa_\nu},
\end{equation}

\noindent where $S_\nu$ is the integrated 870\,\mum\ flux density, $D$ is the distance to the source, $\gamma$ is the gas-to-dust mass ratio (assumed to be 100), $B_\nu$ is the Planck function for a dust temperature $T_{\rm{dust}}$, and $\kappa_\nu$ is the dust absorption coefficient taken as 1.85\,cm$^2$\,g$^{-1}$ (\citealt{schuller2009} and references therein). We use the dust temperatures derived in Sect.\,\ref{sect:seds} and assume that this is a reasonable estimate of the average temperature of the whole clump.

The peak column densities are estimated from the peak flux density of the clumps using:

\begin{equation}
N_{\rm{H_2}} \, = \, \frac{S_{\rm{\nu,peak}} \, \gamma}{B_\nu(T_{\rm{dust}}) \, \Omega \, \kappa_\nu \, \umu\, m_{\rm{H}}},
\end{equation}

\noindent where $S_{\rm{\nu,peak}}$ is the peak flux, $\Omega$ is the beam solid angle, $\umu$ is the mean molecular weight of the gas (which we take to be equal to 2.8), $m_{\rm{H}}$ is the mass of the hydrogen atom, and $\kappa_\nu$ and $\gamma$ are as previously defined. 
 
The clump mass and column density distributions are shown in the left and right panels of Fig.\,\ref{fig:mass_colden}, respectively. The clump masses range from as little as a few solar masses to several times 10$^4$\,\msun, with a peak in the distribution at $\sim$400\,\msun, which is likely to indicate the completeness of the survey (we will use this clump mass value where appropriate in subsequent analysis). The shape of the distribution of clump masses  for the distance-limited sample (upper left panel of Fig.\,\ref{fig:mass_colden}) is significantly different from the full sample and is log-normal; this is in contrast with many other studies (e.g., \citealt{eden2015, moore2015, csengeri2017}) that have tended to report power-law clump mass profiles that have slopes that are similar to the initial mass function (IMF). The column density distributions for the distance-limited and full samples have similar shapes and peak at approximately the same value. This should be a relatively distance-independent parameter, and this is confirmed by the similarity of their shapes: the higher number of sources in each bin is simply a result of the larger volume probed by the full sample.

We have previously noted trends relating increasing temperatures and luminosities to the evolutionary sequence.  We have also seen, however, that the clump radius seems to increase, which is a little harder to interpret. A comparison of distance-limited subsamples illustrates some interesting trends that might provide some insight into the increasing sizes. We note there is a trend for clump masses for the quiescent, protostellar and YSO phases to be higher in earlier stages, however, below a few 100\,\msun\ this is likely to be due to incompleteness; above this threshold the cumulative distribution curves begin to converge. This is also the reason for the differences seen for these three phases in the column density distributions. We do not, therefore, read too much into the differences in the column density and clumps masses for the quiescent, protostellar and YSO phases. We do, however, find that that the clump mass and column densities of the MSF clumps are both significantly higher than for any of the other subsamples. Among the distance-limited samples, almost every clump above 1000\,\msun\ and/or possessing a column density above Log($N$(H$_2$)) = 23\,cm$^{-2}$ is associated with massive star formation: the massive star forming clumps tend to be significantly more massive, have significantly higher column densities and are significantly larger than their non-MSF counterparts. 

There are two possible explanations for the observed properties of the MSF clumps: 1) that they evolve from lower-mass clumps associated with the earlier evolutionary stages by acquiring additional mass and growing in size through global infall, or 2) they are simply larger and more massive at their inception and evolve so rapidly that they are not likely to be observed in the early phases.  

The first of these options seems less likely given the trend we see for decreasing clump masses and column densities between the quiescent and YSO phases. We can estimate the amount of material that a clump might gain via infall from the surrounding lower density envelope. Assuming a clump with a typical radius of 0.5\,pc (see Fig.\,\ref{fig:size_parameter}), infall velocity of 1.5\,\kms\ and particle density of 5$\times 10^3$\,cm$^{-3}$ (e.g., \citealt{wyrowski2016}), the mass infall rate ($\dot M$) is $\sim1.5\times10^{-3}$\,\msun\,yr$^{-1}$. As the time scales for massive-star formation is in the region of several 10$^5$\,yr (\citealt{mottram2011b,davies2011}), the total increase in clump mass due to infall (166-500\,\msun\ for 1-3$\times 10^5$\,yr) is likely to be a relatively small fraction of the mass of the massive-star forming clumps but could potentially be more significant for low-mass clumps where the time scales are likely to be longer.


Furthermore, if infall were having a significant impact, we would expect the mass surface density to increase as the clumps evolve due the influx of material and presumably the contraction of the clumps themselves. In Fig\,\ref{fig:surface_density} we show the mass surface densities for the four evolutionary subsamples. Rather surprisingly, this plot reveals that the pre-stellar clumps actually have the highest surface density, and this decreases with advancing evolutionary stage. So rather than the mass surface density increasing due to contraction and infall, we find the opposite trend. The decreasing mass surface density as a function of evolution $may$ be linked to the ongoing star formation processes within the clumps, which is something that we will investigate in more detail in a later section. 

Given the lack of any evidence that might support the hypothesis that the lower mass clumps can evolve to have larger masses, sizes and column densities of clumps associated with massive star formation, the second option therefore seems more plausible and is something we will revisit in later sections. 

\begin{figure}
\begin{center}

\includegraphics[width=0.49\textwidth, trim= 0 0 0 0]{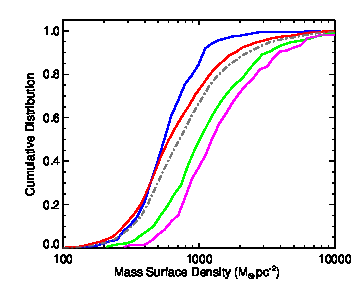}

\caption{\label{fig:surface_density} Cumulative distribution of the mean clump mass surface densities (\mclump/pc$^2$) for the four evolutionary distance-limited subsamples (colours as given in Fig.\,\ref{fig:size_parameter}).} 

\end{center}
\end{figure}

\subsection{Uncertainties in the derived parameters}
\label{sect:uncertainties}

The typical uncertainty on the distances estimated from Bayesian distance algorithm is $\sim$0.3\,kpc (\citealt{reid2016}; this is a 30\,per\,cent error for clumps located at 1\,kpc, but only a few per cent for sources located at distances of 10\,kpc or more. At the mean distance of 5\,kpc, the uncertainty in distance is $\sim$6\,per\,cent. The mean uncertainty in the integrated flux determined from the SED fitting is $\Delta S_\nu/S_\nu \sim$42\,per\,cent. The uncertainty in the luminosity is determined by combining the uncertainties in distance and integrated flux in quadrature, but since the error in the latter is significantly larger than in the former, the typical luminosity uncertainty is totally dominated by the uncertainty in the integrated flux, and effectively has the same level of uncertainty (i.e., $\Delta L/L \sim 42$\,per\,cent).

The uncertainty in the clump mass is estimated from those on the dust temperature ($\Delta T/T \sim 10$\,per\,cent), integrated 870-$\mu m$ flux ($\Delta S_{\rm{\nu,peak}}/S_{\rm{\nu,peak}} \sim 15$\,per\,cent \citealt{schuller2009}) and distance ($\Delta D/D \sim 6$\,per\,cent), added in quadrature. The typical error in clump mass is thus estimated to be of order 20\,per\,cent. The uncertainty on the column density is considered to be similar to that on the clump masses. 

The uncertainties reflected here are the measurement errors; however, all of the calculated parameters are likely to be dominated by inaccuracies in the many assumptions used in determining them (such as the gas-to-dust ratio, the dust opacity, and the value of $\beta$ used in the SED fitting), all of which are poorly constrained. The actual uncertainties are therefore likely to be significantly larger than those determined above. These uncertainties should therefore be considered as lower limits, and the real errors in the measured values could differ by a factor of a few. As noted in the previous papers, however, these additional uncertainties are likely to be systematic and so are unlikely to have a significant impact on the overall distributions or the statistical analysis.   

\subsection{Comparison with previous ATLASGAL results}
\label{sect:revised_masses}

The masses and column densities used in the analyses presented in the previous papers in this series were determined using a temperature of 20\,K. This was necessary because more reliable temperature estimates were not available. Furthermore, a number of distances have been updated and so it is necessary to determine how these improvements have affected the previous values and evaluate whether these have had a significant impact on our previous results. 

The values that are likely to be most affected are the luminosities, clump masses, and column densities. Comparing the luminosities we find them to be in excellent agreement (previous mean value for $L_{\rm{bol}}$ = 10$^{4.01\pm0.03}$, compared to the new value of $L_{\rm{bol}}$ = 10$^{4.08\pm0.03}$). The previous luminosities were determined using the \citet{robitaille2007} models and Hi-GAL fluxes, and given that the photometry has been determined using two different methods and the SEDs have been determined using two independent models the agreement is very reassuring. 

We show the bolometric luminosity distribution of the whole sample and the MSF clumps in the upper panel of Fig.\,\ref{fig:mass_column_den}.  This reveals that the MSF clumps dominate the upper end of the luminosity distribution confirming that our previous efforts of identifying all of the most active sites of massive star formation have been very successful (Papers\,I, II and III). However, we do also note that there are some very luminous clumps that are currently not in our MSF clump sample. There are few reasons why these were missed: 1) these clumps are associated with evolved \hii\ regions and are resolved in the MSX survey and so not included in the MSX point source catalogue (\citealt{price2001}) and consequently not included in the RMS sample (\citealt{lumsden2013}); 2) the clumps are located towards the Galactic centre in a region excluded from the RMS survey due to problems with source confusion (i.e., 350\degr$< \ell <$355\degr\ or 5\degr$ <\ell <$10\degr); 3) they were not located in the region covered in the MMB catalogue used to identify embedded massive protostars in Paper\,I (i.e., 186\degr$< \ell <$20\degr). 

\begin{figure}
\begin{center}

\includegraphics[width=0.49\textwidth, trim= 0 0 0 0]{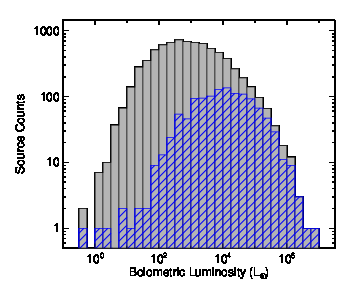}

\includegraphics[width=0.49\textwidth, trim= 0 0 0 0]{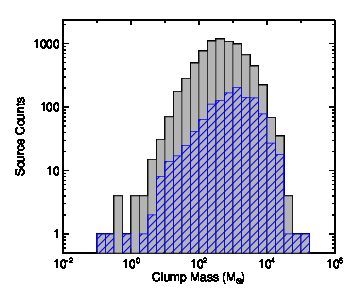}
\includegraphics[width=0.49\textwidth, trim= 0 0 0 0]{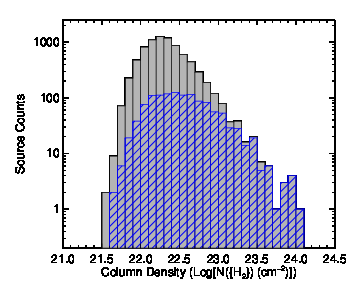}

\caption{\label{fig:mass_column_den} Luminosity, clump mass and peak column density distributions for the whole sample (grey) and the MSF subsample (blue) are presented in the upper, middle and lower panels, respectively. The bin sizes used in the upper, middle and lower panels is 0.25, 0.25 and 0.1\,dex, respectively. } 

\end{center}
\end{figure}

The 20\,K temperature that was previously used to estimate the clump masses and column densities is significantly lower than the mean temperature determined from the SED fits to the MSF clumps ($24\pm0.16$\,K with a standard deviation of 5\,K). We might therefore expect the newly estimated clump masses and column densities to be systematically lower than those previously calculated (difference of 4\,K corresponds to a decrease of $\sim$30\,per\,cent). The prior average clump mass was 10$^{3.29\pm0.02}$\,\msun, a factor of $\sim$2.2 times higher on average than the mean value of 10$^{2.95\pm0.02}$\,\msun\ calculated in this paper. Column densities differ by the same factor. This corresponds to a difference of $\sim$55\,per\,cent, which is slightly higher than expected but is likely to result from that the more massive clumps tend to be more active warmer and these skew the mean mass to a lower value. 

One of the main conclusions from \citet{urquhart2014_csc} was that massive star formation appeared to be associated  with the most massive and highest column density clumps and that there were no clumps with column densities above 10$^{23.5}$\,cm$^{-2}$ that were not already associated with star formation. The question is how has this finding been affected by the change to the column densities. In middle and lower panels of Fig.\,\ref{fig:mass_column_den} we show the clump mass and column densities for the whole sample and the massive star forming clumps, respectively. It is clear from these plots that although the clump masses and column densities have been revised down they are still significantly more massive and have higher column densities than the general population of clumps.




\section{Galactic distribution}
\label{sect:gal_distribution}

\begin{figure*}
\centering
\includegraphics[width=0.98\textwidth, trim= 0 20 0 0]{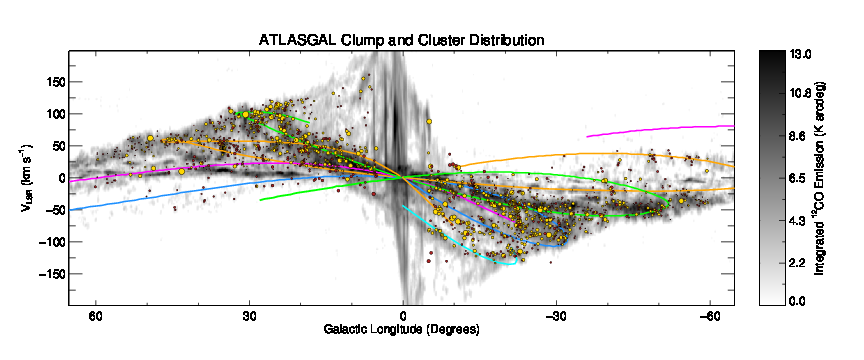}

\caption{Longitude-velocity ($\ell$-$v$) distribution of clumps and complexes. This figure is the same as Fig.\,\ref{fig:vlsr_distribution} overlaid with the positions of the clusters; these are shown by the larger yellow filled circles. The sizes of the individual clumps and clusters give an indication of their masses. }
\label{fig:lv_cluster_distribution}

\end{figure*}

In Sect.\,\ref{sect:vlsr} we compared the position of the ATLASGAL sources with the large-scale molecular gas in the longitude-velocity plane, and found a strong correlation between the locations of the dense gas traced by dust emission and the loci of the main spiral arms. In this section we revisit this analysis using the mass distribution of clumps and clusters as this is less likely to be biased by the large numbers of nearby and predominately lower-mass clumps. 

In Fig.\,\ref{fig:lv_cluster_distribution} we show the longitude-velocity distribution of isolated clumps and clusters. The background greyscale image is the integrated $^{12}$CO (1-0) emission (\citealt{dame2001}), which traces the large-scale distribution of molecular gas. The loci of the spiral arms are overlaid (\citealt{taylor1993}) to facilitate comparison between them and the ATLASGAL sources.  It is clear from Fig.\,\ref{fig:lv_cluster_distribution} that all of the ATLASGAL sources are located within the envelope of molecular gas traced by the $^{12}$CO (1-0) emission. Furthermore, we find that the vast majority of the dense gas is correlated with the spiral-arm loci located within 30\degr\ of the Galactic centre.  Detailed analysis of this correlation is difficult in the inner regions (i.e., $|\ell| < 40$\degr) due to the high density of sources and the overlapping of the spiral arms in $\ell v$-space. Beyond the inner 40\degr\ region there are some clear correlations between the ATLASGAL clumps and some segments of the spiral arms located in less confused parts of the parameter space (such as the far sides of the Perseus and Norma arms in the first quadrant and the far side of the Sagittarius arm in the fourth quadrant). 

Although the vast majority of ATLASGAL sources are seen to be tightly clustered around the four-arm Milky Way model spiral arm loci, there are a few small clusters located between the spiral arm tangents (e.g., between the Scutum-Centaurus and Sagittarius tangents in the first quadrant and between the  Sagittarius tangent and Perseus arm, again in the first quadrant). The  cluster of sources located between the Scutum-Centaurus and Sagittarius arm tangents is correlated with significant amounts of molecular gas that is traced by CO studies (\citealt{lee2001, stark2006, rigby2016}), but the presence of the ATLASGAL clumps in these regions is the first indication that these regions are also associated with significant amounts of dense gas. \citet{rigby2016} presented a very detailed $^{13}$CO (3-2) high-resolution map of this region and discussed the origin for the  emission (minor arm, an extension of the Scutum-Centaurus, a bridging structure or a spur) but at present its nature is unclear. Despite the presence of significant quantities of molecular material and dense clumps, the RMS survey (\citealt{urquhart2014_rms}) finds no MYSOs or UC\hii\ regions in this region, which suggests that massive star formation is lower than in the major arms. 

In Fig.\,\ref{fig:galactic_cluster_distribution} we show the distribution of clusters and clumps with respect to the expected  position of larger-scale Galactic features, primarily the spiral arms and the Galactic bar. A comparison of the distribution of clumps and clusters with the spiral arms reveals some good agreement, particularly in the first quadrant where the rotation curve is more tightly constrained. 

\begin{figure}
\centering

\includegraphics[width=0.49\textwidth, trim= 0 0 0 0]{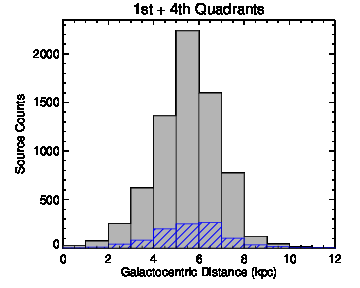}
\includegraphics[width=0.49\textwidth, trim= 0 0 0 0]{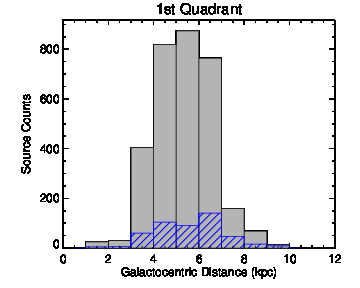}\\
\includegraphics[width=0.49\textwidth, trim= 0 0 0 0]{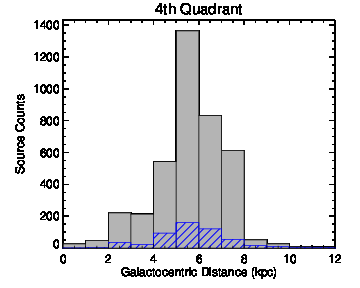}

\caption{\label{fig:galactocentric_dist} Galactocentric distance distribution of all clumps (grey) and the MSF clumps (blue). The upper panel shows the distribution for the whole sample while the middle and lower panels show the galactocentric distribution of the first and fourth quadrants, respectively. The bin size used in all of the plots is 1\,kpc.  } 

\end{figure}

We show the distribution of clumps as a function of Galactocentric distance in the upper panel of Fig.\,\ref{fig:galactocentric_dist}. This plot reveals that the vast majority of sources are located between 2 and 8\,kpc, with a strong peak at $\sim$5.5\,kpc. The sample of MSF clumps is similarly distributed, with the peak of the distribution slightly farther from the Galactic centre. Although the distribution of the whole sample looks simple and suggests that the material in the inner Galaxy is rather smoothly dispersed, the combination of the clumps located in the first and fourth quadrants hides a significant amount of structure. To illustrate this, we have plotted the Galactocentric distributions of the first and fourth quadrant subsamples separately in the middle and lower panels of Fig.\,\ref{fig:galactocentric_dist}. We see two prominent peaks in the distribution of the MSF clumps in the first quadrant; these are at 4.5 and 6.5\,kpc, which correspond to the Scutum-Centaurus arm and its intersection with the end of the Galactic long bar and the Sagittarius arm. Inspecting the fourth quadrant distribution we note a peak located at 5.5\,kpc, which is coincident with a combination of clumps located at the far side of the long bar and the Norma arm, and a second weaker peak at $\sim$2.5\,kpc which is likely to be associated with the near side of the 3\,kpc expanding arm.

The presence of peaks in the Galactocentric distribution at locations where spiral-arm segments and the Galactic bar are predicted to be supports the spiral model of the Galaxy and that there are significant differences in the structure of the first and fourth quadrants. These structural features in the distribution of MSF clumps have been seen and commented on by many other studies (e.g., \citealt{urquhart2014_rms,anderson2009b}) including a previous paper in this series (\citealt{urquhart2014_atlas}); however, the distances presented here have been estimated using a different rotation model, and approximately 20\,per\,cent of the kinematic distance ambiguities have been revised.  The continual presence of these features would suggest they are invariant with respect to the model used.   

\begin{figure}
\begin{center}

\includegraphics[width=0.49\textwidth, trim= 0 0 0 0]{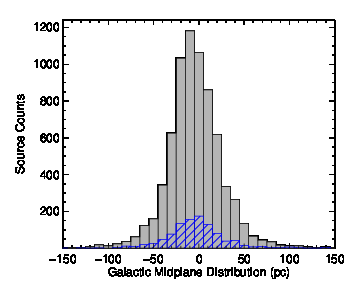}

\caption{\label{fig:scaleheight_dist} Galactic midplane distribution of all clumps (grey) and clumps associated with massive star formation (blue). The bin size used is 10\,pc.  } 

\end{center}
\end{figure}

We show the latitude distribution of all ATLASGAL clumps and those associated with MSF clumps in Fig.\,\ref{fig:scaleheight_dist}. A KS test is unable to reject the null hypothesis that the two samples are drawn from the same parent population ($p$-value  = 0.42). A fit of the absolute values of clump distance from the mid-plane finds that the scale height, determined from a $1/e$ fit to the data, is $26.04\pm0.02$\,pc for the whole sample and $25.15\pm0.13$\,pc for the MSF clumps, in excellent agreement with one another and other survey results (e.g., methanol masers, MYSOs and \hii\ regions; \citealt{green2011b,urquhart2014_rms} and Paper\,II, respectively). We can conclude from this that the dense material is concentrated in a relatively narrow region of the Galactic disk, and that the dense gas and massive star formation are tightly correlated with each other.

\subsection{Correlation with spiral arms}

\begin{figure*}
\centering
\includegraphics[width=0.98\textwidth, trim= 0 0 0 0]{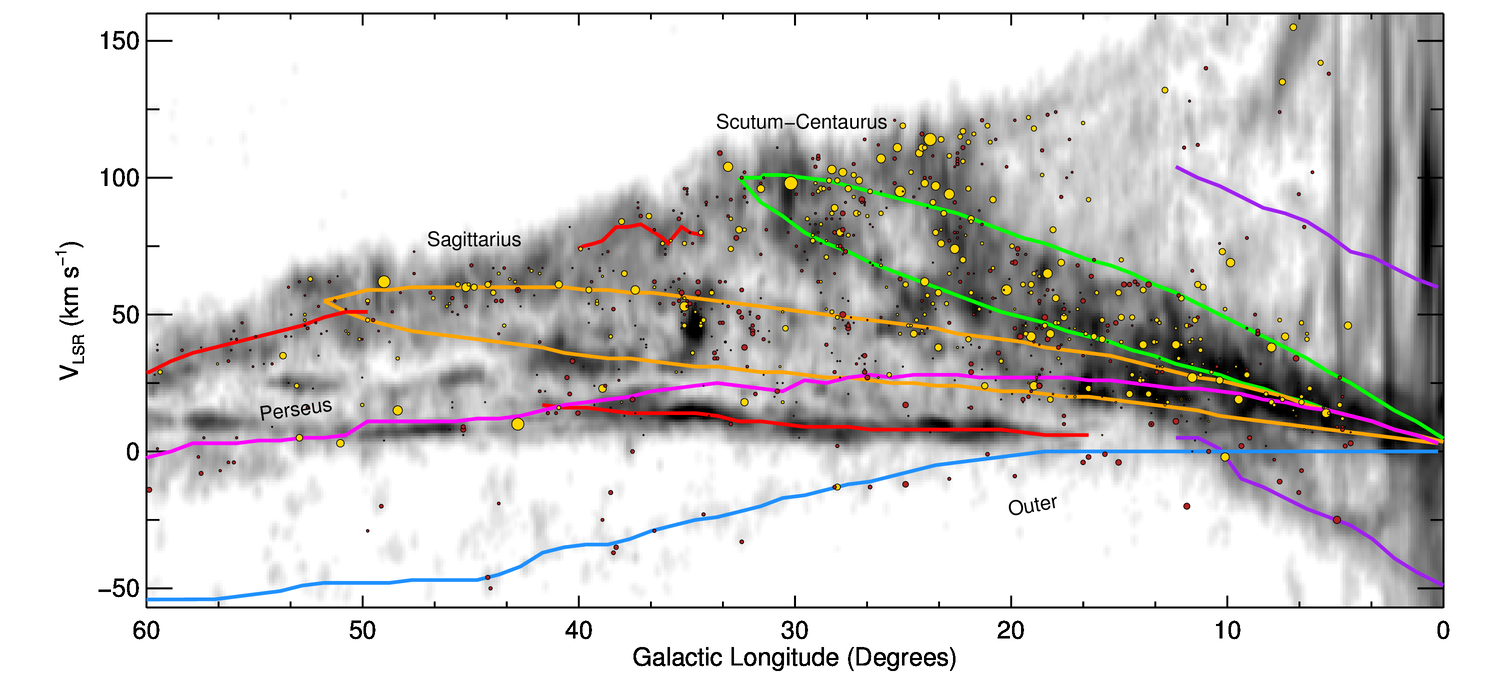}

\caption{Longitude and velocity ($\ell$-$v$) distribution of clumps and complexes located in the first quadrant. The greyscale image is the integrated $^{12}$CO emission from \citet{dame2001}, and the red and yellow circles show the positions of the clumps and clusters, respectively. The sizes of the symbols give an indication of the masses of individual clumps and clusters. This map differs from the one presented in Fig\,\ref{fig:lv_cluster_distribution} in that the spiral-arm loci are taken from the \citet{reid2014} model. This map shows the loci of the four main arms (Outer, \scu, \sag\ and \per, colours are as in Fig.\,\ref{fig:vlsr_distribution}),  but also includes the near and far 3\,kpc arms (lower and upper purple curves, respectively), the local arm (left-most red curve), the Aquila Rift (longest red curve that runs between the Perseus and Outer arms) and the Aquila spur (red curve located between the \sag\ and \scu\ tangents). \label{fig:lv_reid_loci} } 
\end{figure*}

We have overlaid the spiral-arm loci derived by \citet{taylor1993} updated by \citet{cordes2004}  in Fig.\,\ref{fig:lv_cluster_distribution}, as these are readily available for the whole Galaxy. There is a more recent set of loci derived by \citet{reid2014} from maser parallax work, but these are currently poorly constrained for the third and fourth quadrants and so have not been used in this figure. 

Fig.\,\ref{fig:lv_reid_loci} shows a zoom of Fig.\,\ref{fig:lv_cluster_distribution} that focuses on the first quadrant and the spiral-arm loci derived by \citet{reid2014}. In addition to the four main arms (which are also shown in Fig.\,\ref{fig:lv_cluster_distribution}), we include the near and far 3\,kpc arms, the local arm, the Aquila Rift and the Aquila spur. Compared to Fig.\,\ref{fig:lv_cluster_distribution}, this image reveals a significantly better correlation between the ATLASGAL sources and the spiral arm loci, particularly for the Outer and Sagittarius arms. 

\begin{figure}
\centering
\includegraphics[width=0.49\textwidth, trim= 0 0 0 0]{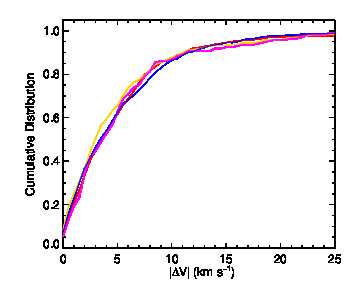}

\caption{ Absolute velocity offset between the ATLASGAL clumps and clusters and their nearest spiral arm locus (red and yellow curves, respectively). The blue curve shows the results obtained by repeating the analysis for the GRS catalogue of molecular clouds.   \label{fig:spiral_arm_offsets} } 

\end{figure}

We have calculated the velocity difference between each source and all spirals arms along that source's line of sight to assess the correlation between the spiral arm loci and the ATLASGAL sources. We then simply assume the source is associated with the spiral arm whose associated velocity difference is lowest.  We show the results of this analysis in the form of a cumulative distribution plot in Fig.\,\ref{fig:spiral_arm_offsets}. It is clear from this plot that the velocities of the vast majority of clumps and clusters are tightly correlated with the spiral arms (almost 90\,per\,cent within 10\,\kms) and that there is no significant difference between the clusters and isolated clumps. We note that the loci of the spiral arms in \emph{lv}-space are not well determined, particularly towards the inner part of the Galaxy, however, the different models do not vary a great deal particularly towards the inner part of the disk (cf \citealt{stark2006}). There is also an issue of non-circular motions due to streaming motions as clouds pass through the spiral arms, however, these tend to be of order $\pm$7\,\kms\ (\citealt{reid2014}), which is close to the mean velocity difference seen in the distribution, and is probably responsible for the spread we find in the velocity differences. 

We have repeated this analysis using the catalogue produced by the GRS (\citealt{rathborne2009}) to compare the distribution of dense gas with larger molecular clouds; the results of this analysis are overlaid in blue. A comparison between the dense gas and larger more diffuse molecular cloud locations reveals no significant difference in their correlation with the spiral arms. We have also compared the distributions of the different samples (clusters, clumps and molecular clouds) broken down by mass and luminosity with the spiral arm loci to determine if the more massive and luminous sources were more tightly correlated with the arms; however, this revealed no significant dependence on either of these two parameters. The distribution of star formation associated with respect to the arms is therefore similar to the distribution of molecular gas, which leads us to conclude that the star formation is not enhanced by the spiral arms and is likely simply the result of source crowding, as suggested by \citet{moore2012}.

\section{Empirical star formation relations}
\label{sect:empirical_relationships}

In the previous sections we have determined the physical sizes, masses, column densities and luminosities for an almost complete population of dense clumps. This is likely to also be complete to the whole embedded evolutionary sequence for massive stars, and allows for robust conclusions to be drawn from a detailed statistical analysis of these data. We have already found that all of the clumps appear to be rather spherical in structure and are centrally condensed (Paper\,III).  We have found no significant differences in the structural properties of the MSF clumps compared with the rest of the clumps, which is perhaps not terribly surprising as $\sim$90\,per\,cent appear to be in the process of forming stars. The clumps typically have masses of 500-1000\,\msun\ and radii of $\sim$0.5-1\,pc, which is similar to the masses and physical sizes expected to form stellar clusters \citep{lada2003}; it is therefore likely that the majority of these are in the process of forming a stellar cluster. 

It follows that the derived clump properties (e.g., mass, density, radius etc.) and the bolometric luminosities of the embedded sources are therefore much more likely to be related to an embedded cluster than to a single star.  The luminosity of a given MSF clump (which tend to be physically larger, more massive and have higher peak column densities than the other evolutionary clump samples) is likely to be dominated by an embedded massive protostellar object traced by the methanol maser, MYSO or \hii\ region.

\subsection{Analysis tools}

In the following subsections we will look at the correlations between the derived properties in an effort to gain some insight into the statistical nature of star formation in the Galaxy. We use a non-parametric partial Spearman correlation test to determine the level of correlation between pairs of parameters, as this removes their mutual dependence on the distance (Marquis bias, e.g., \citealt{yates1986}), of the form $r_{\rm AB,C}$, where
\begin{equation}
r_{\rm AB,C} \; = \; \frac {r_{\rm AB} -  r_{\rm AC} r_{\rm BC}}
{[(1-r^2_{\rm AC})(1-r^2_{\rm BC})]^{1/2}},
\end{equation}
\noindent where A and B  are the two dependent variables and C is the independent variable (in our case the distance), and  $r_{\rm AB}$, $r_{\rm AC}$ and $r_{\rm BC}$ are the Spearman rank correlation coefficients for each pair of parameters. The significance of the partial rank correlation coefficients is estimated using  \mbox{$r_{\rm AB,C} [(N-3)/
(1-r_{\rm AB,C}^2)]^{1/2}$} assuming it is distributed as Student's $t$-statistic (see \citealt{collins1998} for more details). We check the significance of all correlations by their $p$-value, which is the probability that the absolute value of the correlation for uncorrelated data could be equal or higher than the measured value (similar to the KS test $p$-value). We consider a correlation to be significant if the $p$-value is lower than 0.13\,per\,cent (i.e., difference $>$3$\sigma$).

\begin{figure*}
\centering
\includegraphics[width=0.49\textwidth, trim= 0 0 0 0]{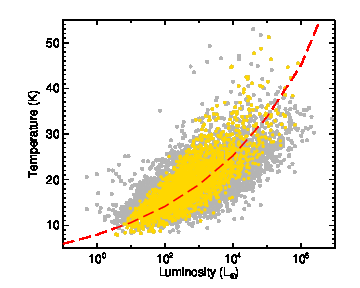}
\includegraphics[width=0.49\textwidth, trim= 0 0 0 0]{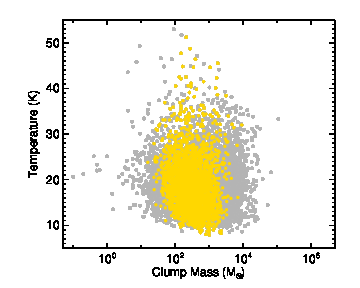}
\includegraphics[width=0.49\textwidth, trim= 0 0 0 0]{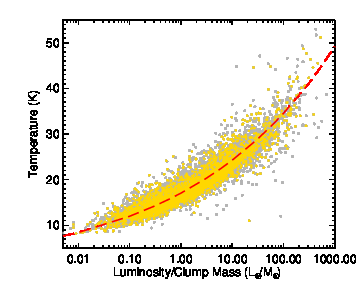}
\includegraphics[width=0.49\textwidth, trim= 0 0 0 0]{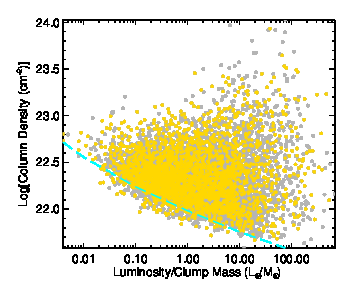}

\caption{Scatter plots of temperature vs.\ luminosity and temperature vs.\ clump mass (upper left and right panels, respectively). The correlation between temperature and \lm\ ratio is shown in the lower left panel while the lower right panel shows the column density and \lm\ ratio relationship. The whole sample is shown as grey circles, while the yellow circles show the distribution of the distance-limited sample. The red dashed lines on the left panels are power-law fits to the data. No significant correlation is found in the distributions in the right-hand panels and so no fits have been made to these data. The dashed cyan line shown in the lower right panel indicates the lower limit to the column density imposed by the observational sensitivity limit. \label{fig:l_m_vs_temp}} 

\end{figure*}

If a significant correlation is found, we fit the data using a linear least-square fit method. This is typically fit in log-log space excluding non-detections and assuming the errors are equal along both axes. These fits capture general trends in the data. As noted in the previous papers in this series, we caution that the mere existence of any correlations between the observed properties does not imply causation. The analysis of the positions of sources in the relevant parameter space can, however, highlight potentially interesting trends that might be present in the data, as well as draw attention to extreme objects, which in turn, may provide some insight into the underlying physics.

\subsection{Relationship between luminosity, mass and temperature}

Previous studies have found a positive correlation between the luminosity of the embedded massive protostars and the gas temperatures of their natal clumps (e.g., \citealt{urquhart2011_nh3}). We are able to extend this analysis to lower-luminosity and less-evolved sources using the ATLASGAL sample. In the upper left panel of Fig.\,\ref{fig:l_m_vs_temp} we show the relationship between temperature and luminosity. This plot shows a strong correlation between these two parameters with an $r_{\rm{AB,C}}$ of 0.80 ($p$-value $\ll$ 0.001). The fit to these data yields a slope of 0.145$\pm$0.0014 for the distance limited sample, which is equivalent to $L=(-6.0 \pm 0.08)\times T^{(6.9\pm0.06)}$. Despite the strong correlation there is a significant amount of scatter that makes it difficult to assign a particular temperature from the luminosity or vice versa. 

In the upper right panel of Fig.\,\ref{fig:l_m_vs_temp} we show the relationship between clump mass and temperature. We note that there is no evidence for a correlation between clump mass and temperature as illustrated in the upper right panel of Fig.\,\ref{fig:l_m_vs_temp}. Given that the clump mass is calculated using the dust temperature we might have expected these parameters should be strongly correlated; however, the flux is also used to determine the mass, and the flux and temperature are tightly correlated with each other. As the temperature rises so to does the emitted flux proportionally and since these are present as the denominator and numerator, respectively, the mass of a clump is effectively invariant to temperature.

\begin{figure}
\begin{center}

\includegraphics[width=0.49\textwidth, trim= 0 0 0 0]{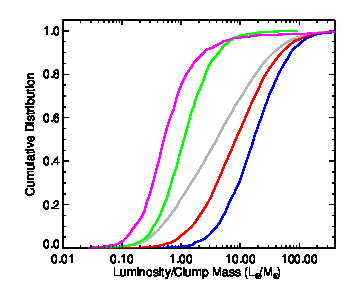}

\caption{\label{fig:l_m_cdf} Cumulative distribution of the \lm\ ratio for the four evolutionary subsamples (colours as given in Fig.\,\ref{fig:size_parameter}).} 

\end{center}
\end{figure}

\begin{figure*}
\centering
\includegraphics[width=0.98\textwidth, trim= 0 0 0 0]{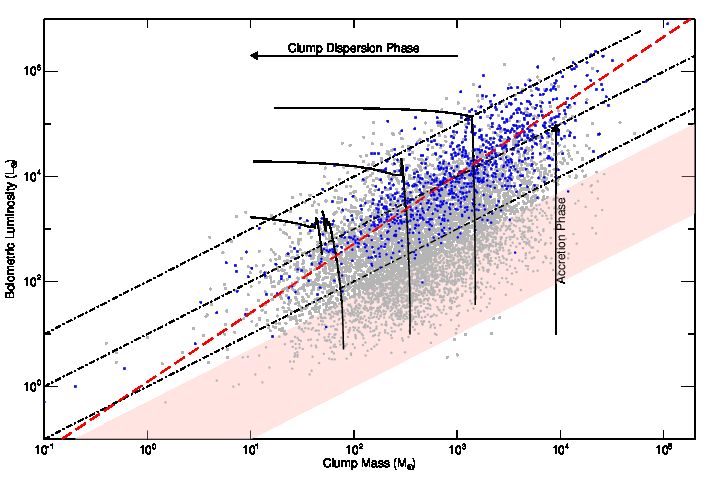}

\caption{Clump mass-bolometric luminosity relationship for non-msf clumps (grey circles) and the MSF clumps (blue). The lower, middle and upper diagonal dash-dot lines indicate the \lbol/\mclump\,=\,1, 10 and 100\,\lsun/\msun, respectively. The solid black curves, running respectively left to right,  show the model evolutionary tracks calculated by \citet{molinari2008} for stars with final masses of 6.5, 13.5 and 35\,\msun, respectively. The light-red shaded area indicates the region of the parameter space we would expect to find quiescent clumps (temperatures between 10-15\,K). The long-dashed diagonal red line show the results of a log-log, outlier-resistant, linear fit to the MSF clumps; this has a slope of $1.314\pm0.0189$ and an intercept of $0.087\pm 0.056$. \label{fig:lum_mass_distribution} } 

\end{figure*}






We show the relation between the temperature and the \lm\ ratio in the lower left panel of Fig.\,\ref{fig:l_m_vs_temp}. It is clear from this plot that these two parameters are very strongly correlated with each other ($r_{\rm{AB,C}}$ of 0.94 and $p$-value $\ll$ 0.01). They are sufficiently well-correlated that it is possible to use one to reliably predict the other, and this relation holds over almost 6 orders of magnitude in \lm\ and the whole range of measured clump temperatures. The \lm\ ratio is a distance-independent parameter, and so the scatter and correlation coefficient are significantly better than that seen in the temperature-luminosity plot. The linear fit to the logs of these data gives an  slope of 0.15235$\pm$0.00059 and an intercept of 1.231$\pm$0.00052; this results in the power-law relation \lm\ $ = 10^{(-8.08\pm0.04)}\times T^{(6.564\pm0.035)}$.

We have found, then, that both luminosity and \lm\ ratio are strongly correlated with the dust temperature, although the large scatter in the data and strong power-law relationship of the luminosity-temperature distribution makes it a somewhat poor indicator of the stellar evolution taking place within the clumps i.e., a particular temperature can correspond to a large range of luminosities ($\sim$3 orders of magnitude). The \lm -temperature relation shows a stronger correlation and has a lower power-law dependence; it is therefore less sensitive to small changes in temperature.   This relation is independent of distance, removing one source of uncertainty. Furthermore, if the total luminosity is a measure of the energy output of an embedded cluster, then \lm\ is a measure of the energy output per unit mass (we are effectively normalising the energy output by the clump mass to obtain an approximate measure of the instantaneous star formation efficiency, which is independent of distance and mass). The results of this \lm\ ratio versus temperature analysis is in excellent agreement with the results of an independent analysis recently reported by the Hi-GAL team (\citealt{elia2017}).

The lower right panel of Fig.\,\ref{fig:l_m_vs_temp} shows the scatter plot of column density and the \lm\ ratio. There is no obvious correlation between these two parameters, which suggests that the column density does not evolve significantly during the star formation process. The test indicates that there is a weak negative correlation ($r_{\rm{AB,C}} = -0.16$ with $p$-value $\ll$ 0.01), suggesting that the column density decreases as the clumps evolve.  This result may be an artifact of a significant observational bias: we are less sensitive to the lower column density pre-stellar clumps.  The dashed cyan line illustrates the minimum column density detectable for the range of temperatures of interest: this limit is estimated using the ATLASGAL survey's 5$\sigma$ sensitivity limit ($\sim$300\,mJy\,beam$^{-1}$), and appears to be a hard boundary at the lower luminosity/lower column density region of the plot. This limit is likely to be skewing the distribution, and so we do not consider the negative correlation to be reliable. 

We have already noted the lack of correlation between the clump mass and temperature.  It follows that if temperature and \lm\ ratio are good measures of evolution that the clump mass is therefore independent of evolution. We can then conclude at this point that neither the clump mass nor the peak column density change significantly as the star formation evolves. This implies that the majority of clumps have not yet reached the envelope-dispersion stage of their evolution, and that our sample is primarily associated with the earliest evolutionary and pre-stellar stages. The utility of the \lm\ ratio as a diagnostic of evolution is illustrated in Fig.\,\ref{fig:l_m_cdf}, which shows that the four evolution stages have clearly distinct profiles.

\subsection{Luminosity-mass relation}
\label{sect:lum-mass-relation}

We present the luminosity-mass  (\lbol-\mclump) distribution of all ATLASGAL sources in Fig.\,\ref{fig:lum_mass_distribution} in order to examine the relationship between the physical properties of the clumps and their associated star formation. This type of diagram has been used in studies of low-mass (\citealt{saraceno1996}) and high-mass star-forming regimes (\citealt{molinari2008, giannetti2013}), and is a useful diagnostic tool for separating different evolutionary stages. This figure shows the MSF and non-MSF clumps as blue and grey circles, respectively. For illustrative purposes we include evolutionary tracks derived by \citet{molinari2008} from a simple model for the formation of massive stars.  This model is based on the observational evidence that their formation is a scaled-up version of the inside-out collapse model (which has been successfully applied to low-mass star formation). These show how the luminosity of the forming massive star and the mass of its natal clump changes as the star evolve over time. These tracks consist of vertical and horizontal components that \citet{molinari2008} refer to as the \emph{main accretion} and the \emph{envelope clean-up} phases; these are indicated by arrows on the figure. The light-red shaded area indicates the region of the parameter space in which we would expect to find quiescent clumps, with temperatures of $\sim$10-15\,K\footnote{The lower end of the temperature range is determined by heating from the interstellar radiation field and cosmic ray heating ($\sim$10\,K), and the upper temperature from the mean temperature determined here from the SED fits to the quiescent clumps and from the mean ammonia temperature measured for IRDCs (e.g., \citealt{wienen2012}).}. The sample therefore covers all evolutionary stages from the quiescent starless/pre-stellar phase through to the formation of an \hii\ region, when feedback from the massive star halts further accretion and begins to disperse the natal cloud. We also include three lines of constant \lm\  (i.e., 1, 10 and 100\,\lsun/\msun) in Fig.\,\ref{fig:lum_mass_distribution}.

The MSF clumps are clearly the most luminous for any given clump mass: this is particularly true for the higher-mass clumps. It is also clear that the most massive stars are forming in the most massive and dense clumps. As noted in Paper\,III, nearly all of the MSF clumps are found within a relatively broad range of \lm\, but the upper and lower envelopes of this range are well described by values of 100 and 1\,\lsun/\msun, respectively. It is particularly interesting to note the upper bound to the \lm\ range, as this may indicate that there is a mechanism that works to limit the conversion of mass into stars: this could be linked to the feedback from the embedded proto-clusters. In Paper\,III we compared the luminosity for the MSF clumps to what would be expected assuming a star formation efficiency (SFE) of 10\,per\,cent, and found that the most massive clumps were under-luminous while the lower mass clumps were more luminous than expected: this suggests that the SFE naturally decreases with the mass of the cluster.  

The MSF clumps form a continuous distribution that is concentrated towards a \lm\ $\sim$ 10 \lsun/\msun\ locus that extends over approximately 4 orders of magnitude in both axes. This clustering around 10 \lsun/\msun\ is likely to trace the transition between the main accretion phase and the clump dispersion phase. If the accretion increases with the mass of the protostar  (as favoured by the population synthesis model; \mbox{\citealt{davies2011}}), the protostar will accelerate towards the end of its vertical track, reaching the end of its main accretion phase with increasing rapidity. Once the main accretion phase has ended, it takes a long time before the embedded star begins to significantly disrupt its natal clump, retarding its progression along its horizontal track.

Similar thresholds in \lm\ are discussed by \citet{giannetti2017} in the context of an evolutionary sequence for the process of high-mass star formation, confirming the above results. The authors compare different molecular tracers of physical conditions, and find that after an initial compression phase of the material in the clump (\lm\ $\lesssim$ 2 \lsun/\msun), the YSOs accrete material and grow in mass, reaching the ZAMS at \lm\ $\sim$ 10 \lsun/\msun, and start dispersing the parent clump. Compact \hii\ regions become common at \lm\ $\gtrsim$ 40 \lsun/\msun, however, we also note that few regions are found with \lm\ in excess of 100 \lsun/\msun.

In the previous papers we used a temperature of 20\,K to estimate the masses for the MSF clumps, and fitting these data resulted in a relatively constant \lm\ of 10\,\lsun/\msun\ (slope $1.03\pm0.05$) for all clump masses. We have repeated this analysis with the updated masses calculated using the dust temperatures, and find a slightly steeper relationship:   (${\rm{log}}(L) = 1.314\times {\rm{log}}(M) + 0.0874$). The updated value for the slope is in better agreement with the value reported by \mbox{\citet{molinari2008}} ($\sim$1.27) from a comparable fit to a sample of 27 infrared-bright sources, and consistent with the results of a similar analysis presented by \citet{giannetti2013} and \citet[][1.32$\pm$0.08]{urquhart2015_nh3}. The luminosity and clump masses are strongly correlated, with $r_{\rm{AB,C}}$ = 0.69 and $p \ll 0.001$, verifying that the correlation is statistically significant.

\subsubsection{Dependence of $L/M$ on clump mass}

\begin{figure}
\centering
\includegraphics[width=0.49\textwidth, trim= 0 0 0 0]{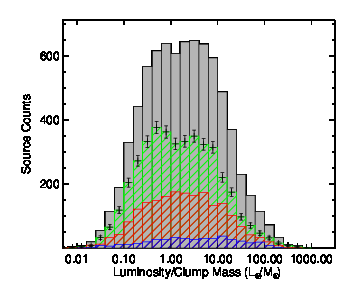}
\includegraphics[width=0.49\textwidth, trim= 0 0 0 0]{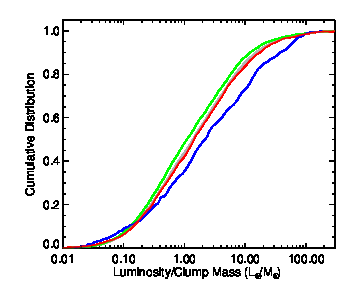}

\caption{Luminosity to clump mass ratio. In the upper and lower panels we show the frequency distribution of the \lm\ ratio for clumps with masses  M $<$ 1000\,\msun, 1000\,\msun $<$  M $<$ 5000\,\msun\ and M $>$ 5000\,\msun\ in green, red and blue, respectively. The distribution of the whole clump population is shown in grey.
\label{fig:lum_mass_dist} } 

\end{figure}

We have previously noted that the MSF clumps are generally the most luminous and have higher \lm\ ratios: this is illustrated in the lower panel of Fig.\,\ref{fig:luminosities} and Fig.\,\ref{fig:l_m_cdf}. We also observe from Fig.\,\ref{fig:lum_mass_distribution} that the ratio of the numbers of MSF to non-MSF clumps increases noticeably as the clump mass increases. This indicates that the \lm\ distribution is not the same for the whole mass range and may increase with clump mass. In Fig.\,\ref{fig:lum_mass_dist}, we compare the \lm\ distribution for three mass ranges (i.e., M $<$ 1000\,\msun, 1000\,\msun $<$  M $<$ 5000\,\msun\ and M $>$ 5000\,\msun) for the whole population of clumps. This figure clearly shows that the \lm\ ratios are significantly different for the three mass bins, and this is confirmed by KS tests ($p \ll 0.001$). The mean values and error for the Log(\lm) ratios for the increasing mass bins are  0.60$\pm$0.012, 0.80$\pm$0.022 and 1.10$\pm$0.052\,\lsun/\msun, respectively. Closer inspection of the lowest mass range reveals a bimodal distribution with peaks a few tenths of a dex to either side of \lm\ = 1\,\lsun/\msun\ (approximately 0.5 and 3.0\,\lsun/\msun). The upper panel of Fig.\,\ref{fig:lum_mass_dist} includes error bars calculated assuming \poi\ counting statistics (i.e., $\sqrt{N}$), and these confirm that the dip is significant. The distributions of the other two mass bins show only a single peak, although the distribution for the largest mass range is significantly flatter. 

We show the cumulative distributions for the three mass ranges in the lower panel of Fig.\,\ref{fig:lum_mass_dist}. This plot reveals that the \lm\ ratio for the most massive clumps (i.e.,  M $>$ 5000\,\msun) is significantly higher than for the other clump mass ranges. This suggests that the massive clumps ($>5000$\,\msun) have a very short pre-stellar lifetime and evolve more rapidly than lower mass clumps. This shift in the \lm\ ratio may also be due to the steeper-than-linear slope in the $L-M$ plot, and is therefore not necessarily evidence of evolution. In Fig.\,\ref{fig:SFF} we show the fractions of clumps found to be quiescent (70\,\mum\ dark), protostellar (24\,\mum\ dark) and YSOs (mid-infrared bright). This plot reveals a clear trend for an increasing fraction of YSO-forming clumps with increasing mass, and a decrease in the fraction of quiescent and protostellar clumps. The fraction of clumps harbouring a YSO increases from 60\,per\,cent to nearly 90\,per\,cent over a range of clump masses that extends over almost 2 orders of magnitude, while the fraction of quiescent clumps decreases from $\sim$15\,per\,cent to 5\,per\,cent over the same mass range. Furthermore, the fractions derived for the quiescent clumps are likely to be upper limits, as a significant number may host low-mass protostars that are not yet sufficiently luminous to be detected at 70\,\mum. This supports the hypothesis that the statistical lifetime of the earliest stages decreases with the mass of the clump (e.g. \citealt{mottram2011b} and \citealt{davies2011}).

\begin{figure}
\centering
\includegraphics[width=0.49\textwidth, trim= 0 0 0 0]{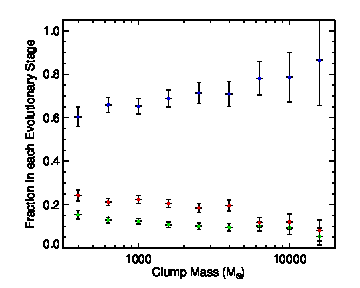}

\caption{Fractions of clumps sorted by evolutionary stage. The green, red and blue circles indicate the fractions of quiescent, protostellar and mid-infrared bright clumps (YSOs and MSF clumps), respectively, as a function of clump mass. The uncertainties are determined from \poi\ counting statistics. This plot only includes mass bins above the nominal survey completeness ($\sim$400\,\msun), and averages the source counts in bins of 0.2\,dex. 
\label{fig:SFF} } 

\end{figure}

It is also clear that the fraction of protostellar clumps diminishes with increasing clump mass: initially the decrease is quite modest but becomes noticeable for clump masses over a few times 10$^{3}$\,\msun\ where it drops from $\sim$20\,per\,cent to $\sim$10\,per\,cent. This suggests that once star formation begins, it evolves more rapidly in the more massive clumps than it does in the lower-mass clumps. 

At this point it is worth noting that the fraction of star-forming clumps identified by \citet{elia2017} is significantly lower than that found with our sample ($\sim$25\,per\,cent; see also \citealt{ragan2016}). They are, however, approximately 100 times more sensitive than ATLASGAL, and many of the 100,000 sources detected by HiGAL are likely to be lower-mass clumps. The trend seen in Fig.\,\ref{fig:SFF} of lower star formation fraction with decreasing clump mass would likely explain the difference in the star formation fraction observed between the results reported here and those reported by \citet{elia2017}. 

Ordinarily, the dip seen in the full-sample histogram in the upper panel of Fig.\,\ref{fig:lum_mass_dist} might not have been viewed as notable.  As already shown, this dip is significant in the lowest-mass clump distribution, and therefore suggests that this may be connected to the statistical lifetimes of the different evolutionary stages. We have already noted that the lifetimes of the quiescent and protostellar stages decrease with clump mass, and the bimodal distribution seen in the distribution of lower mass clumps might be related to the pre-stellar phases. Conversely, the smoothness of the distributions of the higher-mass clumps may indicate that there is effectively no observable pre-stellar stages for these larger clumps. This may seem to contradict the pre-stellar lifetime determined from Fig.\,\ref{fig:SFF}, but this quantity is estimated from the proportion of clumps that are dark at 70\,\mum:  as mentioned earlier, not all of these are necessarily pre-stellar as some have been associated with outflows.  

The lifetimes of massive stars has been investigated by \citet{davies2011} and \citet{mottram2011b}, both of which determined the lifetimes for \hii\ regions to be several 10$^5$\,yr. Mottram fitted the empirical data and derived the following relationship for the lifetime as a function of the source luminosity: (log($t_{\rm{H{\sc II}}}$/yr = ($-0.13\pm0.16) \times {\rm{log}}(L/{\rm{L\odot}}) + (6.1\pm0.8) $). We can obtain an estimate of the clump lifetimes by combining this with the relationship we derived for the clump mass and luminosity discussed in Sect.\,\ref{sect:lum-mass-relation}. If we assume that the \hii\ region lifetime dominates the timescales of all of the embedded stages (\citealt{motte2007}), we can use the fraction of sources in each stage to estimate the statistical lifetime of each stage.

\setlength{\tabcolsep}{1pt}

\begin{table}
\begin{center}\caption{Statistical lifetime of clumps as a function of clump mass.}
\label{tbl:statistical_lifetimes}
\begin{minipage}{\linewidth}
\small
\begin{tabular}{.......}
\hline \hline

 \multicolumn{1}{c}{log[\mclump]}&  \multicolumn{1}{c}{log[Luminosity]}&\multicolumn{5}{c}{Statistical lifetimes (10$^5$\,yr)}   \\
\cline{3-7}
  \multicolumn{1}{c}{(\msun)}&  \multicolumn{1}{c}{(\lsun)}&\multicolumn{1}{c}{$t_{\rm{HII}}$}  &\multicolumn{1}{c}{Quiescent}  &\multicolumn{1}{c}{Protostellar}  &\multicolumn{1}{c}{YSO}  &\multicolumn{1}{c}{MSF}  \\
\hline
2.60 &3.50& 4.41& 0.68& 1.07& 2.25& 0.41\\
2.80 &3.77& 4.08& 0.53& 0.86& 2.08& 0.61\\
3.00 &4.03& 3.77& 0.47& 0.84& 1.80& 0.66\\
3.20 &4.29& 3.48& 0.37& 0.71& 1.68& 0.71\\
3.40 &4.56& 3.22& 0.32& 0.60& 1.58& 0.72\\
3.60 &4.82& 2.98& 0.28& 0.58& 1.20& 0.91\\
3.80 &5.08& 2.75& 0.28& 0.32& 1.19& 0.96\\
4.00 &5.34& 2.54& 0.24& 0.31& 1.01& 0.99\\
4.20 &5.61& 2.35& 0.13& 0.19& 1.14& 0.89\\
4.40 &5.87& 2.17& 0.08& 0.50& 0.42& 1.17\\
4.60 &6.13& 2.01& \multicolumn{1}{c}{$\cdots$}& \multicolumn{1}{c}{$\cdots$}& 1.34& 0.67\\
4.80 &6.39& 1.86& \multicolumn{1}{c}{$\cdots$}& \multicolumn{1}{c}{$\cdots$}& 0.93& 0.93\\
5.00 &6.66& 1.72& \multicolumn{1}{c}{$\cdots$}& \multicolumn{1}{c}{$\cdots$}& \multicolumn{1}{c}{$\cdots$}& 1.72\\

\hline\\
\end{tabular}\\

\end{minipage}

\end{center}
\end{table}
\setlength{\tabcolsep}{6pt}

In Table\,\ref{tbl:statistical_lifetimes} we present the statistical lifetimes for the various evolutionary stages as a function of clump mass. The lifetime of the quiescent stage is approximately $5\times 10^4$\,yr for clumps with masses $\sim$1000\,\msun, and decreases to  $2\times 10^4$\,yr for clumps with masses of $\sim$ $1\times 10^4$\,\msun. These values are broadly consistent with the statistical lifetimes of massive starless clumps as determined by a number of studies reported in the literature (e.g., \citealt{motte2017,csengeri2014,tackenberg2012, motte2007}). We also note that the lifetime of the quiescent stage for clump masses above a few times $10^4$\,\msun\ decreases to negligible values.

\subsubsection{Dependence on $L/M$ on Galactocentric distance}

In Sect.\,\ref{sect:gal_distribution} we investigated the Galactocentric distribution of the ATLASGAL clumps, and this revealed a number of peaks that coincided with the expected positions of various spiral arms. These correlations suggest that the spirals play a significant role in the star formation process, although the nature of this role is still unclear. 

As mentioned in Sect.\,\ref{sect:intro}, previous efforts by \citet{moore2012} to evaluate the role of the spiral arms reported higher average luminosities and molecular cloud masses at well-defined Galactocentric distances in the first quadrant. They found that L$_{\rm{MYSO}}$/M$_{\rm{cloud}}$ was relatively flat for the inner 5\,kpc, and although two significant peaks were detected at $\sim$6\,kpc and $\sim$8\,kpc, these peaks were attributed to the presence of W51 and W49 at these distances. The work of \citet{moore2012} has been extended to investigate the SFE along lines of sight centred at $\ell$\,=\,30$\degr$ and  $\ell$\,=\,40$\degr$ (\citealt{eden2013,eden2015}) using GLIMPSE 8\,$\upmu$m, WISE 12\,$\upmu$m and 22\,$\upmu$m, and Hi-GAL 70\,$\upmu$m to estimate the infrared luminosity and the clump masses calculated with the BGPS. These studies estimated the SFE for spiral arms and inter-arm regions, and found that there was no significant variation either between the different arms or the inter-arm regions on the kiloparsec scale. 

These studies suggest that the spiral arms play a role in aggregating material but that the increase in star formation is perhaps the result of source crowding within the spiral arms, and not due to the influence of the spiral arms themselves. It is worth noting that these studies have been limited to either small regions  or limited sample sizes, and all have been conducted within the first quadrant. We are able to test these results and extended them to a larger fraction of the Galactic plane. 

Fig.\,\ref{fig:l_m_vs_rgc} presents the \lm\ distribution as a function of Galactocentric distance for clumps (grey circles), clusters (blue circles) and for all sources averaged over kiloparsec scales (red circles). This plot reveals that although there are significant variations in the clump to clump \lm\ ratios, this decreases when we move to the larger clusters and becomes almost constant when  averaged over the largest scales; the clumps have typical sizes of 0.5-1 parsecs, while the clusters have typical sizes of a few tens of parsecs. 

We observe two significant peaks at $\sim$6 and 8\,kpc in the 1\,kpc bins, coincident with those previously identified by \citet{moore2012} as W51 and W49, the two most active star-forming regions in the Galaxy. To check that the peaks seen in our data are also due to the presence of these extreme regions we have recalculated  the \lm\ ratio in these two distance bins (these values are shown as green circles on Fig.\,\ref{fig:l_m_vs_rgc}). A comparison between the red and green circles at $\sim$6 and 8\,kpc clearly shows that the exclusion of these two extreme star forming regions results in a reduction of the \lm\ ratio which brings these more in line with the \lm\ ratios on either side. We therefore find that the \lm\ ratio is relatively flat between 2 and 9\,kpc when evaluated over kiloparsec scales.  There is, however, significant variation on smaller scales, and this variation increases as the physical scales decrease. This is fully consistent with the results of the previous studies (\citealt{moore2012,eden2013,eden2015}).

\begin{figure}
\centering
\includegraphics[width=0.49\textwidth, trim= 0 0 0 0]{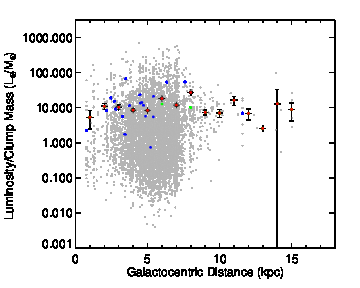}
\caption{\lm\ as a function of Galactocentric distance. The grey  and blue circles show the distribution of all of the individual ATLASGAL clumps and the 30 most massive clusters, respectively. The red circles and the error bars show the mean values and the standard error of the mean. The two green circles show the \lm\ ratios for the 6 and 8\,kpc bins after the W51 and W49 have been excluded. These red and green circles are evaluated in bins of 1\,kpc. 
\label{fig:l_m_vs_rgc}
}  
\end{figure}

\begin{figure*}
\centering
\includegraphics[width=0.98\textwidth, trim= 0 0 0 0]{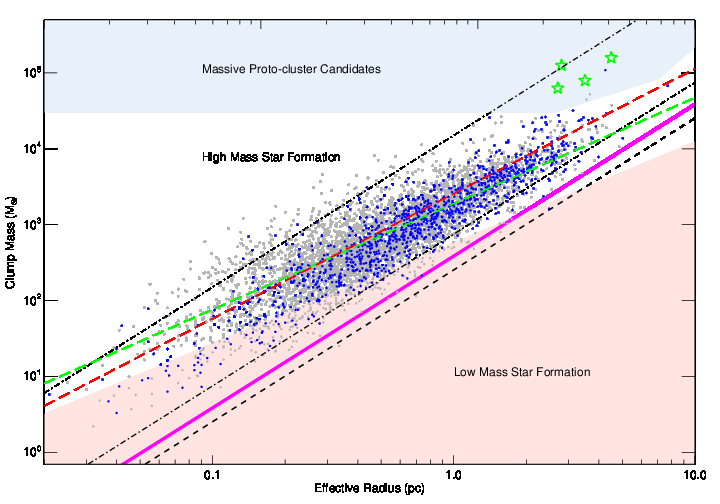}
\caption{The mass-size relationship of all ATLASGAL clumps (grey) and those associated with MSF (blue). The green stars show the distribution of the MPC candidates found towards the Galactic centre (\citealt{longmore2012a,immer2012}). The light-red shaded region shows the part of the parameter space found to be devoid of massive star formation that satisfies the relationship $m(r) \le 580$\,\msun\, $(R_{\rm{eff}}/{\rm{pc}})^{1.33}$ (cf. \citealt{kauffmann2010c}). The light-blue shaded area towards the top of the diagram indicates the region of parameter space where young massive cluster progenitors are expected to be found (i.e., \citealt{bressert2012}). The long-dashed red line shows the result of a linear power-law fit to the whole sample of associated clumps while the green line shows the fit to the distance-limited sample. The dashed black line shows the sensitivity of the ATLASGAL survey ($N_{\rm{H_2}} \sim 10^{22}$\,cm$^{-2}$) and the upper and lower dot-dashed lines mark surface densities of 1\,g\,cm$^{-2}$ and 0.05\,g\,cm$^{-2}$, respectively. The diagonal magenta band indicates the gas surface density ($\Sigma(\rm{gas})$) parameter space between 116-129\,\msun\,pc$^{-2}$, suggested by \citet{lada2010} and \citet{heiderman2010}, respectively, to be the threshold for ``efficient'' star formation. 
\label{fig:mass_radius_dist} } 
\end{figure*}

\subsection{Mass-radius distribution}\label{subsec:massRadiusDist}

We present a mass-radius (\mclump-\reff) diagram of the whole distance-determined sample of ATLASGAL sources in Fig.\,\ref{fig:mass_radius_dist}. We have shown similar diagrams in the previous papers in this series and found that the MSF clumps exhibit a strong correlation between these parameters. We extend our previous analysis by significantly increasing the sample size, and including lower-luminosity and less-evolved clumps as well as providing more reliable measurements for the MSF clumps resulting from the improved dust temperature estimates. We find the populations of the MSF clumps and non-MSF clumps to be similarly distributed, forming a continuous spread over almost three orders of magnitude in radius and almost 5 orders of magnitude in clump mass.  The diagonal upper and lower dashed-dotted lines shown in Fig.\,\ref{fig:mass_radius_dist} indicate lines of constant surface density, $\Sigma({\rm{gas}})$, of 1\,g\,cm$^{-2}$ and 0.05\,g\,cm$^{-2}$, respectively; these provide a reasonable empirical fit to the upper and lower range of the whole sample of clumps (this range encompasses $\sim$90\,per\,cent). We also show the minimum threshold of $\sim$116-129\,\msun\,pc$^{-2}$ (thick magenta band) proposed by \citet{lada2010} and \citet{heiderman2010} (hereafter referred to as the ``LH threshold'') for ``efficient'' star formation in nearby molecular clouds (d $\le 500$\,pc). 

The LH threshold seems to represent a stronger constraint for the lower limit for the ATLASGAL clumps, although the vast majority of  clumps have significantly higher mean surface densities. If we expect the clumps to form gradually through global accretion from larger scales, and to undergo a significant amount of evolution before star formation, we might expect to find that the population of MSF clumps dominates the higher-surface-density region of the narrow range and the non-MSF clumps dominate the region near the 0.05\,g\,cm$^{-2}$ threshold; this does not, however, appear to be the case. We have already found that there is no correlation between clump mass and evolution, and as there is no clear separation between the different evolutionary stages we may surmise that the surface density does not change significantly either. 

As seen in the previous papers, there is a strong correlation between these parameters for the MSF clumps ($r_{\rm{AB,C}} = 0.85$) and they are located within a relatively narrow range of surface densities. The correlation coefficient for the complete sample shown in Fig.\,\ref{fig:mass_radius_dist} is $r_{\rm{AB,C}}$ = 0.73 with $p \ll 0.01$). This is slightly lower than that found for the MSF clumps alone, but there is noticeably more scatter associated with the non-MSF clumps, which may include clumps that will remain starless or are destined to form only lower-mass stars. We have fitted the whole sample and a distance-limited subset of the sample (between 2 and 5\,kpc) to determine the slopes: these are plotted on Fig.\,\ref{fig:mass_radius_dist} as red and green dashed lines, respectively. The slope obtained for the whole sample is 1.647$\pm$0.012, which is in excellent agreement with the slope determined for the MSF clumps reported in Paper\,III (1.67$\pm$0.025).  The slope found for the distance-limited sample is shallower (1.397$\pm$0.014) and more reliable. 

The parameter space covered in Fig.\,\ref{fig:mass_radius_dist} can be separated into three regions of interest: the region where low-mass stars are found (light red shaded region), the region where efficient high-mass star formation is expected to be found (unshaded region; \citealt{kauffmann2010b}) and the region where massive proto-cluster (MPC) candidates  would be located (light blue shaded region -- compact clumps with a minimum mass of $3\times 10^4$\,\msun; \citealt{bressert2012}). We find that the vast majority of our sources are located above the empirical threshold required for efficient massive star formation, and so our expectation that most of these clumps are capable of forming massive stars is justified. We also note that only three sources have sufficient mass and are compact enough to fulfill the criteria to be considered as viable MPC candidates (we will discuss these further in Sect.\,\ref{sect:mpc}). 

\subsubsection{Galactic population of young massive proto-clusters}
\label{sect:mpc}
\setlength{\tabcolsep}{3pt}

\begin{table*}

\begin{center}\caption{Derived parameters for massive proto-cluster (MPC) candidates identified here and in Paper\,III. Masses have been estimated assuming dust temperatures derived from the SED fits for all sources except G002.53+00.016, for which a range of temperatures between 19 and 27\,K has been estimated from SED fits to individual pixels across the clump (see \citealt{longmore2012a} for details).}
\label{tbl:MPC_derived_para}
\begin{minipage}{\linewidth}
\begin{tabular}{lcc.......c}
\hline \hline
  \multicolumn{1}{c}{ATLASGAL}&  \multicolumn{1}{c}{Complex}&\multicolumn{1}{c}{Association}&	\multicolumn{1}{c}{Distance}&\multicolumn{1}{c}{\reff}  &\multicolumn{1}{c}{$T_{\rm{dust}}$} & 	\multicolumn{1}{c}{Log[\mclump]} & \multicolumn{1}{c}{Log[N(H$_2$)]}& \multicolumn{1}{c}{Log[\lbol]} & \multicolumn{1}{c}{\lbol/\mclump} & \multicolumn{1}{c}{Reference}\\

  \multicolumn{1}{c}{name}& \multicolumn{1}{c}{name}&\multicolumn{1}{c}{type}& \multicolumn{1}{c}{(kpc)}&	\multicolumn{1}{c}{(pc)} & \multicolumn{1}{c}{(K)}  &	\multicolumn{1}{c}{(\msun)}&	\multicolumn{1}{c}{(cm$^{-2}$)}&\multicolumn{1}{c}{(\lsun)}&\multicolumn{1}{c}{(\lsun/\msun)}&\\
 \hline
 \multicolumn{10}{c}{Galactic Disc}\\
  \hline
AGAL043.166+00.011	&	W49	&	MSF	&	11.1	&	4.255	&	33.3	&	5.04	&	23.892	&	6.91	&	74.0	&	1,2,4,5	\\
AGAL341.932$-$00.174	&	G341.982$-$00.125	&	YSO	&	12.4	&	3.549	&	14.9	&	4.65	&	22.944	&	4.75	&	1.3	&	1	\\
AGAL341.942$-$00.166	&	G341.982$-$00.125	&	YSO	&	12.4	&	3.670	&	14.4	&	4.72	&	23.214	&	4.82	&	1.2	&	1	\\
\hline  
  
AGAL010.472+00.027	&	G010.47+00.02	&	MSF	&	8.5	&	2.320	&	25.1	&	4.41	&	23.803	&	5.65	&	17.4	&	2,3,5	\\
AGAL012.208$-$00.102	&	G012.203$-$00.117	&	MSF	&	13.4	&	2.724	&	24.4	&	4.44	&	23.340	&	5.52	&	11.8	&	2,5$^a$	\\
AGAL019.609$-$00.234	&	G019.649$-$00.239	&	MSF	&	12.6	&	2.749	&	29.6	&	4.36	&	23.447	&	6.01	&	44.6	&	2,4,5$^a$	\\
AGAL032.797+00.191	&	G032.768+00.192	&	MSF	&	13.0	&	2.450	&	34.2	&	4.15	&	23.171	&	6.10	&	87.8	&	2,5$^a$	\\
AGAL043.148+00.014	&	W49	&	MSF	&	11.1	&	2.316	&	31.2	&	4.26	&	22.894	&	5.93	&	46.5	&	2,4	\\
AGAL043.164$-$00.029	&	W49	&	MSF	&	11.1	&	3.124	&	31.2	&	4.50	&	23.265	&	6.21	&	50.3	&	2,4,5	\\
AGAL043.178$-$00.011	&	W49	&	MSF	&	11.1	&	2.155	&	31.9	&	4.23	&	22.882	&	6.60	&	236.5	&	2	\\
AGAL049.472$-$00.367	&	W51	&	MSF	&	5.3	&	2.124	&	30.3	&	4.34	&	23.420	&	6.38	&	109.7	&	2,4,5	\\
AGAL049.482$-$00.402	&	W51	&	MSF	&	5.3	&	2.176	&	28.6	&	4.26	&	23.372	&	6.17	&	81.8	&	2	\\
AGAL049.489$-$00.389	&	W51	&	MSF	&	5.3	&	1.408	&	31.6	&	4.41	&	24.025	&	6.29	&	76.2	&	2,4,5	\\
AGAL328.236$-$00.547	&	RCW99	&	MSF	&	2.7	&	1.178	&	\multicolumn{1}{c}{$\cdots$}	&	\multicolumn{1}{c}{$\cdots$}	&	\multicolumn{1}{c}{$\cdots$}	&	\multicolumn{1}{c}{$\cdots$}	&	\multicolumn{1}{c}{$\cdots$}	&	2,3,8\\
AGAL329.029$-$00.206$^c$	&	G329.111$-$00.278	&	MSF	&	2.9	&	0.807	&	20.6	&	3.45	&	23.358	&	4.18	&	5.5	&	2,3,8	\\
AGAL330.954$-$00.182	&	G331.394$-$00.125	&	MSF	&	5.3	&	1.238	&	32.2	&	3.95	&	23.790	&	5.78	&	67.5	&	2	\\
AGAL337.704$-$00.054	&	G337.651$-$00.048	&	MSF	&	12.1	&	2.403	&	23.6	&	4.42	&	23.429	&	5.41	&	9.8	&	2,8	\\
AGAL350.111+00.089	&	G350.219+00.081	&	MSF	&	10.5	&	1.940	&	27.3	&	4.29	&	23.080	&	5.83	&	34.2	&	2,3,9	\\
\hline
\multicolumn{10}{c}{Galactic Centre}\\
  \hline
G000.253+00.016 & ``The Brick'' & Starless? & 8.4 & 2.8 &	\multicolumn{1}{c}{$\cdots$} & 5.1  &23.39&\multicolumn{1}{c}{$\cdots$}&\multicolumn{1}{c}{$\cdots$}& 6,7 \\
AGAL000.411+00.051 (d)$^c$ & \multicolumn{1}{c}{$\cdots$} & Starless? & 8.4 & 3.5&	\multicolumn{1}{c}{$\cdots$} & 4.9  &23.23&\multicolumn{1}{c}{$\cdots$}&\multicolumn{1}{c}{$\cdots$}& 7 \\
AGAL000.476$-$00.007 (e)$^c$ & \multicolumn{1}{c}{$\cdots$} & Starless? & 8.4 & 4.5&	\multicolumn{1}{c}{$\cdots$} &5.2 &23.53&\multicolumn{1}{c}{$\cdots$}&\multicolumn{1}{c}{$\cdots$}& 7 \\
AGAL000.494+00.019 (f)$^c$ & \multicolumn{1}{c}{$\cdots$} & MM & 8.4 & 2.7&	\multicolumn{1}{c}{$\cdots$} & 4.8   &23.26&\multicolumn{1}{c}{$\cdots$}&\multicolumn{1}{c}{$\cdots$}& 7 \\
\hline\\
\end{tabular}\\
References: (1)\,this work, (2)\,Paper\,III, (3)\,Paper\,I, (4)\,Paper\,II, (5)\,\citet{ginsburg2012}, (6)\,\citet{longmore2012a}, (7)\,\citet{immer2012}, (8)\,\citet{contreras2017}, (9)\,\citet{longmore2017}. \\
Notes: $^a$ Identified by \citet{ginsburg2012} as a massive clump but the derived mass was slightly less than the $3\times 10^4$\,\msun\ required for inclusion in their sample of MPC candidates.  $^b$ The distance to this source is uncertain and should be used with caution. $^c$ The source names have been taken from the ATLASGAL CSC (\citealt{contreras2013}) while the letters in parentheses are the nomenclature used by \citet{lis1999}. 

\end{minipage}

\end{center}
\end{table*}
 
\setlength{\tabcolsep}{6pt}

Young massive clusters (YMCs, e.g., Arches, Quintuplet) are the most active clusters in the Galaxy. These can be broadly defined as being younger than 100\,Myr  and having stellar masses $\gtrsim$10,000\,\msun\ (\citealt{portegies_zwart2010}). Identifying the precursors to these rare and extreme clusters will provide some insight into their initial conditions and the impact of the environment on their formation and evolution.  \cite{bressert2012} suggested a set of threshold criteria that a clump would need to satisfy in order to form a YMC (i.e., massive and compact enough to ensure that the gravitational force was sufficiently strong to counteract the enormous feedback long enough for the YMC to form). Assuming a star formation efficiency of 30\,per\,cent, \cite{bressert2012} estimated the progenitors of these YMCs would need a minimum mass of $3\times10^4$\,\msun\ and have a radius of no more than a few parsecs: clumps that satisfied these criteria are referred to as massive proto-clusters (MPCs).

There have been a number of studies that have used survey data to search for MPCs both towards the Galactic centre region (\citealt{longmore2012a, immer2012}) and in the disk (\citealt{ginsburg2012, longmore2017, contreras2017}). In the previous papers in this series we have also identified 16 MPC candidates located in the disk that are associated with MSF tracers,  many of which have been independently identified in these other studies (see Table\,\ref{tbl:MPC_derived_para} for MPC candidates and references to studies that have discovered them). Our previous search for these objects used a temperature of 20\,K to estimate clump masses, and was further biased to clumps associated with MSF tracers and so excluded the majority of clumps ($\sim$80\,per\,cent).  Here we are able to extend and improve on our previous work by using the dust temperatures and distances determined.

As described in the introduction, the ATLASGAL CSC is likely to include $\sim$90\,per\,cent of all massive clumps in the Galaxy and is sensitive to all clumps with a surface density $>$ 3600\,\msun\,beam$^{-1}$. Our sample will therefore include all MPC candidates within the Galactic disk that are located away from the Galactic centre ($355\degr < \ell < 5\degr$). As we have determined distances to 97\,per\,cent of the sample (7770 of 8002 clumps), we are very likely to be able to identify all MPC candidates in the region and perhaps in the whole Galaxy. As mentioned in subsection \ref{subsec:massRadiusDist}, the comparison of our data and the MPC parameter space has only resulted in the identification of 3 MPC candidates:   AGAL043.166+00.011, AGAL341.932$-$00.174 and AGAL341.942$-$00.166. Only one of the previously identified clumps remains in the sample: three of the other sources have been moved to the near distance (AGAL328.236$-$00.547, AGAL329.029$-$00.206 and AGAL330.954$-$00.182), but the dust temperatures for the others are significantly higher than the 20\,K previously assumed, and this has resulted in lower calculated clump masses which no longer satisfy the mass criterion.

\begin{figure*}
\centering
\includegraphics[width=0.49\textwidth, trim= 0 0 0 0]{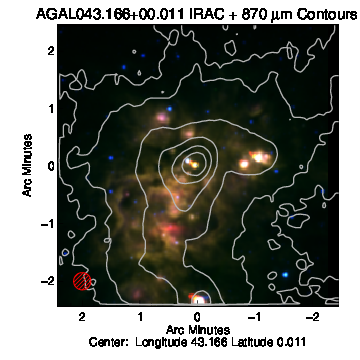}
\includegraphics[width=0.49\textwidth, trim= 0 0 0 0]{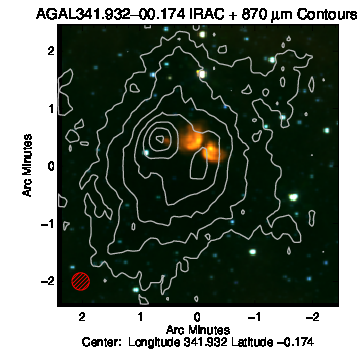}
\caption{Three-colour mid-infrared GLIMPSE image of the two regions associated with MPC candidates discussed in the text. The contours trace the dust emission detected by ATLASGAL and the red circle shown in the lower left corner of each map indicates the resolution of the ATLASGAL survey ($\sim$19\arcsec). In the left panel we show  AGAL043.166+00.011, which is one of most active star formation regions in the Galaxy. In the right panel we show the AGAL341.942$-$00.166 and AGAL341.932$-$00.174 (left and right contour peaks, respectively). \label{fig:MPC} } 
\end{figure*}

The two new clumps (AGAL341.932$-$00.174 and AGAL341.942$-$00.166) are part of the same larger-scale molecular structure, as can be seen in Fig.\,\ref{fig:MPC}. Although neither of these sources has been previously associated with a MSF tracer, the spatial correlation between the extended mid-infrared emission and peaks in the dust emission seen in the image suggest that star formation is already underway. These clumps have been placed at the far distance from the analysis presented in this paper: as we have not been able to find any confirmation of this distance in the literature, their classification as MPC candidates is not considered robust. The remaining clump, AGAL043.166+00.011, is the central part of W49, which is one of the most active star formation regions in the Galaxy, and has a well-constrained maser parallax distance. All three MPCs identified in the disk appear to be actively star-forming, although AGAL043.166+00.011 has an \lm\ ratio $\sim$50 times larger than the other two candidates and therefore is much more evolved. 

We have found no starless MPC candidates in the disk and, given the completeness of the survey, it seems unlikely that there are any others to be found. This is consistent with the studies of \citet{ginsburg2012,tackenberg2012,svoboda2016_bgps} and \citet{longmore2017}, who also failed to find any starless MPC candidates, although one potential starless candidate (AGAL331.029$-$00.431) has been identified by \citet{contreras2017}. This source is located inside the Solar circle and is therefore subject to the kinematic distance ambiguity as pointed out by \citet{contreras2017}.  They place this at the far distance due to the absence of a correlation with an IRDC, but point out that this source is placed at the near distance by \citet{wienen2015}. Our independent assessment of the \hi\ data has associated this source with the G331.104$-$00.413 cluster (see  upper panel of Fig.\,\ref{fig:MPC_clusters}), which consists of 53 clumps, 44 of which are placed at the near distance, 3 at the far distance and 6 are uncertain. As $\sim$90\,per\,cent of the distance solutions point to a near distance solution, the near distance assignment has been adopted for this cluster. This region was also included in a survey of \hii\ regions reported by \citet[331.314$-$0.336]{caswell1987} who also placed it at the near distance. We are therefore relatively certain that this source is located at the near distance, indicating that this clump is not a viable MPC candidate. 

\begin{figure}
\centering
\includegraphics[width=0.49\textwidth, trim= 0 0 0 0]{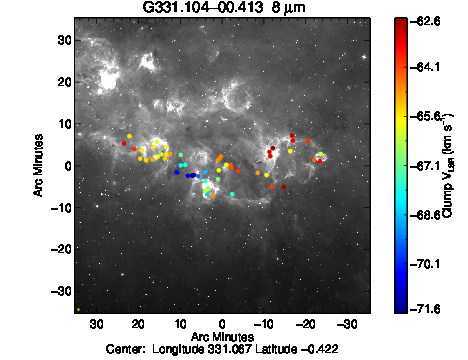}
\includegraphics[width=0.49\textwidth, trim= 0 0 0 0]{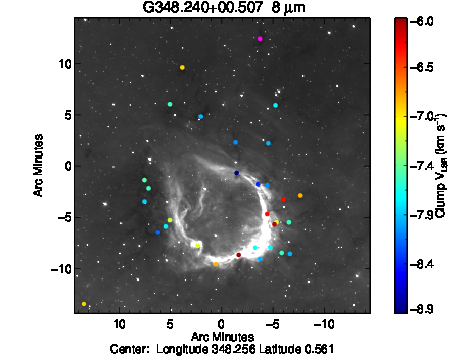}
\includegraphics[width=0.49\textwidth, trim= 0 0 0 0]{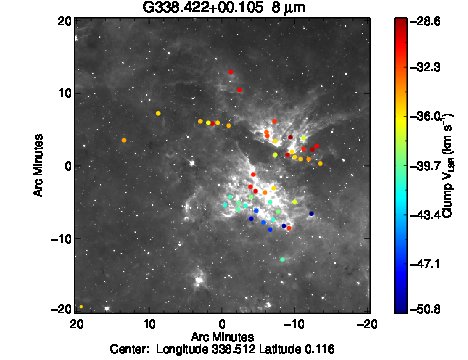}

\caption{ATLASGAL clusters associated with MPC candidates identified by \citet{contreras2017} and \citet{longmore2017} identified by the Friends-of-Friends analysis. The background image is the 8\,\mum\ emission and filled circles show the positions of ATLASGAL CSC objects, the colours of which provide an indication of their velocities (see colour bars for values).  \label{fig:MPC_clusters} } 

\end{figure} 
 
Three of the other four MPC candidates identified by \citet{contreras2017} (private comm.) had already been identified in our previous papers (i.e., Paper\,III; AGAL328.236$-$00.547, AGAL329.029$-$00.206 and AGAL337.704$-$00.054). The other MPC identified in the \citet{contreras2017} study that was not already identified is AGAL348.183+00.482. This is placed at the far distance by \citet{contreras2017} and the MALT90 team, but we find this source is associated with the RCW120 (SH 2-3) region (see middle panel of Fig.\,\ref{fig:MPC_clusters}; this is identified as ATLASGAL cluster G348.240+00.507 in this paper). This cluster is associated with 29 clumps, all but one of which have been placed at the near distance in agreement with \citet[348.225+0.459]{caswell1987}. It seems likely, therefore, that  AGAL348.183+00.482 is located at the near distance and so can also be discarded as a MPC candidate.

\citet{longmore2017} has used the HOPS study to identify a sample of seven MPCs, but they discard four as either being located at the near distance or being too diffuse to be able to collapse to form a MPC on a realistic timescale. This leaves three possible candidates from their sample: G330.881$-$0.371, G338.464+0.034 and G350.170+0.070. The first of these is associated with our cluster G331.104-00.413 (see upper panel of Fig.\,\ref{fig:MPC_clusters}), which we have already discussed and is located at the near distance ($\sim$4\,kpc). The second is associated with ATLASGAL cluster G338.422+00.105 (see lower panel of Fig.\,\ref{fig:MPC_clusters}), which we have placed at a distance of 2.7\,kpc; these two MPC candidates are therefore less probable. The one remaining MPC candidate identified by \citet[G350.170+0.070]{longmore2017} was included in the sample we presented in Paper\,III.

We note that \citet{elia2017} have identified 22 MPC candidates between $\ell = 16\degr$ and 67\degr, however, these result from a default assignment at the far distance for all sources where the kinematic distance ambiguity has not been resolved.  Until reliable distances have been derived for these sources their identification as MPC candidates is somewhat speculative.

Between them, the \citet{longmore2017} and \citet{contreras2017} studies have identified 12 MPC candidates.  For the reasons discussed in the previous paragraphs only four are likely to be viable, all of which were identified in our previous analysis. We  include all of the MPC candidates that were identified in Paper\,III with their updated parameters  in Table\,\ref{tbl:MPC_derived_para}. We have updated the references to indicate those that have been confirmed by \citet{longmore2017} and \citet{contreras2017}. Only the top three are still considered to be reliable MPC candidates (i.e., Log(\mclump) $>$ 4.48\,\msun), but we include the others to provide updated physical parameters for them.  Although they do not satisfy the mass criterion, these still represent the most massive and compact clumps in the Galaxy, and some of these are close enough to the threshold that a small change in the distance or dust emissivity solution might bring some back into play.

Our hunt for MPC candidates through a large fraction of the inner Galactic disk has only identified 3 MPC candidates, two of which are associated with the same cloud. We have also assumed a SFE of $\sim$30\,per\,cent, which is an upper limit to the SFEs that have been determined in nearby cloud studies (\citealt{lada2003}).  If a more modest SFE of 10\,per\,cent is assumed then the number of MPC candidates in the disk reduces to one (i.e., AGAL043.166+00.011). These objects are therefore extremely rare in the disk, and given that they all are associated with mid-infrared emission it is likely that star formation is already ongoing in all of them. Assuming the lifetime of a YMC to be of order 10\,Myr and the number of clusters with masses between 10$^4$\,\msun\ and 10$^5$\,\msun\ to be $\sim$25, \citet{longmore2012a} estimated the YMC formation rate to be one every  $\sim$2.5\,Myr, which agrees well with the number we have detected. Given the low number found in the disk it is extremely unlikely that one will be detected in a pre-stellar stage, which is also consistent with what we have found here. 

The situation in the disk stands in contrast to that found in the Galactic centre region, where twice as many MPC candidates have been identified including three of which currently appear to be starless (\citealt{longmore2012a, immer2012}): the parameters for these MPC candidates are included at the bottom of Table\,\ref{tbl:MPC_derived_para}. We also note that the MPC candidates found in the Galactic centre also tend to be smaller and more massive than those found in the disk, but the environmental conditions found in the Galactic centre region are the most extreme in the Galaxy and finding differences between these two regions is perhaps not all that surprising.

\section{Evolution of clumps}
\label{sect:clump_evolution}

\begin{figure}
\centering

\includegraphics[width=0.49\textwidth, trim= 0 0 0 .0]{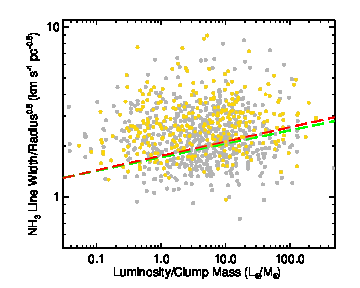}
\includegraphics[width=0.49\textwidth, trim= 0 0 0 0]{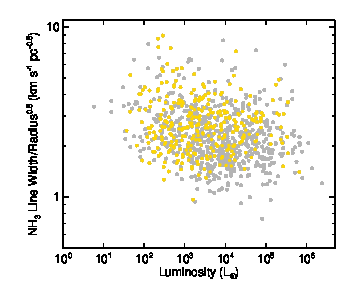}

\caption{NH$_3$ (1,1) line-width normalised by the square-root of the clump radius as a function of the \lm\ ratio (upper panel) and as a function of the bolometric luminosity (lower panel). The grey circles show the distributions of all clumps while the yellow shows the distribution of the distance-limited sample. The distribution in the upper panel is weakly correlated but the distribution in the lower panel is uncorrelated. The long-dashed red and green lines shown in the upper panel shows the result of a linear power-law fit to the whole sample of associated clumps and to the distance-limited sample, respectively . \label{fig:nh3_dvlsr_l_m}} 

\end{figure}

The analysis presented thus far has revealed trends for increasing temperatures and luminosities with the evolutionary stage of the embedded stars as they move towards the main sequence. These two parameters can be attributed to the result of feedback from the forming proto-cluster on its natal clump. We can investigate the feedback by looking at changes in the line-widths of molecular transitions. 

We show the correlation between the NH$_3$ line-width  normalised by the square-root of the clump radius and \lm\ ratio in the upper panel of Fig.\,\ref{fig:nh3_dvlsr_l_m}. The partial-Spearman test reveals a moderate correlation between these parameters ($r_{\rm{AB,C}}$ = 0.12, $p$-value = 0.0002), suggesting that the feedback from the evolving proto-cluster is resulting in an increase in the line-width over time, although this increase is rather modest. The power-law fits to the full and distance-limited samples reveal slopes of 0.085 and 0.079, respectively. The increase in line-width may not be due to feedback from the embedded proto-cluster as there may also be a contribution from globally infalling material from the extended envelope. We find no significant correlation between the normalised line-width and bolometric luminosity (Fig.\,\ref{fig:nh3_dvlsr_l_m}, lower panel), the lack of which, suggest that radiative feedback has little impact on the clump and that the increase in the normalised line-width with evolution may be due to mechanical energy injected into the clump and/or changes in the global infall over the evolutionary timescale.

The virial parameter measures the balance between gravity and the internal energy that can support the clump against gravitational collapse. The virial parameter is defined as:

\begin{equation}
\alpha_{\rm{vir}}= \frac{5\sigma_v^2R_{\rm{eff}}}{GM_{\rm{clump}}},
\end{equation}

\noindent

\noindent where $R_{\rm{eff}}$ is the effective radius of the clumps, $\sigma_v$ is the velocity dispersion of the ammonia (1,1) inversion transition and $G$ is the gravitational constant. 

We show the virial parameter as a function of clump mass in the upper panel of Fig.\,\ref{fig:virial_plots}. A value of less than 2 is generally taken as indicating that a particular clump is unstable to gravity and is likely to be globally collapsing in the absence of significant magnetic support. It is clear from this plot that the most massive clumps are less gravitationally stable, which probably explains their rapid evolution and short pre-stellar lifetimes. For the distance-limited sample we find $\alpha_{\rm{vir}} \propto M^{-0.4}$: this implies $\sigma^2 R \propto M^{0.6}$}. In the upper panel of Fig.\,\ref{fig:nh3_dvlsr_l_m} we see a correlation between increasing line-with and evolution and naively we might expect the virial mass to increase. It is not possible, however, to separate the turbulent component from the contribution from infall and outflow motions that are essentially random and do not contribute to the stability of the clump, and so we cannot say with any certainty how this will affect the virial mass. We can, however, calculate how the virial parameter would be affected if all of the increase in line-width did contribute to stabilising the clump.    

The lower panel of Fig.\,\ref{fig:virial_plots} shows a positive correlation between virial parameter and the \lm\ ratio where the $\alpha_{\rm{vir}}$ increase by a factor of 2-3. The virial mass is $M_{\rm{vir}} \propto R \Delta v^2$, and so the increase in line-width will result in an increase in the virial mass by a factor of 4-9, i.e., almost an order of magnitude. However, even assuming the feedback leads to a direct increase in the virial parameter, which is unlikely to be the case, the clumps will still be unstable to gravitational collapse (even the more evolved clumps have $\alpha_{\rm{vir}}$ significantly below 2) and it seems improbable that feedback could be driving an expansion of the clumps as they evolve. 

\citet{elia2017} have also investigated how the surface density changes as a function of evolution and found no evidence to support this; however, their analysis focused on more compact objects than are considered here. In fact, their surface density vs. \lm\ ratio plot appears very similar to our peak column density vs. \lm\ ratio plot (cf. the middle left panels of their Fig.\,22 and the lower-right panel of Fig.\,\ref{fig:l_m_vs_temp}), and so when we look at similar size scales we find similar behaviour for these parameters for both data sets. 

\begin{figure}
\centering
\includegraphics[width=0.49\textwidth, trim= 0 0 0 0]{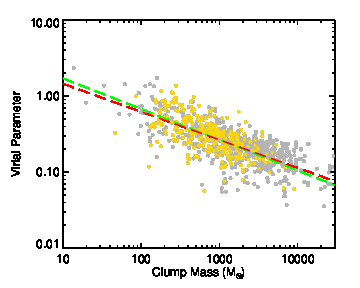}
\includegraphics[width=0.49\textwidth, trim= 0 0 0 0]{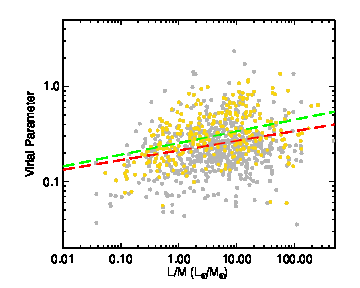}

\caption{Virial parameter as a function of clump mass and \lm\ ratio (upper and lower panels, respectively).  The long-dashed red line shows the result of a linear power-law fit to the whole sample of associated clumps while the green line shows the fit to the distance-limited sample.\label{fig:virial_plots} } 

\end{figure}

We have also seen trends for increasing source size and decreasing mass surface density as a function of evolutionary stage (i.e., Fig.\,\ref{fig:size_parameter} and \ref{fig:surface_density}, respectively) and so perhaps the feedback from the proto-cluster or accretion onto the cluster itself are having a direct impact on the structure of the clumps. We present a plot of the mass surface density as a function of the \lm\ ratio in the upper panel of Fig.\,\ref{fig:surface_density_l_m}: this plot would seem to confirm the trend for a significant decrease in the mass surface density as a function of evolution. The change in mass surface density covers two orders of magnitude and so the fraction of initial clump masses that would need to be depleted over the star formation time scale is $\sim$99\,per\,cent. Although this is a larger fraction than perhaps we would expect, it is at least feasible given that eventually all of a clump's mass needs to be dissipated. However, if the clump mass was being so dramatically depleted we would expect to find it and a clump's  peak column density to be inversely correlated with the \lm\ ratio, which is not the case (see lower right panel of Fig.\,\ref{fig:l_m_vs_temp}). Furthermore, if clumps were in that main dispersal phase, $T$ should be correlated with $M$ and we would expect the range of $T$ to be much larger and the upper-left region of the $L$-$M$ plot would be populated, neither of which is observed.

\begin{figure}
\centering

\includegraphics[width=0.49\textwidth, trim= 0 0 0 0]{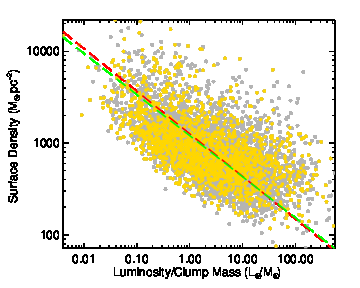}
\includegraphics[width=0.49\textwidth, trim= 0 0 0 .0]{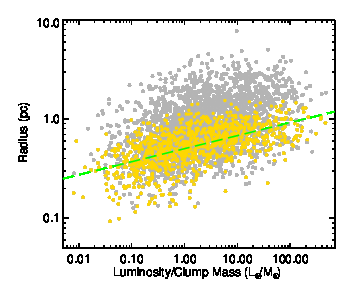}

\caption{Upper panel: the mass surface density as a function of the \lm\ ratio. Lower panel: clump radius as a function of \lm\ ratio.  The grey circles show the distributions of all clumps while the yellow shows the distribution of the distance-limited sample. The dashed red and green lines are linear fit to the log-log values of the full and distance-limited samples, respectively. \label{fig:surface_density_l_m}} 

\end{figure}

If the clump mass is not decreasing significantly as a function of evolution then the only other way to produce a decrease in the mass surface density is to increase the radius. There are two questions that need to be addressed:  1) is the increase in radius sufficient to explain the decrease in the mass surface density and 2) what is driving the increase in clump size? To address the first, we plot the clump radius as a function of evolutionary stage in the lower panel of Fig\,\ref{fig:surface_density_l_m}, and the increase in size is approximately a factor of 10 and so this can explain the decrease in mass surface density. 

A similar set of trends have been recently reported from a study of a large sample ATLASGAL CSC sources by the MALT90 team (\citealt{contreras2017}). They found that as the temperature increased, the volume density decreased and the virial parameter increased, and concluded that the clumps had started to expand, increasing their size and decreasing their volume density. However, as previously mentioned, even if the virial parameter increases as the star formation evolves, the clumps would still be gravitationally unstable and so it is unlikely that the feedback processes could drive an expansion of the clumps (virial parameters significantly lower than 2). 

An alternative explanation is that, as the temperature of the clump increases, so to does the column density sensitivity of the ATLASGAL survey, and as a result, we are simply more sensitive to the extended envelopes of the evolved clumps due to their higher temperatures than we are for the less evolved clumps. We have tested this hypothesis using RATRAN to investigate the expected change in source size assuming a similar flux sensitivity of 200\,mJy\,beam$^{-1}$, a distance of 4\,kpc and a radial density profile of $n \propto R^{-1.5}$ for the same clump over a temperature range of 12 to 45\,K; this results in a change in the observed size of the clumps of a factor of 4.2 and so this provides an explanation for approximately half of the difference in size between the quiescent and MSF clumps, and if we were to also consider the impact of infalling material this would be larger still. Although both explanations are plausible, we think the scheme of clump expansion at the same time as the increase in star formation is significantly less likely than the observed increase in radius being the result of an observational bias resulting from a temperature increase.

\setlength{\tabcolsep}{1pt}

\begin{table*}

\begin{center}\caption{Derived cluster parameters for the thirty most massive clusters identified, ordered by their total dense gas mass. We include the standard deviations for the longitude, latitude, velocity and temperature.}
\label{tbl:derived_cluster_para}
\begin{minipage}{\linewidth}
\small
\begin{tabular}{lccccc.c.....}
\hline \hline
  \multicolumn{1}{c}{Cluster}& \multicolumn{1}{c}{Literature} & \multicolumn{1}{c}{\# of}  &\multicolumn{1}{c}{$\ell$}  &\multicolumn{1}{c}{$b$}  &\multicolumn{1}{c}{\vlsr}  &\multicolumn{1}{c}{Distance}   &	\multicolumn{1}{c}{T} &\multicolumn{1}{c}{Log($L_{\rm{bol}}$)}&\multicolumn{1}{c}{$L_{\rm{bol}}$} & \multicolumn{1}{c}{Log($M_{\rm{clump}}$)} &	\multicolumn{1}{c}{$M_{\rm{clump}}$} &\multicolumn{1}{c}{\lm} \\
  
  \multicolumn{1}{c}{name}& \multicolumn{1}{c}{name} &\multicolumn{1}{c}{members}  &\multicolumn{1}{c}{(\degr)}  &\multicolumn{1}{c}{(\degr)}  &\multicolumn{1}{c}{(\kms)}  &\multicolumn{1}{c}{(kpc)}   &	\multicolumn{1}{c}{(K)} &\multicolumn{1}{c}{(\lsun)}&\multicolumn{1}{c}{(\%)} & \multicolumn{1}{c}{(\msun)} &	\multicolumn{1}{c}{(\%)} &\multicolumn{1}{c}{(\lsun\,/\msun)} \\
\hline
G030.650$-$00.015	& W43 &	245	&	30.651$\pm$0.472	&	$-$0.014$\pm$0.170	& 	97.44$\pm$8.90	&	5.2	&	9.5$\pm$6.09	&	6.608	&	3.12	&	5.530	&	3.34	&	12.0	\\
G043.141$-$00.018	& W49 &	14	&	43.141$\pm$0.053	&	$-$0.018$\pm$0.035	&	9.47$\pm$3.96	&	11.1	&	21.2$\pm$4.58	&	7.181	&	11.69	&	5.354	&	2.23	&	67.1	\\
G024.319+00.191	&&	61	&	24.313$\pm$0.165	&	0.189$\pm$0.091	&	112.77$\pm$4.66	&	7.8	&	10.9$\pm$5.30	&	5.750	&	0.43	&	5.349	&	2.20	&	2.5	\\
G049.261$-$00.318	& W51&	88	&	49.256$\pm$0.220	&	$-$0.316$\pm$0.099	&	60.92$\pm$6.80	&	5.3	&	12.9$\pm$4.98	&	7.045	&	8.56	&	5.320	&	2.06	&	53.2	\\
G350.219+00.081	& &	31	&	350.222$\pm$0.081	&	0.082$\pm$0.048	&	-64.61$\pm$4.37	&	10.5	&	13.2$\pm$4.55	&	6.477	&	2.31	&	5.295	&	1.95	&	15.2	\\
G337.228$-$00.065	&&	30	&	337.228$\pm$0.071	&	$-$0.065$\pm$0.065	&	-70.13$\pm$3.40	&	11.0	&	11.5$\pm$5.01	&	6.407	&	1.97	&	5.262	&	1.80	&	14.0	\\
G331.394$-$00.125	&&	99	&	331.394$\pm$0.243	&	$-$0.125$\pm$0.124	&	-90.03$\pm$5.65	&	5.3	&	11.5$\pm$6.05	&	6.562	&	2.81	&	5.195	&	1.54	&	23.3	\\
G333.125$-$00.348	& G333&	240	&	333.123$\pm$0.425	&	$-$0.346$\pm$0.233	&	-50.75$\pm$4.11	&	3.6	&	10.4$\pm$5.70	&	6.524	&	2.57	&	5.192	&	1.53	&	21.5	\\
G347.650+00.210	&&	22	&	347.650$\pm$0.064	&	0.210$\pm$0.049	&	-86.91$\pm$9.73	&	9.8	&	10.3$\pm$7.03	&	6.428	&	2.07	&	5.150	&	1.39	&	19.0	\\
G349.168+00.118	&&	32	&	349.168$\pm$0.188	&	0.120$\pm$0.032	&	-68.53$\pm$5.38	&	10.5	&	13.6$\pm$3.48	&	6.108	&	0.99	&	5.129	&	1.33	&	9.5	\\
G349.130$-$00.003	&&	6	&	349.130$\pm$0.033	&	$-$0.003$\pm$0.021	&	15.25$\pm$1.34	&	19.7	&	18.6$\pm$2.83	&	5.933	&	0.66	&	5.089	&	1.21	&	7.0	\\
G337.017$-$00.001	&&	37	&	337.012$\pm$0.189	&	$-$0.001$\pm$0.077	&	-116.38$\pm$6.66	&	7.7	&	15.5$\pm$5.10	&	5.889	&	0.60	&	5.043	&	1.09	&	7.0	\\
G025.656$-$00.087	&&	27	&	25.656$\pm$0.168	&	$-$0.087$\pm$0.080	&	93.50$\pm$5.91	&	10.2	&	10.3$\pm$5.59	&	6.166	&	1.13	&	5.036	&	1.07	&	13.5	\\
G337.651$-$00.048	&&	9	&	337.677$\pm$0.026	&	$-$0.037$\pm$0.033	&	-48.53$\pm$1.57	&	12.1	&	15.4$\pm$3.18	&	5.773	&	0.46	&	5.034	&	1.07	&	5.5	\\
Bania Clump 1	&&	12	&	354.759$\pm$0.067	&	0.335$\pm$0.056	&	87.92$\pm$10.35	&	8.4	&	12.4$\pm$5.98	&	5.376	&	0.18	&	5.030	&	1.06	&	2.2	\\
G342.533+00.179	&&	9	&	342.533$\pm$0.054	&	0.179$\pm$0.008	&	-41.91$\pm$0.49	&	12.5	&	11.5$\pm$4.06	&	4.881	&	0.06	&	5.010	&	1.01	&	0.7	\\
G020.739$-$00.136	&&	22	&	20.739$\pm$0.049	&	$-$0.136$\pm$0.105	&	57.68$\pm$2.67	&	11.7	&	11.8$\pm$5.37	&	5.784	&	0.47	&	4.982	&	0.95	&	6.3	\\
G353.511$-$00.053	&&	21	&	353.511$\pm$0.079	&	$-$0.053$\pm$0.038	&	-53.86$\pm$6.00	&	10.2	&	18.3$\pm$5.43	&	5.916	&	0.64	&	4.972	&	0.92	&	8.8	\\
G023.375$-$00.101	&&	52	&	23.374$\pm$0.111	&	$-$0.098$\pm$0.114	&	93.09$\pm$10.98	&	5.9	&	12.0$\pm$5.68	&	6.026	&	0.82	&	4.970	&	0.92	&	11.4	\\
G351.598+00.189	&&	18	&	351.598$\pm$0.076	&	0.189$\pm$0.033	&	-42.02$\pm$2.30	&	11.5	&	14.8$\pm$6.78	&	6.745	&	4.28	&	4.962	&	0.90	&	60.6	\\
G305.453+00.065	& G305&	98	&	305.459$\pm$0.269	&	0.062$\pm$0.198	&	-37.34$\pm$3.82	&	3.8	&	12.3$\pm$6.40	&	6.296	&	1.52	&	4.916	&	0.81	&	24.0	\\
G338.586+00.043	&&	11	&	338.586$\pm$0.024	&	0.043$\pm$0.050	&	-23.67$\pm$2.20	&	13.6	&	13.1$\pm$3.63	&	5.614	&	0.32	&	4.870	&	0.73	&	5.6	\\
G354.479+00.087	&&	4	&	354.480$\pm$0.018	&	0.087$\pm$0.006	&	16.80$\pm$1.49	&	23.3	&	20.2$\pm$4.05	&	5.822	&	0.51	&	4.869	&	0.73	&	9.0	\\
G320.403+00.131	&&	10	&	320.403$\pm$0.024	&	0.131$\pm$0.027	&	-5.11$\pm$1.52	&	12.5	&	12.1$\pm$5.74	&	5.728	&	0.41	&	4.860	&	0.71	&	7.4	\\
G012.203$-$00.117	&W33&	4	&	12.203$\pm$0.009	&	$-$0.117$\pm$0.010	&	26.12$\pm$1.29	&	13.4	&	22.8$\pm$3.06	&	6.175	&	1.15	&	4.828	&	0.66	&	22.2	\\
G018.929$-$00.343	&&	57	&	18.934$\pm$0.121	&	$-$0.349$\pm$0.205	&	63.94$\pm$3.29	&	5.0	&	10.7$\pm$4.84	&	5.592	&	0.30	&	4.812	&	0.64	&	6.0	\\
G048.651+00.131	&&	13	&	48.651$\pm$0.084	&	0.131$\pm$0.097	&	13.85$\pm$4.29	&	10.8	&	12.6$\pm$5.22	&	5.869	&	0.57	&	4.801	&	0.62	&	11.7	\\
G019.649$-$00.239	&&	6	&	19.649$\pm$0.039	&	$-$0.239$\pm$0.032	&	40.85$\pm$2.21	&	12.6	&	18.9$\pm$3.52	&	6.117	&	1.01	&	4.789	&	0.61	&	21.3	\\
G035.553+00.008	&&	11	&	35.553$\pm$0.040	&	0.008$\pm$0.063	&	52.08$\pm$3.14	&	10.4	&	15.4$\pm$4.52	&	5.653	&	0.35	&	4.789	&	0.61	&	7.3	\\
G033.527$-$00.011	&&	25	&	33.521$\pm$0.195	&	$-$0.010$\pm$0.028	&	103.25$\pm$2.21	&	6.5	&	10.4$\pm$5.58	&	5.175	&	0.12	&	4.778	&	0.59	&	2.5	\\

\hline\\
\end{tabular}\\
Notes: A more complete table that includes parameters for all 776 is available in electronic form at the CDS via anonymous ftp to cdsarc.u-strasbg.fr (130.79.125.5) or via http://cdsweb.u-strasbg.fr/cgi-bin/qcat?J/MNRAS/.
\end{minipage}

\end{center}
\end{table*}
\setlength{\tabcolsep}{6pt}

We have found that the feedback from the evolving proto-cluster is warming the clumps and increasing the molecular line-widths.  Although the material towards the centre of the clump is depleted as the embedded proto-cluster evolves through a combination of accretion and outflows, this is unlikely to have a significant impact on the mass of the clumps given that there is no correlation with evolution for either the clump mass or peak column density. The lack of any observed correlation between mass and evolution suggests, additionally, that global infall has little impact on the mass of the clumps (as calculated in Sect.\,\ref{sect:mass}): if there was an impact, we would have expected to find a positive correlation between the two parameters. It is of course possible that the amount of material gained through infall is almost exactly matched by material accreted on the proto-cluster and removed by outflows, although the existence of such a delicate balance over such large ranges in clump mass and evolution is difficult to imagine.

\section{Properties of most active complexes in the Galaxy}
\label{sect:active_regions}

We present the derived physical parameters for the 30 most massive clusters identified in Table\,\ref{tbl:derived_cluster_para}. A comparison of the mass and luminosity contained in these 30 regions with those of the whole sample ($\sim 1 \times 10^7$\,\msun\ and $\sim 1.3\times 10^8$\,\lsun, respectively) shows that they collectively contribute 36\,per\,cent of all of the mass and 52\,per\,cent of all of the luminosity, but only 16.4\,per\,cent of the total population of clumps. This rather limited number of regions is therefore responsible for a majority share of the current star-formation rate in the Galaxy. 

This sample includes all of the previously well-known star forming complexes such as W31, W33, W43, W49, W51, G305 and G333. All clusters contribute between $\sim$0.5-3\,per\,cent of the total mass and no cluster stands out in this regard. A comparison of luminosities is similar, although clusters W49 and W51 do stand out from the others.  These two clusters together contribute $\sim$20\,per\,cent of the total luminosity found in the inner part of the Galactic disk.

We previously used the \lm\ ratio as a means to evaluate the current evolutionary state of the embedded star formation, where higher values are associated with more-evolved objects. Here we estimate the average \lm\ ratio for the clusters as a whole to obtain a measure of their star formation activity: these values are given in the last column of Table\,\ref{tbl:derived_cluster_para}. The average \lm\ ratio for the sample is 15.9\,\lsun/\msun\ with a standard deviation of 16.3\,\lsun/\msun. While most clusters are within one standard deviation of the mean, W51 and G351 are both nearly three sigma from the mean and W49 is over three sigma away from the mean. These three clusters have \lm\ ratios at least twice as large as the next-highest cluster, and are the most-evolved star formation regions (have the highest instantaneous SFE) in the Galaxy.  These values put them in the wings of the \lm\ distribution: they are therefore rather extreme compared to all of the other regions, and are probably the best examples of ``mini-starbursts'' we have in the Galaxy. We note that as pointed out by \citet{eden2015}, these sources are not outliers from the general (log)normal \lm\ distribution: while certainly extreme, they are not abnormal, and sources like these are expected to arise statistically when the sample is large enough.

\section{Summary and conclusions}
\label{sect:conclusions}

The ATLASGAL survey has identified $\sim$10,000 dense clumps located across the inner mid-plane of the Galaxy. We concentrate in this paper on a subsample of 8002 clumps located within the disk and away from the Galactic centre region. We have avoided the inner-most 5 degrees of the Galactic plane due to problems with source confusion and difficulty obtaining reliable distances. We have used a combination of archival molecular line data sets, line surveys reported in the literature, and dedicated follow-up observations of ATLASGAL sources (some of which are reported here) to determine a radial velocity (\vlsr) for 7807 clumps (corresponding to $\sim$98\,per\,cent of the sample). 

These velocities have been used in conjunction with a Galactic rotation model to estimate kinematic distances, and analysis of \hi\ spectra has been conducted to resolve the distance ambiguities that affect all sources located inside the Solar circle. These kinematic distance solutions have been improved using maser parallax and spectroscopic distances reported in the literature where these quantities have been available. We use a friend-of-friends algorithm to identify clusters of clumps that are coherent in position and velocity (i.e., $\ell$, $b$ and $v$). This identifies groups of sources that are likely to be part of larger-scale structures such as giant molecular clouds and giant molecular filaments. This analysis has resulted in the identification of 776 clusters, and has recovered all of the most active and well-known star forming complexes in the Galaxy. 

The identifications of clumps as parts of larger complexes allows the properties of star forming complexes to be compared to each other, and the role of the spiral arms and other environmental conditions to be explored. This can additionally reduce the inherent uncertainties associated with resolving the kinematic distance ambiguity. We have been able to assign a distance to 7770 clumps ($\sim$97\,per\,cent of the whole sample) but many of the remaining sources are weak and often diffuse. 

We have performed aperture photometry on mid-infrared to submillimetre survey data, and used these flux measurements to fit the spectral energy distribution of the clumps. These provide a measure of the dust temperatures and bolometric fluxes. This analysis has also been used to separate the clumps into four evolutionary types: quiescent, protostellar, young stellar objects and massive star forming. The distances, temperatures and bolometric fluxes are used to derive the physical parameters for the clumps (luminosities, sizes, mass and column densities), and comparisons have been made between the different evolutionary stage subsamples to identify trends that may provide some insight to the star formation process and the evolution of the clumps. These comparisons have revealed that dust temperatures, bolometric luminosities, and \lm\ ratios increase with advanced evolutionary stage.\\

\noindent Our main findings are as follows:
\begin{enumerate}

\item A comparison of the mass-radius distribution to empirically-determined thresholds for the efficient formation of massive stars finds that the vast majority of all clumps are capable of forming massive stars. Furthermore, the vast majority of clumps appear to be unstable against gravity and can therefore be considered to be in a pre-stellar state as opposed to starless.\\

\item We find that there is a very strong positive correlation between column density and the fraction of clumps associated with massive star formation; this saturates at $\log(N) \sim$ 23\,cm$^{-2}$,  above which the proportion is 100\,per\,cent. We do not find a correlation between column density and evolution, and so we can exclude the possibility that the column density increases over time and conclude that the reason we do not find any quiescent, protostellar or YSO-forming clumps with similar column densities is because the evolution timescale is inversely correlated to the column density. A consequence of this is that the highest column density clumps evolve so rapidly and are so rare that the early stages are simply not observed. \\

\item We find that the clump mass and evolutionary stage are uncorrelated: the clump mass is not significantly altered by infalling mass from the larger scale environment over its lifetime.  Clumps are widely thought to form over long periods of time from infalling material either funneled along filaments or from global collapse from larger scales and although there have been some studies that have found evidence to support this hypothesis, it does not seem to apply to the majority of clumps. \\

\item The fractions of clumps in the quiescent and protostellar stages are $\sim$10 and $\sim$20\,per\,cent for clumps up to $4\times 10^{3}$\,\msun\ but drop to $\sim$5 and 10\,per\,cent, respectively, for more massive clumps. The statistical lifetime of the early quiescent and protostellar stages are therefore relatively brief and are negatively correlated with the mass of the clump. As the most massive clumps are associated with the most massive stars (which have the shortest star formation timescales) the actual lifetimes for the most massive clumps will be very short indeed. \\

\item The short quiescent and protostellar statistical lifetimes are consistent with the hypothesis that the clumps form rapidly with most of their mass in place, and are initially very unstable to gravity so that star formation begins almost immediately. The mass infall rate is relatively modest, as is the mass accreted onto the proto-cluster over the star formation timescales, resulting in the clump mass being independent of evolution. The key conclusion here is that the formation of the clump and ensuing star formation must be very rapid, and once it has begun, is little influenced by the larger scale environment. \\

\item We have compared the distribution of clumps and clusters to the large scale features of the Galaxy and find a good correlation between the sources and the spiral arms ($\sim$90\,per\,cent found within 10\,\kms\ of a spiral arm loci). There is also a good correlation between the expected peaks in the Galactocentric distance distribution of the spiral arms and the MSF clumps. We find that all of the clumps are tightly correlated with the mid-plane of the Galaxy with a scale height of $\sim$26\,pc: we find no difference between the different subsamples. \\

\item We find little evidence for variations in the \lm\ ratio or mean clump temperature as a function of Galactocentric distance ($\sim$10\,\lsun/\msun\ and $\sim$20\,K, respectively) within the Solar circle when averaged over kiloparsec scales. The mean temperature increases slightly outside of the Solar circle, possibly due to the lower density of material and decreased shielding from the interstellar radiation field, but the \lm\ ratio appears to be relatively unchanged. There are more significant variations on smaller size scales of clusters and clumps (several tens of parsecs and parsec scales, respectively).     \\

\item We do not find any evidence for any enhancements in the \lm\ ratio at the expected galactocentric distances of the spiral arms. From this we conclude that although the spiral arms play a role in concentrating the molecular material into large scale structures they do not play a significant role in the star formation process. This finding is consistent with other recent studies.\\

\item We have combined the properties of the clumps grouped into clusters to estimate the properties of these larger-scale structures. We find that the 30 most massive of these structures consists of only 16\,per\,cent of all of the clumps but contain 36\,per\,cent of the mass and 52\,per\,cent of the total bolometric luminosity. These clusters therefore are responsible for a significant fraction of the Galactic star formation rate and energy budget.
\\

\item The majority of clusters identified have \lm\ ratios  $<$25\,\lsun/\msun, however, three clusters have significantly higher values ($>$50\,\lsun/\msun) and so are much more active and evolved regions (these are W49, W51 and G351.598+00.189). Combined, these three clusters produce almost 25\,per\,cent of the total luminosity produced by the ATLASGAL sources (11.7\,per\,cent for W49, 8.56\,per\,cent for W51 and 4.3\,per\,cent for G351.598+00.189). The most extreme of these three regions is W49, the most luminous complex in the Galaxy.  W49 has the highest \lm\ ratio and is one of only three objects found in the disk that satisfies the criteria to be considered a massive proto-cluster candidate.\\

\end{enumerate}

This paper is the fourth in a series that is focusing on exploiting the ATLASGAL survey to provide a catalogue of properties for all dense clumps located within the inner Galaxy. This paper completes this work and in subsequent papers we will use this catalogue to further investigate the role of the spiral arms and Galactic environment and investigate the difference between clump properties and star formation between the inner and outer Galaxy.

\section*{Acknowledgments}

 We thank the referee for a timely report and helpful comments and suggestions that have improved the clarity of this paper. The ATLASGAL project is a collaboration between the Max-Planck-Gesellschaft, the European Southern Observatory (ESO) and the Universidad de Chile. This research has made use of the SIMBAD database operated at CDS, Strasbourg, France. This work was partially funded by the Collaborative Research Council 956, sub-project A6, funded by the Deutsche Forschungsgemeinschaft (DFG). This paper made use of information from the Red MSX Source survey database at http://rms.leeds.ac.uk/cgi-bin/public/RMS\_DATABASE.cgi which was constructed with support from the Science and Technology Facilities Council of the UK.




\bibliographystyle{mn2e_new}
\bibliography{urquhart2016,temp}


\appendix

\section{APEX $^{13}$CO and C$^{18}$O observations}

\subsection{CO observations and data reduction}\label{sect:obs}

We targeted the $^{13}$CO and C$^{18}$O (2--1) transitions at $\sim$220\,GHz to measure the source LSR velocity. The data were obtained between June and November 2016 at the 12~m diameter Atacama Pathfinder Experiment telescope (APEX, \citealt{guesten2006}), using the low frequency module of the Swedish Heterodyne Facility Instrument (SHeFI, \citealt{vassilev2008}) receiver. Back-ends consisted of two wide-band Fast Fourier Transform Spectrometers (XFFTS; \citealt{klein2012}) each covering 2.5\,GHz instantaneous bandwidth with 32768 spectral channels.  We observed the transitions with a velocity resolution of $\sim$0.1\,\kms\ in position-switching mode with an on-source integration time of 1-4\,mins depending on the line strength. 
 
We obtained spectra towards 1177 sources at an average precipitable water vapor column of PWV = $2.9\pm1.2$\,mm. No emission was detected towards 4 sources (AGAL354.814+00.121, AGAL005.476$-$00.391, AGAL006.461$-$00.389 and AGAL338.306$-$00.522) and the spectra towards another 8 sources were discarded due to contamination in the off-source position. These observations therefore provided useful velocity information for the remaining 1165 sources.

\begin{table}
\begin{center}
\caption{Summary of the APEX observational parameters.}
\begin{tabular}{lc}
\hline
Parameter & Value \\
\hline\hline
Galactic longitude range      &      300\degr\ $< \ell <$  60\degr\ \\
Galactic latitude range      &      $-$1.54\degr\  $< b <$ 1.58\degr\ \\
Number of observations & 1177 \\
Number of usable spectra & 1165 \\
Frequency & 220\,GHz \\
Angular resolution & 30\arcsec \\
Spectral resolution & 0.1\,\kms\ \\
Smoothed spectral resolution & 1\,\kms\ \\
Medium noise (T$^{*}_{\rm{A}}$) & $\sim$95\,mK\,channel$^{-1}$ \\
Medium system temperatures (T$_{\rm{sys}})$ & $\sim$220\,K\\
Integration time (on-source) & 1-4\,mins\\
\hline
\end{tabular}

\label{tab:obs_parameters}
\end{center}
\end{table}

\subsubsection{Data reduction and line fitting}

\begin{figure}

\includegraphics[width=\linewidth, trim= 20 0 0 0]{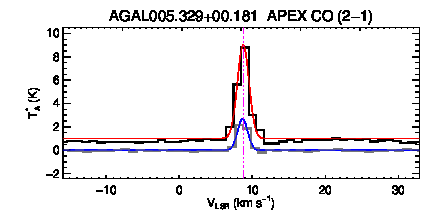}
\includegraphics[width=\linewidth, trim= 20 0 0 0]{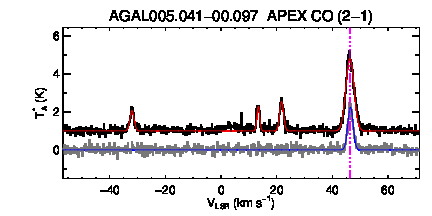}
\includegraphics[width=\linewidth, trim= 20 0 0 0]{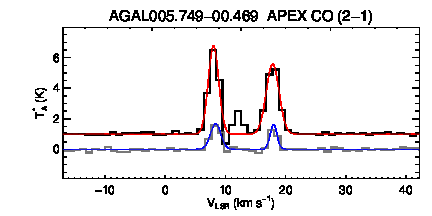}

\caption{Example APEX CO (2--1) spectra towards three sources. The $^{13}$CO spectra is offset from the C$^{18}$O spectra by 1\,K and the fits are shown by the red and blue lines, respectively. The vertical dashed lines shown in the upper two panels indicates the preferred source velocity. The two components shown in the lower panel have comparable intensities and optical depths and so no velocity has been assigned to the source.}

\label{fig.veloComponents}
\end{figure}

The data are analyzed for peaks within a velocity range between $\pm$200\,\kms\ of the rest frequency of the transition to obtain the major velocity components for each position.  We first added all scans for each source using the Continuum and Line Analysis Single-dish Software (CLASS)\footnote{\url{http://www.iram.fr/IRAMFR/GILDAS}} from within our Python code through the PyCLASS module of the PyGILDAS\footnote{\url{http://cdsarc.u-strasbg.fr/doc/man/gildas/html/gildas-python-html/}} package. The resulting spectra are then smoothed to a velocity resolution of $\sim$\,1\,\kms, resulting in a mean noise level of $\sigma$=$95\pm 50$\,mK. The spectra are subsequently flattened using a first-order polynomial and the velocities of the emission components are determined using a Python code. Here the peaks are identified iteratively by first de-spiking the spectrum and then searching a contiguous window for emission above the $3\sigma$ noise level. The number of maxima within a window is determined and these peaks are simultaneously fit with Gaussian profiles. The resulting sum of these profiles is subsequently subtracted from the spectrum, and the resulting spectrum is then used in the next iteration to find the next strongest emission window. Only peaks separated by at least twice the width of the fitted Gaussian are considered as major velocity components in order to avoid cluttering with multiple components per window that might be all associated with the same cloud. Negative features likely to originate from a contaminated off position were automatically identified in a similar way.

Figure\,\ref{fig.veloComponents} presents a few examples of the spectra obtained. We have not converted these to the main beam temperature scale ($T_{\rm{mb}}$) as we are only concerned with the velocity information these provide and so the spectra are plotted using the antenna temperature scale ($T^*_{\rm{a}}$). In many cases a single component is detected and the source velocity can be determined without any ambiguity (upper panel). However, in approximately 40\,per\,cent of cases multiple components are detected (middle and lower panels); in these cases we applied the criteria described in Sect.\,\ref{sect:archival_surveys}. In addition, we also used the peaks of the $^{13}$CO and C$^{18}$O components to estimate their optical depths and used this as an additional constraint --- if the $^{13}$CO components have similar integrated intensities but one has a significantly higher optical depth (effectively selecting the component with the lowest peak $^{13}$CO/C$^{18}$O ratio) as this is also likely to have the highest column density and therefore more likely to be associated with the high-column clumps identified by ATLASGAL.

In total, 3142 $^{13}$CO and 1271 C$^{18}$O components are detected towards 1165 clumps. As discussed in Sect.\,\ref{sect:archival_surveys}, in case of two velocity components, we associated the velocity of the brightest component to the ATLASGAL peak. We have been able to assign a velocity to 1115 clumps from these data. The fitted line parameters are given in Table\,\ref{tbl:apex_results}. For the other 50 sources  multiple emission components are detected with similar intensities and we are unable to assign a velocity with any confidence.



\setlength{\tabcolsep}{4pt}
\begin{table*}


\begin{center}
\caption{CO components detected by APEX. In this table we indicate the velocity assigned to the source by appending a $\star$ to the source name. We give the noise per channel of the $^{13}$CO in Col.\,2, however, the $^{13}$CO and C$^{18}$O noise agree to within a few per\,cent.}
\label{tbl:apex_results}
\begin{minipage}{\linewidth}
\small
\begin{tabular}{lr....c.....}
\hline \hline

\multicolumn{1}{c}{} & 	\multicolumn{5}{c}{$^{13}$CO (2-1)}  &\multicolumn{4}{c}{C$^{18}$O (2-1)} & \multicolumn{1}{c}{}\\
\cline{3-6}\cline{8-11}

\multicolumn{1}{c}{ATLASGAL} & \multicolumn{1}{c}{RMS} &	\multicolumn{1}{c}{\vlsr}  &\multicolumn{1}{c}{Peak}  & \multicolumn{1}{c}{Width} & \multicolumn{1}{c}{Intensity} & & \multicolumn{1}{c}{\vlsr}  &\multicolumn{1}{c}{Peak}  & \multicolumn{1}{c}{Width} & \multicolumn{1}{c}{Intensity} & \multicolumn{1}{c}{Optical}\\

\multicolumn{1}{c}{CSC name}&  \multicolumn{1}{c}{(mK)} &	\multicolumn{1}{c}{(\kms)}  &\multicolumn{1}{c}{(K)}  & \multicolumn{1}{c}{(\kms)} & \multicolumn{1}{c}{(K\,\kms)} & & \multicolumn{1}{c}{(\kms)}  &\multicolumn{1}{c}{(K)}  & \multicolumn{1}{c}{(\kms)} & \multicolumn{1}{c}{(K\,\kms)} & \multicolumn{1}{c}{depth}\\
\hline
AGAL005.001+00.086$\star$	&	61	&	2.1	&	5.8	&	1.38	&	20.2	&&	2.0	&	1.8	&	0.93	&	4.2	&	0.35	\\
AGAL005.041$-$00.097$\star$	&	118	&	46.3	&	3.8	&	1.53	&	14.4	&&	46.5	&	2.1	&	1.04	&	5.5	&	0.82	\\
AGAL005.049$-$00.192$\star$	&	72	&	6.1	&	3.2	&	1.53	&	12.4	&&	6.1	&	0.9	&	0.97	&	2.2	&	0.32	\\
AGAL005.139$-$00.097$\star$	&	76	&	44.1	&	5.0	&	1.50	&	18.9	&&	44.2	&	1.9	&	1.07	&	5.1	&	0.47	\\
AGAL005.184+00.159	&	168	&	182.1	&	1.0	&	1.26	&	3.1	&&	\multicolumn{1}{c}{$\cdots$}	&	\multicolumn{1}{c}{$\cdots$}	&	\multicolumn{1}{c}{$\cdots$}	&	\multicolumn{1}{c}{$\cdots$}	&	\multicolumn{1}{c}{$\cdots$}	\\
AGAL005.192$-$00.284$\star$	&	70	&	8.0	&	3.4	&	2.21	&	18.6	&&	8.6	&	0.6	&	1.09	&	1.7	&	0.16	\\
AGAL005.202$-$00.036$\star$	&	68	&	0.6	&	4.8	&	1.99	&	23.7	&&	0.6	&	1.4	&	1.78	&	6.1	&	0.32	\\
AGAL005.329+00.181$\star$	&	69	&	8.8	&	8.0	&	0.79	&	15.9	&&	8.7	&	2.7	&	0.69	&	4.6	&	0.40	\\
AGAL005.371+00.319$\star$	&	61	&	18.7	&	5.7	&	1.51	&	21.4	&&	18.7	&	1.7	&	1.31	&	5.6	&	0.35	\\
AGAL005.387$-$00.551$\star$	&	92	&	8.4	&	1.1	&	1.61	&	4.5	&&	\multicolumn{1}{c}{$\cdots$}	&	\multicolumn{1}{c}{$\cdots$}	&	\multicolumn{1}{c}{$\cdots$}	&	\multicolumn{1}{c}{$\cdots$}	&	\multicolumn{1}{c}{$\cdots$}	\\
AGAL005.389$-$00.384	&	82	&	21.4	&	1.1	&	9.66	&	26.0	&&	\multicolumn{1}{c}{$\cdots$}	&	\multicolumn{1}{c}{$\cdots$}	&	\multicolumn{1}{c}{$\cdots$}	&	\multicolumn{1}{c}{$\cdots$}	&	\multicolumn{1}{c}{$\cdots$}	\\
AGAL005.392$-$00.409$\star$	&	164	&	140.8	&	1.1	&	7.93	&	22.8	&&	\multicolumn{1}{c}{$\cdots$}	&	\multicolumn{1}{c}{$\cdots$}	&	\multicolumn{1}{c}{$\cdots$}	&	\multicolumn{1}{c}{$\cdots$}	&	\multicolumn{1}{c}{$\cdots$}	\\
AGAL005.399$-$00.237$\star$	&	53	&	4.8	&	1.9	&	1.34	&	6.3	&&	5.0	&	0.4	&	1.01	&	1.1	&	0.24	\\
AGAL005.409$-$00.302$\star$	&	60	&	14.2	&	4.6	&	0.61	&	7.1	&&	\multicolumn{1}{c}{$\cdots$}	&	\multicolumn{1}{c}{$\cdots$}	&	\multicolumn{1}{c}{$\cdots$}	&	\multicolumn{1}{c}{$\cdots$}	&	\multicolumn{1}{c}{$\cdots$}	\\
AGAL005.437$-$00.314	&	74	&	22.5	&	1.1	&	1.35	&	3.7	&&	22.2	&	0.4	&	1.00	&	1.0	&	0.43	\\
\hline\\
\end{tabular}\\

Notes: Only a small portion of the data is provided here, the full table is available in electronic form at the CDS via anonymous ftp to cdsarc.u-strasbg.fr (130.79.125.5) or via http://cdsweb.u-strasbg.fr/cgi-bin/qcat?J/MNRAS/.
\
\end{minipage}

\end{center}
\end{table*}
\setlength{\tabcolsep}{6pt}

\section{Distance determination}
\label{sect:append_distances}

A key element required for determining the physical properties of dense clumps identified by ATLASGAL is their heliocentric distance. The most reliable way to do this is to measure the parallax of bright, compact objects such as methanol and water masers (e.g., \citealt{reid2014}). These are time-consuming and complex measurements, and since they require the presence of a bright source (which is not always available, particularly for the very earliest evolutionary stages) this method is not universally applicable to all sources. The difficulties involved with these measurements and the poor frequency coverage of telescopes in the southern hemisphere has limited the number of maser parallax measurements reported in the literature ($\sim$150 measurements; \citealt{reid2014}) with nearly all them located in the first and second quadrants of the Galaxy. A consequence of this is that the structure of half of the Milky Way is poorly constrained. This situation is improving with the number of maser parallax measurements steadily increasing, and there has been significant progress in developing the Australia Long Baseline Array (e.g., \citealt{krishnan2017}) to provide distance measurements in the third and fourth quadrants. However, it will be a long time before this method can provide distances for the majority of the ATLASGAL sample and we therefore need another method to determine distances.

Kinematic distances, although less reliable than maser parallax distance, can be determined from a source's radial velocity with respect to the local standard of rest (\vlsr) and a model of the rotation of the Milky Way. The radial velocity can be obtained from molecular line observations towards the clumps as has been described in the previous section. There are a number of descriptions of the rotation curve of the Galaxy (e.g., \citealt{clemens1985, brand1993, reid2014}).  In most cases the kinematic distances obtained from all of them agree within their associated uncertainties (typically $\pm$0.3-1\,kpc; the largest variations are associated with sources located close to the Solar circle) and so the choice of rotation curve is not particularly critical.    

\citet{wienen2015} has performed a detailed kinematic distance study of several thousand ATLASGAL sources using the rotation curve of \citet{brand1993}. Here we extend that work to an almost complete sample of ATLASGAL sources and refine some of their distances with an improved rotation curve model. We have used the rotation curve determined by \citet{reid2014} to estimate the kinematic distances as it takes advantage of all of the maser parallax measurements to constrain the model, and has been shown to provide kinematic distances that are comparable with the maser distances. An inherent problem with kinematic distances is that for sources located within the Solar circle there are two possible distance solutions for each velocity; these are evenly spaced on  either side of the tangent position and are commonly referred to as the near and far distances. These kinematic distance ambiguities (KDA) need to be resolved before a unique distance can be assigned for a particular source. 

\begin{figure*}
\centering 
\includegraphics[width=0.49\textwidth, trim= 20 0 30 0, clip]{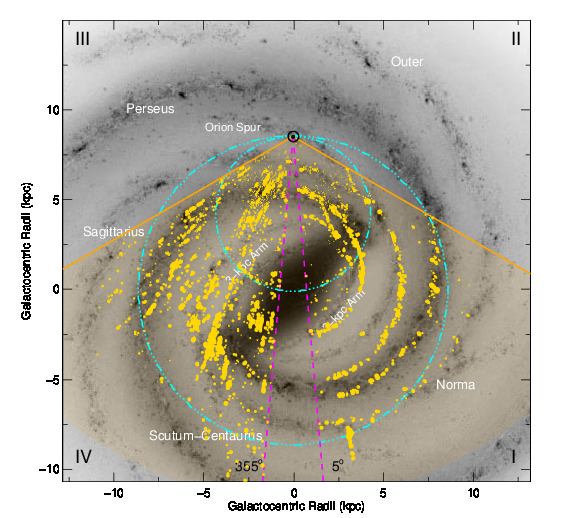}
\includegraphics[width=0.49\textwidth, trim= 20 0 30 0, clip]{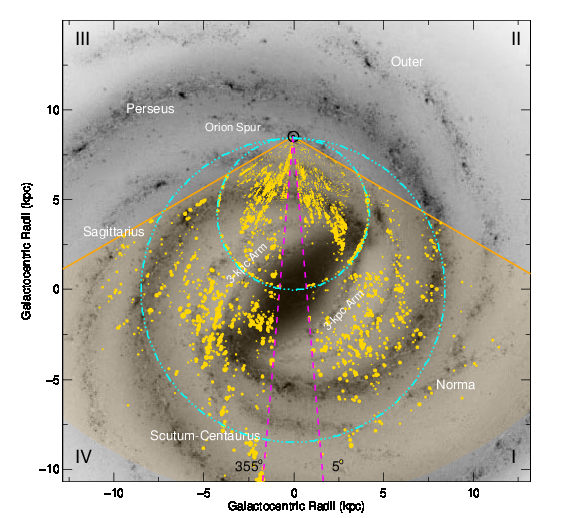}

\caption{As Fig.\,\ref{fig:galactic_cluster_distribution}. Left panel: distribution of all ATLASGAL sources given by the Bayesian maximum likelihood tool developed by \citet{reid2016}. Right panel: distribution of all ATLASGAL sources determined using the \citet{reid2014} Galactic rotation curve. }
\label{fig:galactic_distribution}
\end{figure*}

\citet{reid2016} have also developed a Bayesian maximum likelihood method that takes into account the relative position of the spiral arms along the line of sight and the latitude of the source to resolve the KDA and determine the distance (the application of this method to all of the ATLASGAL clumps can be seen in the left panel of Fig.\,\ref{fig:galactic_distribution}). This method assumes that all sources are likely to be associated with a spiral arm, which may be a reasonable assumption for star-forming regions; however, it is to be avoided, as one of the main purposes of studying Galactic structure is to determine the effect of spiral arms on the star-formation process and this requires distinguishing arm from inter-arm sources, as much as possible. In the right panel of Fig.\,\ref{fig:galactic_distribution} we show the distribution of clumps using distances calculated using only the rotation curve. We find that in the vast majority of cases ($\sim$95\,per\,cent), the difference between the rotation-curve distance and the Bayesian maximum likelihood distance is smaller than 1\,kpc (see Fig.\,\ref{fig:delta_distance}). This difference is relatively modest and is almost negligible for sources located in the fourth quadrant as the position of the spiral arms are poorly constrained and parallax distances virtually non-existent. The Bayesian method does provide more reliable distances for source located near the Solar circle, where the kinematic distance uncertainties are significantly larger. As a result of these concerns, we  decided to resolve the KDA via alternate methods (these will be discussed in the following subsection), and then adopt the Bayesian value if it is in agreement with the kinematically determined solution.  Otherwise, we adopt the kinematic value.

\begin{figure}
\centering
\includegraphics[width=0.49\textwidth, trim= 0 0 0 0]{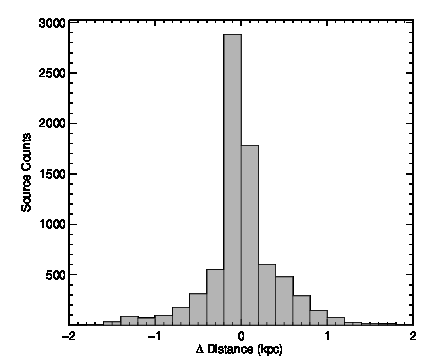}

\caption{Histogram showing the difference between the distance determined from using the Bayesian method presented by \citet{reid2016} and the near/far distances determined using the \citet{reid2014} rotation curve (Note that the $x$-axis has been truncated). 98.4\,per\,cent of the sample have velocity differences within the plotted range, and $\sim$95\,per\,cent are within 1\,kpc. \label{fig:delta_distance} } 

\end{figure}

We present the source names, radial velocities, near and far distances determined from the rational curve (\citealt{reid2014}), the Bayesian distance (\citealt{reid2016}), the kinematic solution, and the assigned distance to the sources in Table\,\ref{tbl:source_vlsr}.

\subsection{Resolving the kinematic distance ambiguities}
\label{sect:flowchart_description}

Kinematic distance ambiguities affect all sources located within the Solar circle (i.e., Galactocentric radius ($R_{\rm{GC}}$) $ < 8.35$\,kpc; \citealt{reid2014}) and these need to be resolved before a unique distance can be determined for a particular source.  We present a flow chart in Fig.\,\ref{fig:distance_flowchart} to illustrate the various steps we have used to resolve these distance ambiguities. 

\begin{enumerate}

\item We match clumps to reliable distances reported in the literature (maser parallax, e.g., \citealt{reid2014} and spectroscopic measurements, e.g., \citealt{moises2011}). This is done by comparing the longitude, latitude and velocities of the sources with positions of the given these studies and where a correlation is found the distance is adopted. \\

\item Sources located outside the Solar circle are not affected by this twofold distance ambiguity and so the kinematic distance is adopted; however, this only applies to a relatively small fraction of the sources ($\sim$300).\\

\item We can exclude sources located close to the tangent velocity ($|v_{\rm{clump}}-v_{\rm{tangent}}|<$ 10\,\kms) from the KDA tests, as the difference between the near and far distance becomes smaller than their associated uncertainties: this eliminates $\sim$10\,per\,cent of the sample.\\

\item The latitude distribution of molecular material and star formation is tightly correlated with the Galactic mid-plane. High-mass stars found within the Solar circle have a scale height of $\sim$30\,pc (e.g., \citealt{reed2000,green2011b,urquhart2014_rms}), and since most high-mass star-forming clumps have masses on the order of $10^3$\,\msun\ we would expect them to have a similar scale height distribution. We can therefore use this as an additional constraint on the possible distance for clumps located inside the Solar Circle. We have calculated the distance of all sources from the Galactic mid-plane assuming they are located at the far distance; if this distance is greater than 120\,pc (i.e., 4 times the scale height) then the far distance is considered significantly less likely and the source is placed at the near distance.\\

\item Distances to many \hii\ regions have been determined by looking for \hi\ absorption features against the strong radio continuum emission associated with the ionized gas (e.g., \citealt{wienen2015,urquhart2012_hiea,anderson2009a,Kolpak2003}). The envelope of cooler \hi\ gas surrounding a molecular cloud located between the \hii\ region and the observer will produce an absorption feature in the \hii\ continuum emission at the velocity of the intervening clump. If the \hii\ region is located at the far distance we would expect to see absorption at higher velocities than the source, and these should extend all the way up to the velocity of the tangent position. However, if the \hii\ region is located at the near distance we would expect to see absorption at the same velocity as the \hii\ region but not at higher velocities. We have matched our clumps with \hii\ region studies reported in the literature and if a reliable distance has been found then this has been adopted (e.g., \citealt{Kolpak2003, urquhart2012_hiea, urquhart2013_cornish}). If no distance is available the source has been returned to the rest of the sample for further analysis. \\

\item \hi\ self-absorption (\hi\ SA) is then employed to resolve the distance ambiguity for any remaining sources. Clumps located at the near distance are likely to be associated with an absorption feature in the \hi\ spectra at the same velocity as the source (due to the cold clump absorbing emission from the warmer diffuse \hi\ gas located behind it). Conversely, if the clump is located at the far distance and the warm \hi\ gas is distributed throughout the Galactic plane there would be no corresponding dip in the \hi\ spectra at the same velocity of the source. This method for resolving the distance ambiguities has been widely used in a number of studies (e.g.,  \citealt{roman2009,anderson2009a,jackson2006}). 

We have extracted \hi\ spectra from the Southern Galactic Plane Survey (SGPS; \citealt{mcclure2005}) and the VLA Galactic Plane Survey (VGPS; \citealt{stil2006}) archives towards all ATLASGAL clump. The resolution of these \hi\ surveys is comparable to the typical sizes of the clumps ($\sim$1\arcmin; \citealt{contreras2013}) and so we sum emission over one resolution element of the respective surveys (typically 9 pixels). We make no attempt to remove background emission as these clumps are likely to be embedded in larger more diffuse giant molecular cloud structures which will also absorb \hi\ emission at a similar velocity as the sources, and the boundaries of these larger structures are unknown.

The \hi\ spectra and the source velocity were combined in the same plot to facilitate comparison of the \hi\ profile around the velocity of the source (see Fig.\,\ref{fig:example_HI_spectra} for some examples). A review of a representative sample of the spectra identified four assignments (near, far,  ambiguous and problem) that were found to cover all of the possible outcomes. The first three  options are self-explanatory, while the last one was used to flag \hi\ spectra that are contaminated by strong continuum emission from evolved \hii\ regions located nearby. All plots were visually inspected by two of the lead authors (JSU and AG) independently to resolve the distance ambiguities. Once the analysis of the \hi\ data was completed, the two sets of results were compared. If the solutions agreed then the distance solution was adopted: these were considered to be high-reliability distances. If one of the solutions was near or far and the other was classed as ambiguous then the source was placed at the appropriate near or far distance but given a lower reliability flag. All other sources where the solutions disagreed were considered ambiguous and no distance was concluded from the \hi\ analysis. We present a key showing the distance solution adopted depending on the possible combinations of assigned solutions in Table\,\ref{tbl:hi_solution_key}.\\

\item If the \hi\ data were inconclusive, we searched for evidence of extinction towards the clumps. If a clump is associated with an infrared dark cloud (IRDC; e.g., \citealt{rathborne2006}), it suggests the source is in the foreground with respect to the relatively bright infrared emission that fills the inner part of the Galaxy. A recent study of a small sample of IRDCs by \citet{giannetti2015} found approximately 11\,per\,cent were actually located at the far distance and so placing these at the near distance is reasonable for the majority of clumps. We have searched for associations using the IRDC catalogue compiled by \citet{peretto2009} to identify sources associated with IRDCs. This catalogue does not include the inner 10\degr\ of the Galactic mid-plane and so we have visually inspected the mid-infrared images to identify possible IRDC associations for sources in this region. Any clumps found to be associated with an IRDC are placed at the near distance (the justification for this is discussed in Sect.\,\ref{sect:distance_summary}).\\

\item For any remaining sources, we searched the literature to see if a distance had been previously assigned and whether is was considered to be reliable (i.e., an evidence-based distance and not simply assumed to be at the near distance). If so, the corresponding kinematic distance solution was adopted (e.g., \citealt{svoboda2016_bgps,battisti2014}).  Spectroscopic distances (e.g., \citealt{moises2011,stead2010}) and parallax distances (e.g., \citealt{reid2014} and references therein) from the literature were simply adopted. 

\end{enumerate}

\begin{table}
\begin{center}
\caption{Possible kinematic distance options applied by the two authors and how these have been used to obtain a final solution. The question marks (?) indicate that although a solution has been adopted it is considered to be at a lower level of confidence.}
\label{tbl:hi_solution_key}
\begin{minipage}{\linewidth}
\small
\begin{tabular}{ccc}
\hline \hline
\multicolumn{1}{c}{Author 1}&  \multicolumn{1}{c}{Author 2}&\multicolumn{1}{c}{Adopted solution}  \\
\hline
Near & Near & Near \\
Far & Far & Far \\
Far & Ambiguous & Far?\\
Near & Ambiguous & Near?\\
Near & Problem & No solution \\
Far & Problem & No solution \\
Ambiguous & Ambiguous & No solution \\
\hline
\end{tabular}
\end{minipage}
\end{center}
\end{table}
\setlength{\tabcolsep}{6pt}

\begin{figure}
\centering 
\includegraphics[width=0.49\textwidth]{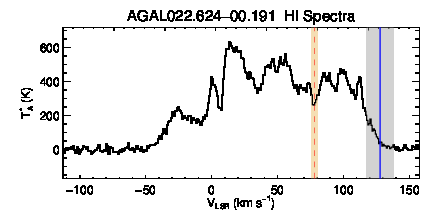}
\includegraphics[width=0.49\textwidth]{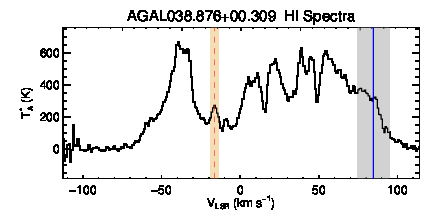}
\includegraphics[width=0.49\textwidth]{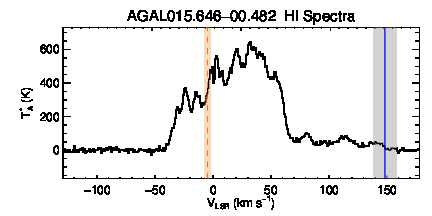}
\includegraphics[width=0.49\textwidth]{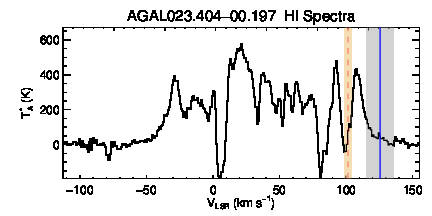}
\caption{Example \hi\ spectra extracted towards dense clumps identified from the ATLASGAL survey. The velocity of the tangent position is determined from a fit to \hi\ data (e.g.\, \citealt{mcclure2007}) and is indicated by the blue line; the grey shaded region covers a velocity range of $\pm$10\,\kms\ centred on the tangent velocity. Sources in this region are placed at the distance of the tangent point. The dashed red vertical line shows the velocity of the source with the yellow shaded region showing the typical FWHM line-width of the molecular lines. From top to bottom, panels show examples of clumps located at the near distance, the far distance, or identified as ambiguous distance and problem sources, respectively.}
\label{fig:example_HI_spectra}
\end{figure}

\subsection{Sources located near the Solar circle}

Assigning distances to sources located near the Solar circle (i.e., $|$\vlsr$|$ $<$ 10\,\kms) is a little more complicated because we also need to consider the additional uncertainty due to streaming motions in the Galaxy ($\pm$10\,\kms). These are already taken into account when estimating the errors in the derived distance. However, for sources located near the Solar circle this can result in a source actually located within the Solar circle (associated with a near/far distance ambiguity) appearing to be in the outer Galaxy (with no distance ambiguity) or vice versa. For sources located within the Solar circle we can simply try to resolve the distance ambiguity in the normal way using the \hi SA method; however, for sources with velocities that place them outside the Solar circle we only obtain a single distance although due to streaming motions they may actually be located inside the Solar circle. We therefore apply the following additional criteria:  

\begin{enumerate}

\item If the kinematic solution suggests that a near distance is likely, then no distance is assigned. These sources are likely to be located very close by ($\sim$1\,kpc) and are therefore probably relatively low-mass clumps.  Excluding these clumps is unlikely to impact our statistical analysis of the properties of the sample  and its Galactic distribution. \\ 

\item By similar rationale, if the kinematic solution suggests a far distance, then this source is likely to be at the far distance and we assign the source to the far distance. 

\end{enumerate}

\subsection{Summary of the kinematic distance analysis and comparison with the literature}
\label{sect:distance_summary}

In total we resolved the distance ambiguity for 7091 sources, of which we rate 6292 as `reliable'; these values correspond to 88.6\,per\,cent and 78.6\,per\,cent of the ATLAGAL CSC, respectively. We were unable to determine a distance for 911 sources; however, velocities were not available for  240 of these (as discussed in Sect.\,\ref{sect:completenss}). The distribution of all of these clumps determined using the \citet{reid2014} model is shown in the right panel of Fig.\,\ref{fig:galactic_distribution}. In Table\,\ref{tbl:distance_solutions} we present a breakdown of how the distance have been assigned following the application of the various steps outlined in Sect.\,\ref{sect:flowchart_description}. 

We placed 4817 clumps at the near distance, 1430 at the far distance and 760 sources at the tangent position.  The proportion of sources placed at the near and far distances is therefore $\sim$77 and 23\,per\,cent, respectively. This is similar to the findings of other studies (e.g., \citealt{eden2012} assigned $\sim$75\% of sources to the near distance).

\setlength{\tabcolsep}{6pt}

\begin{table}
\begin{center}
\caption{Summary of the kinematic distance solutions. The roman numerals (i) to (viii) given in Column\,1 refer to the various steps described in Sect.\,\ref{sect:flowchart_description}, while the (ix) and (x) are used to identify sources for which the \hi SA is ambiguous and clumps where no velocity is available.}
\label{tbl:distance_solutions}
\begin{minipage}{\linewidth}
\begin{tabular}{clc}
\hline 
\hline
Step & Description   & Total number distances  \\
 & of method   & assigned  \\
\hline
 (i) & Parallax/Spectroscopic  & 26 \\
(ii) & Outer Galaxy  & 84  \\
(iii)&  Tangent  & 760  \\
(iv) & Z distance   & 1295   \\
(v) & \hi EA   & 140   \\
(vi) &  \hi SA Near (?) & 3126 (559) \\
(vi) & \hi SA Far  (?)& 1280  (612) \\
(vii) & IRDC Associations  & 269  \\
(viii) & Literature  & 111  \\
\hline
(ix) & Ambiguous (Solar Circle) & 671 (43) \\
(x) & No \vlsr\ available  & 240  \\
\hline
\end{tabular}
\end{minipage}
\end{center}
\end{table}
\setlength{\tabcolsep}{6pt}

We  cross-correlated the positions of our sample of dense clumps with the position of infrared dark clouds (IRDCs; e.g., \citealt{peretto2009}), thereby identifying 1912 IRDCs in our sample that can be used to check the consistency of our distance assignments. IRDCs are thought to be located in the foreground between us and the bright diffuse infrared emission that fills much of the inner Galaxy. They are so dense that the background infrared emission is totally absorbed towards these objects resulting in these objects appearing in extinction with respect to the background field stars. If our distance solutions are reliable we should expect to find the vast majority of IRDCs will have been placed at either the near or tangent distances. 

Excluding step (vii) we have resolved the distance ambiguity towards 1606 of the matched IRDCs, of which 1457 have been placed at the near or tangent distances. This corresponds to 90.7\,per\,cent of the IRDC sources. Detailed studies of the distances of large and representative samples of IRDCs have not yet been performed and so the actual ratio of near/far sources is not well constrained. However, a recent study of $\sim$40 IRDCs associated with the ATLASGAL Top100 sample (\citealt{konig2017}) found 11\,per\,cent to be located at the far distance (\citealt{giannetti2015}). Our results are therefore consistent with what has been previously found and thereby increases confidence in the reliability of our results. This also provides strong support for automatically placing all remaining IRDCs for which we were unable to resolve the distance ambiguity at the near distance (i.e., step (vii)).

\setlength{\tabcolsep}{2pt}

\begin{table}
\begin{center}
\caption{Comparison of derived distance solutions with recent reports reported in the literature.}
\label{tbl:dist_literature_comparion}
\begin{minipage}{\linewidth}
\begin{tabular}{lccccc}
\hline 
\hline
Survey & Number  & Agree & Disagree & Agreement & Reference  \\
&  of Matches & &  &  \% &  \\
\hline
 BGPS&  902 &  698 &  204 &  77.4 & 1\\
 Reid Bayesian &  7130 &  5160 &  1970 &  72.4 &   2\\
 ATLASGAL &  1882 &  1410 &  472 &  74.9 &   3\\
 RMS  &  695 &  557 &  138 &  80.1 & 4\\
 BGPS  &  91 &  61 &  30 &  67.0 &   5\\
 MMB &  376 &  226 &  150 &  60.1 &   6\\
\hline
\end{tabular}
\end{minipage}
References: (1)  \citet{svoboda2016_bgps}; (2) \citet{reid2016}; (3) \citet{wienen2015}; (4) \citet{urquhart2014_rms}; (5) \citet{battisti2014}; (6) \citet{green2011b}.  
\end{center}
\end{table}
\setlength{\tabcolsep}{6pt}

We have also compared our distance solutions  with a number of other recent studies. We present a summary of this analysis in Table\,\ref{tbl:dist_literature_comparion}; this reveals a high level of agreement between our distances and nearly all of the comparison samples. Given that the reliability of the \hi\ analysis methods used here is only $\sim$80\,per\,cent (e.g., \citealt{busfield2006,roman2009}), the level of agreement with most studies considered here and provides further confidence in the reliability of our distance assignments. We have not performed a direct comparison with the recent work present by \citet{elia2017} as only $\sim$40\,per\,cent of their distances were the result of distance ambiguity resolution: the remainder were arbitrarily  placed at the far distance and so the samples are not comparable.

\begin{figure*}
\centering

\includegraphics[width=0.7\textwidth, angle=-90, trim= 0 0 0 0]{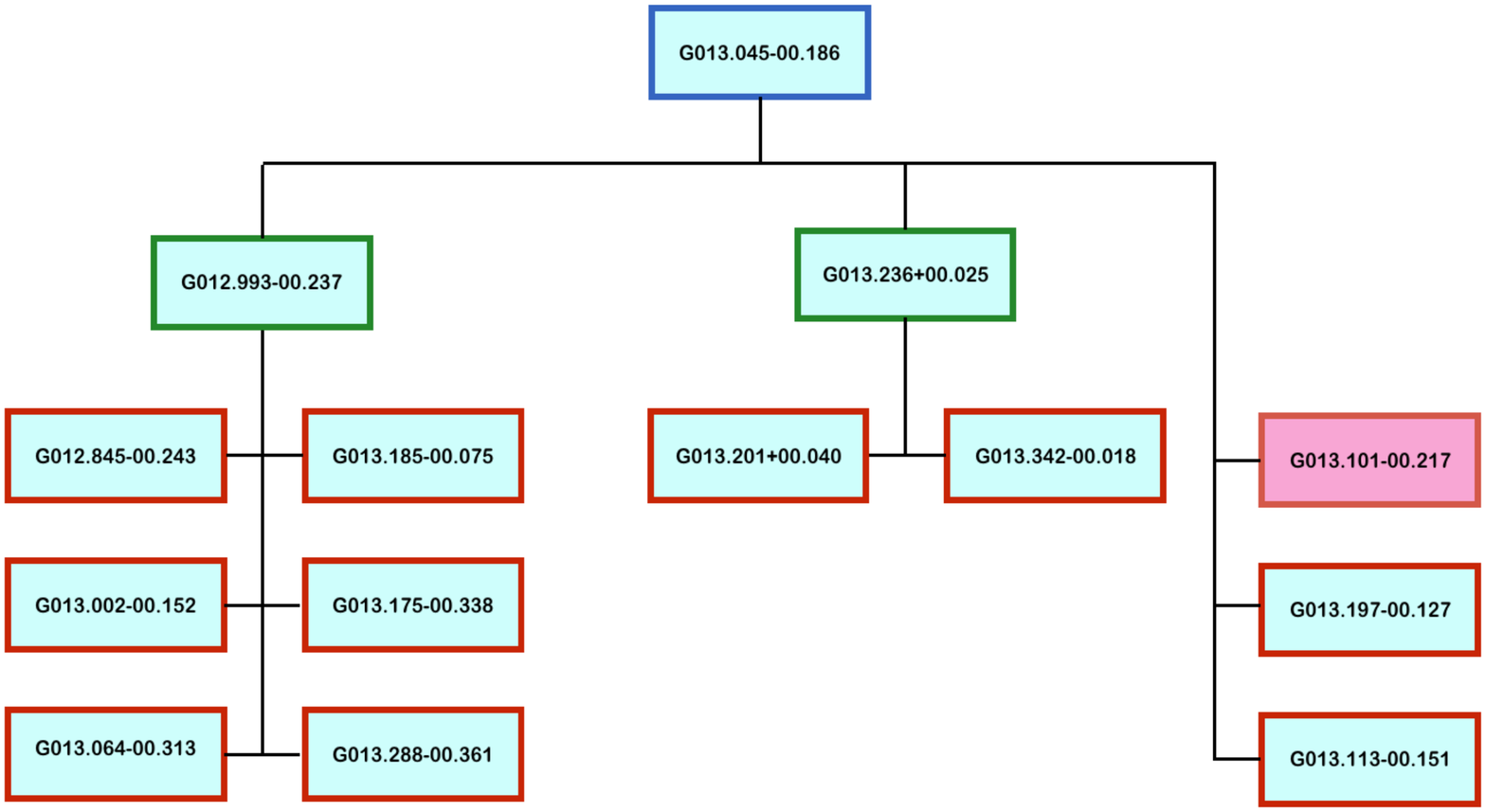}

\caption{Schematic diagram showing the hierarchical structure of a cluster identified from the friends-of-friends analysis. At the top of the diagram (outlined in blue) we have the structure identified by using the largest set of search parameters (i.e., $R_{\rm fof}<8$\arcmin\ and $\Delta v_{\rm fof} <8$\,\kms), the second row (outlined in green) shows the sub-regions identified by the using the intermediate search parameters (i.e., $R_{\rm fof}<6$\arcmin\ and $\Delta v_{\rm fof} <6$\,\kms), while the remaining rows show the smallest sub-regions identified using the smallest search parameters (i.e., $R_{\rm fof}<4$\arcmin\ and $\Delta v_{\rm fof} <4$\,\kms); these are outlined in red. Regions located at the near distance are shown with a cyan background while those found at the far distance are shown in with pink background. The distances associated with the smallest sub-regions are considered to be the most reliable.  \label{fig:cluster_structure} } 

\end{figure*}

\section{Identifying larger scale structures: GMCs and complexes}
\label{sect:complexes}

In the previous section we described how we have determined kinematic distance solutions to $\sim$7000 ATLASGAL clumps; however, there are still a significant number ($\sim$10\,per\,cent) of sources for which we have not been able to assign a distance. Furthermore, the reliability of many of the criteria we applied to resolve the distance ambiguities have an inherent uncertainty of $\sim$20\,per\,cent. We know that much of the star formation taking place in the Galaxy lies within the Solar circle ($R_{\rm{GC}} < 8.35$\,kpc) and is tightly concentrated to a relatively narrow range in the mid-plane (scale height $\sim$30\,pc) -- this has already been used to help resolve the distance ambiguities to many sources. We also know that a large fraction of the star formation is also further concentrated in large star forming complexes that can be associated with numerous GMCs (\citealt{murray2010,urquhart2014_rms}). 

We have thus far treated the ATLASGAL clumps as isolated/individual sources and determined their velocities and distances on an individual basis. Given that these clumps are simply the highest column density regions of GMCs we can group them together in $\ell b v$-space to identify the large-scale structures they are part of. This has many potential advantages, e.g.,:

\begin{enumerate}

\item it allows the distance solutions of clumps associated with the same GMCs and complexes to be assessed in a statistical way and for potential erroneous solutions to be identified and excluded, which will improve the overall reliability of the distances.\\

\item it allows a distance to be applied to sources that are found to be associated with a particular GMC but for which we were unable to resolve the ambiguity using the \hi\ analysis. \\

\item many GMCs are associated with strong velocity gradients and so measuring this for the individual clumps can result in significant differences between the component clumps. Although the differences in velocity between individual clumps might actually be quite modest ($<5$\,\kms) this can result in large differences in their kinematic distances ($\sim$0.5\,kpc) which may impact source properties and introduce significantly more scatter in their Galactic distribution. \\

\item many of the most prominent complexes in the Galaxy are already well studied (e.g., W31, W43 and G305) and there are very reliable distances available: identifying sources that are likely to be associated with each other allows us to adopt these distances.

\end{enumerate}

\subsection{Friends-of-friends analysis}
\label{sect:fof_analysis}

We have implemented the friends-of-friends method as a first step to identify possible larger-scale associations of clumps. We employed three different sets of parameters beginning with a matching angular radius ($R_{\rm fof}$) of 4\arcmin\ and velocity dispersion of ($\Delta v_{\rm fof}$) 4\,\kms\ and successively increasing the search parameters by 2\arcmin\ and 2\,\kms.  The selection of the first set of search parameters is based on the analysis by \citet{svoboda2016_bgps} on the reliability of assigning kinematic distance ambiguity solutions to groups of sources that are spatially and kinematically correlated sources (i.e., $\ell b v$); they refer to this as {\em Distance Resolution Broadcasting}. These parameters were determined by comparing the KDA solutions in groups of sources as the search parameters were increased and evaluating the number of disagreements within the group. 

The broadest set of search parameters (i.e., $R_{\rm fof}<8$\arcmin\ and $\Delta v_{\rm fof} <8$\,\kms) were selected as to include sources or groups of sources, identified using the smaller search parameters, that are associated with the same giant molecular clouds (these parameters correspond to size scales of $\sim$30\,pc at a distance of 15\,kpc, which is the typical cloud radius (\citealt{miville-deschenes2017}), and FWHM velocity dispersion of 10\,\kms, typical of GMCs found in the Large Magellanic Cloud (\citealt{hughes2010}). The intermediate set of search parameters roughly corresponds to the properties of smaller molecular clouds. 

Applying these search parameters produces a hierarchy of correlated sources that correspond to different scale structures in a GMC. At the smallest scale these identify tightly grouped clusters of clumps associated with one of perhaps many dense sub-regions within a cloud but where a common distance solution can be considered reliable. The second set of search parameters associates the coherent subgroups with their large scale molecular clouds, while the last set of search parameters links the molecular clouds with their host GMCs and star forming complexes. 

We are primarily interested in the largest scale structures identified with the largest search parameters. However, the distances for all of the sub-regions within these larger scale structures have been determined independently and so considering the distances of substructures provides a strong consistency check on the viability of the GMCs. We therefore begin at the top of the hierarchical tree (at the largest scale) and apply the following criteria to determine the reliability of a particular structure:

\begin{enumerate}

\item the agreement in distance solutions for all constituent clumps within a structure must be better than  $70$\,per\,cent. If this is not the case, then we look at the distances of the individual sub-regions. If a particular sub-region that has a distance resolution that is out of step with that of the larger scale structure and the sub-region can be removed without impacting the integrity of the larger structure then this is done. If this cannot be done then the top level structure is discarded and we perform the same analysis on the next level down in the hierarchical structure.  \\ 

\item a visual inspection of the distribution of the structure with respect to the large scale structure traced by the mid-infrared emission images and maps of the combined ATLASGAL+PLANCK emission (\citealt{csengeri2016_planck}) reveal good correlation. In some cases inspection of the mid-infrared images revealed that two or more clusters identified are actually part of larger scale structures and in such cases these have been combined into a single entity.\\  

\item As a final step we identify additional possible members by correlating the spatial extent of the clusters with the positions of clumps for which we were unable to determine a velocity. If a clump was found to fall within the footprint of a single cluster on the sky then it has been assigned to that cluster. If a clump fell within the footprint of 2 or more clusters along the line of sight then cluster membership is ambiguous and no cluster was assigned.

\end{enumerate}

In Fig.\,\ref{fig:cluster_structure} we present a schematic diagram showing the results of this analysis for one such cluster identified. The largest structure (G013.045$-$00.186) consists of two sub-regions (G012.993$-$00.237 and G013.236+00.025), each of which themselves consist of between 2-6 smaller sub-regions, and three smaller sub-regions (G013.101$-$00.217, G013.197$-$00.127 and G013.113$-$00.151). The distances for the smallest sub-regions are determined independently, and as can be seen in Fig.\,\ref{fig:cluster_structure}, 10 of the 11 of these have been placed at the near distance (cyan backgrounds); this includes all of the sub-regions associated with the two larger regions identified by the intermediate set of search parameters and these are therefore considered to be reliable structures. In this case we are able to exclude the one sub-region without affecting the integrity of the top level structure. In Fig.\,\ref{fig:apend_clusters} we show an image of the cluster towards W33 identified by this analysis.

\begin{figure}
\centering

\includegraphics[width=0.49\textwidth, trim= 30 0 0 0]{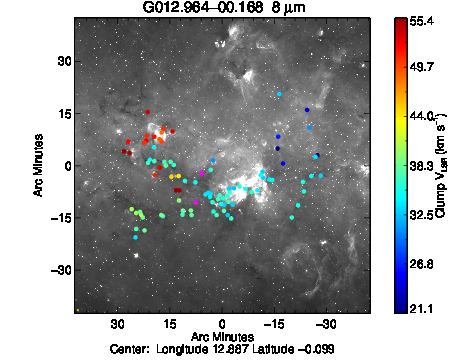}

\caption{Example of clustering analysis towards W33 complex. For description of image refer to Fig.\,\ref{fig:clusters}. The hierarchical structure of W33 is used as an example of the association method discussed in Sect.\,\ref{sect:fof_analysis} and shown in the flow chart presented in Fig.\,\ref{fig:cluster_structure}.  \label{fig:apend_clusters} } 

\end{figure}

\begin{figure}
\centering
\includegraphics[width=0.49\textwidth, trim= 0 0 0 0]{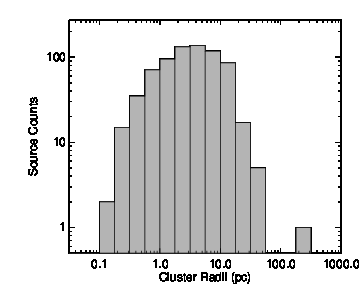}
\includegraphics[width=0.49\textwidth, trim= 0 0 0 0]{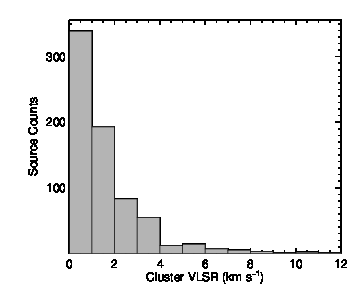}
\caption{Distribution of cluster sizes and velocity dispersions. The bin size used in the upper panel is 0.25\,dex and in the lower panel it is 1\,\kms.   \label{fig:cluster_sizes} } 

\end{figure}

In Fig.\,\ref{fig:cluster_sizes} we present plots of the physical size and velocity dispersion of the clusters as determined from the standard deviation of their associated clumps. The size of the clusters are from a fraction of a parsec to several tens of parsecs, but sizes of a few parsecs are more typical. One cluster stands out as being exceptionally large with a size of several 100 parsecs; this source is the \emph{Wisp} structure identified by \citet{li2013}, which is one of the largest velocity coherent molecular structures detected so far. The velocity dispersion of the clumps associated with each cluster are typically lower than a few \kms\ and in line with the velocity dispersions reported for Galactic and extragalactic GMCs (5-20\,\kms; \citealt{schneider2006, fukui2009}).

\section{SED: Photometry and fitting procedure}
\label{sect:append_sed}

\subsection{Photometry}\label{sect.photometry}

We use an aperture-and-annulus geometry to measure flux densities about each source.  We initially set a circular aperture to a spatial position that is optimal across all observed bands for the given source.  This aperture is then re-centred to within one ATLASGAL beam radius (i.e. $r_\mathrm{search}=19.2\arcsec$) of the peak emission found in the 250, 350 or 870\,$\mu$m band, using the first band where no saturation is found\footnote{Pixels were the detector suffered from saturation are found as NaN values in the reduced maps} within the source aperture. We set the aperture radius to twice the major axis reported in \citet{contreras2013}  and \citet{urquhart2014_csc} in order to ensure the aperture to be large enough that most of the source emission lies within the aperture, while still being small enough to avoid cutting into the background for more complex sources.  Saturated pixels present within the aperture are set to the maximum pixel value of the $5\times5$\,arcmin image and the flux is thereafter regarded only as a lower limit. 

The background is determined over an annulus about the aperture with inner and outer radii of 1.5 times and 2.5 times the aperture size, respectively. In the case that the annulus width is smaller than the ATLASGAL beam size, the outer annulus radius is increased to be at least 3 times the ATLASGAL pixel size (i.e.  18\arcsec) larger than the inner radius, to allow for a statistical analysis of the background annulus flux. Any pixels within the background annulus with a flux above the source aperture's peak-pixel flux are omitted, as we assume these pixels represent sources within the background aperture. We take the median background flux (rather than the mean) to reduce any influence of a fainter source within the  annulus. The background flux is then normalized to the area of the source aperture and subtracted from the latter, yielding the background-corrected source flux.

A negative background-corrected flux may be indicative of either the background flux being overestimated as the background annulus is cutting into a nearby source or of absorption in the source aperture. We take care of the former issue as described in the last paragraph by ignoring pixels above the maximum aperture flux and taking the median value of the background pixels, but the latter issue may result in negative flux measurements even when a point source is clearly visible within the aperture. To automatically identify a point source within the source aperture, we have empirically determined that a point source in an arbitrary group of pixels has at least 80\,per\,cent, 50\,per\,cent and 20\,per\,cent of the pixels above the $1\sigma$, $2\sigma$ and $3\sigma$ noise levels, respectively. In case a point source is identified in this way, we restrict the pixels taken into account for photometry to those with a value above the median background pixel level. Any pixels below this limit are likely to be caused by absorption within the aperture and are therefore neglected, resulting in a positive background-corrected flux. Applying this method allowed us to obtain 2727 fluxes over all bands, affecting a total of 1814 sources, successfully recovering the flux for a visible point source that otherwise would have been missed. 

The flux uncertainties are estimated from the pixel noise level of the image and a general flux measurement uncertainty added in quadrature. We assume a rather conservative flux measurement uncertainty, as we not only consider the absolute calibration error but also take into account the uncertainty involved in determining the source aperture (which is applied to all bands and might not always be optimal). We therefore assume a flux measurement uncertainty of 15\,per\,cent for ATLASGAL, 20\,per\,cent for the 70, 160, 250 and 350\,$\mu$m Herschel bands, and an uncertainty of 50\,per\,cent for the 500\,$\mu$m Herschel band due to the large pixel size. We assume a general measurement uncertainty of 30\,per\,cent for the mid-IR bands, although the 8\,$\mu$m MSX band earns an increased measurement uncertainty of 50\,per\,cent due to the possible influence of PAHs within the band. The pixel noise level is determined over the full $5\arcmin\times5\arcmin$ image, where pixels within a beam around local maxima are blanked. The noise level is then determined over the filtered image, where the $1\sigma$ level is determined from a Gaussian fit to the histogram of the remaining pixel fluxes.

\subsection{Overview of the fitting procedure}\label{sect.seds}

As in \citet{konig2017}, the measured SED is fit simultaneously by a combination of a greybody and a blackbody, where the greybody is fit to wavelengths upwards of 20\,$\mu$m, fitting the emission of the cold dust. The blackbody component is added to reflect the presence of a more evolved, optically-thick embedded hot component and is only added if at least 2 flux measurements are present downward of 70\,$\mu$m. The fitting is performed using a Levenberg-Marquardt least-squares minimization, allowing us to estimate the fitting parameter uncertainties from the covariance matrix as calculated by the algorithm. To reduce the number of free parameters, we keep the dust emissivity spectral index of the greybody fixed to 1.75, which is the mean value as calculated over all dust models from \citet{ossenkopf1994}, allowing a fit with fewer available data points, as well as allowing better comparison with the literature \citep[e.g.][]{thompson2004a, nguyen2011}.

We take a different approach for emission in the 70\,$\mu$m band in contrast to \citet{konig2017}. The emission in this band is generally interpreted as being contaminated with the emission of small grains \citep{compiegne2010a, compiegne2010b} in addition to having the previously addressed contributions from both the cold dust and a more evolved hot component within the clump \citep[e.g.][]{beuther2010}. For these reasons, we take the emission at 70\,$\mu$m as an upper limit in the fitting process. When no emission is found at the 70\,$\mu$m band, we use a conservative upper limit of 5 times the noise determined over an aperture including the background annulus or the point source sensitivity level as determined for the instrument (i.e. 21\,mJy\,beam$^{-1}$)\footnote{\url{herschel.esac.esa.int/Docs/PMODE/html/ch02s03.html}}, whichever is higher. Fitting the SEDs provides us with reliable measurements of the dust temperature of the cold component that characterises the emission from dust and the bolometric flux; these will be used in the following section together with source distances to derive the physical properties of the clumps.

\subsection{Consistency tests}

\begin{figure}
\centering
\includegraphics[width=0.49\textwidth, trim= 0 0 0 0]{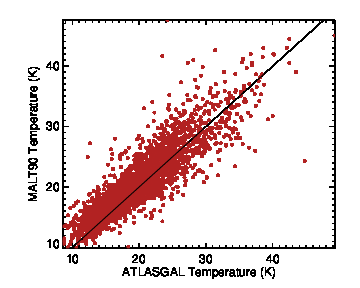}
\includegraphics[width=0.49\textwidth, trim= 0 0 0 0]{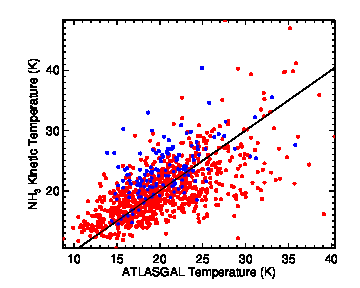}

\caption{Upper panel: comparison between dust temperatures derived here and those determined by the MALT90 team (\citealt{guzman2015}). This subsample of $\sim$2500 includes a significant fraction of all of the sources observed as part of the MALT90 survey (a total of $\sim$3000 ATLASGAL clumps were observed as part of this programme of follow-up observations). The solid line indicates the line of equality. Lower panel: comparison between the dust temperatures and kinematic gas temperatures determined for $\sim$1500 clumps from NH$_3$ (1,1) and (2,2) inversion transitions (\citealt{wienen2012,urquhart2011_nh3}, red and blue circles, respectively). Least-square fits to the data presented in both plots result in slopes that are within 1$\sigma$ of being linear. \label{fig:bolo_temp_comparison} } 

\end{figure}

In the upper panel of Fig.\,\ref{fig:bolo_temp_comparison} we compare the temperatures derived from the greybody fits with those derived by a recent study by \citet{guzman2015} who performed a similar method of aperture photometry on a sample of $\sim$2500 ATLASGAL clumps observed as part of the MALT90 survey (\citealt{jackson2013}). This plot illustrates the excellent agreement between the greybody temperatures determined from our work and the work reported by \citet{guzman2015} to estimate the fluxes and to fit the SEDs. The mean difference between the two sets of temperature measurements is $0.8\pm0.1$\,K with a standard deviation of 2.8\,K. In the lower panel of Fig.\,\ref{fig:bolo_temp_comparison} we show the correlation between the kinetic dust temperatures derived from the ammonia (NH$_3$) (1,1) and (2,2) transitions (\citealt{urquhart2011_nh3,wienen2012}). This plot also shows a strong correlation between the temperature of the gas and the dust (the Spearman rank coefficient $r$ is 0.64 with a $p$-value $\ll$ 0.01), although we note the scatter is clearly more significant (the mean difference is 0.7$\pm$0.14\,K with a standard deviation is 4.8\,K). 

The bolometric fits derived here are consistent with the results of a similar analysis of dust emission reported by \citet{guzman2015} and with independently estimated kinetic temperatures of the dense molecular gas. The strong agreement with these two previously published studies provides confidence in our fitting method and the derived results.

\begin{figure}
\centering
\includegraphics[width=0.49\textwidth, trim= 0 0 0 0]{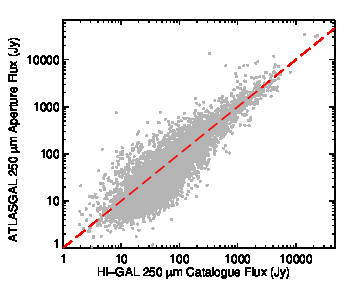}
\caption{Comparison of flux densities obtained from the Hi-GAL  compact source catalogue with the fluxes obtained through aperture photometry in the present work. The dashed red line indicates equality between the two sets of measurements.}
\label{fig.higal_agal_fluxes}
\end{figure}

Finally, we also compared the fluxes derived in the present work with those found in the Hi-GAL compact-source catalogue for the inner Galaxy \citep{molinari2016}. To obtain the fluxes from the Hi-GAL CSC, we summed up the fluxes of all compact sources found within our source aperture. The result can be seen in Fig.\,\ref{fig.higal_agal_fluxes}, for the 250-$\mu$m band. Although there is a significant amount of scatter for individual sources ($\sim$10), a strong correlation between the fluxes obtained with the two significantly different methods is found ($r = 0.86$ with a $p$-value $< 0.0013$), showing the general consistency between the two photometric approaches. 

In total, we were able to successfully fit the SEDs of 7861 sources ($\sim$98\% of the full CSC sample). 2586 of these sources ($\sim$33\%) are fit with a single-component greybody, and 5275 sources ($\sim$67\%) with the two-component model.


\bsp	
\label{lastpage}
\end{document}